\documentclass[11pt]{myArticle}       

\usepackage{amsmath,amssymb,amsfonts} 
\usepackage{graphicx}
\usepackage{epsfig}
\usepackage{color}                    
\usepackage{hyperref}                 
\usepackage{myFancyhdr}               
\usepackage{makeidx}                  
\usepackage{helvet}                   
\usepackage{setspace}                 
\usepackage{fancybox}                 
\usepackage{float}                    
\usepackage{xspace}                   
\usepackage{wasysym}                  
\usepackage[numbers,sort&compress]{natbib}
\usepackage{hypernat}
\definecolor{mygray}{rgb}{0.3,0.32,0.35}
\definecolor{darkblue1}{rgb}{0,0,.2}
\definecolor{darkblue}{rgb}{0,0,.3}
\definecolor{darkred}{rgb}{0.5,0,0}
\pagecolor{white} 
\color{black}     
\hypersetup{breaklinks=true, 
            colorlinks=true, 
            linkcolor=darkblue1, 
            menucolor=darkblue1, 
            urlcolor=darkblue1,
            citecolor=darkblue1,
            pdftitle={The SM EW fit Revisited},
            pdfauthor={Gfitter Group},
            pdfsubject={The SM EW fit Revisited},
            pdfkeywords={},
            pdfproducer={Gfitter Group}
}
\newcommand\defaultFigureScale{0.6}
\bibstyle{plain}
\renewcommand{\headrulewidth}{0.5pt}    
\renewcommand{\footrulewidth}{0.5pt}    

\newcommand\allFontSize{\small}

\newenvironment{myquote}
               {\list{}{\leftmargin0cm}%
                \item\relax}
               {\endlist}

\usepackage[bf]{caption}
\renewcommand{\captionfont}{\allFontSize}

\newcommand\detailsSize{\allFontSize}
\newenvironment{details}%
{\begin{myquote}\vspace{-0.2cm}\detailsSize}{\end{myquote}\vspace{-0.2cm}}

\newfloat{codeexample}{H}{loc}
\floatname{codeexample}{\sf\allFontSize Code Example}

\newlength{\gfitterboxwidth}
\definecolor{DarkGray}{rgb}{0.4,0.42,0.45}
\definecolor{LightGray}{rgb}{0.97,0.98,0.98}

{\VerbatimEnvironment\normalsize
\setlength{\gfitterboxwidth}{\textwidth}\addtolength{\gfitterboxwidth}{-2.3\fboxsep}%
\begin{Sbox}\begin{minipage}[t]{\gfitterboxwidth}\begin{Verbatim}}%
{\end{Verbatim}\end{minipage}\end{Sbox}\vspace{1ex} \fcolorbox{DarkGray}{LightGray}{\TheSbox}}

\newfloat{option}{H}{loc}
\floatname{option}{\sf\allFontSize Option Table}

\usepackage{tabularx}
{\setlength{\gfitterboxwidth}{\textwidth}\addtolength{\gfitterboxwidth}{-2.3\fboxsep}%
\begin{Sbox}\begin{tabular*}{\gfitterboxwidth}{>{\rule[-0.5ex]{.0ex}{3.0ex}\tt\small}p{3cm}>{\tt\small}p{4.5cm}>{\small}p{5.7cm}}
&&\\[-0.7cm]
\rm\small Option  & \rm\small Values    & \rm\small Description\\[0.1cm]\hline
&&\\[-0.45cm]}%
{\\[-0.1cm]\end{tabular*}\end{Sbox}\vspace{1ex} \fbox{\TheSbox}}

{\setlength{\gfitterboxwidth}{\textwidth}\addtolength{\gfitterboxwidth}{-2.3\fboxsep}%
\begin{Sbox}\begin{tabular*}{\gfitterboxwidth}{>{\rule[-0.5ex]{.0ex}{3.0ex}\tt\small}p{3cm}>{\tt\small}p{3.5cm}>{\small}p{6.7cm}}
&&\\[-0.7cm]
\rm\small Option  & \rm\small Values    & \rm\small Description\\[0.1cm]\hline
&&\\[-0.45cm]}%
{\\[-0.1cm]\end{tabular*}\end{Sbox}\vspace{1ex} \fbox{\TheSbox}}

{\setlength{\gfitterboxwidth}{\textwidth}\addtolength{\gfitterboxwidth}{-2.3\fboxsep}%
\begin{Sbox}\begin{tabular*}{\gfitterboxwidth}{>{\rule[-0.5ex]{.0ex}{3.5ex}\tt\small}p{3.0cm}>{\tt\small}p{3.0cm}>{\small}p{7.2cm}}
&&\\[-0.7cm]
\rm\small Option  & \rm\small Values    & \rm\small Description\\[0.1cm]\hline
&&\\[-0.45cm]}%
{\\[-0.1cm]\end{tabular*}\end{Sbox}\vspace{1ex} \fbox{\TheSbox}}

{\setlength{\gfitterboxwidth}{\textwidth}\addtolength{\gfitterboxwidth}{-2.3\fboxsep}%
\begin{Sbox}\begin{tabular*}{\gfitterboxwidth}{>{\rule[-0.5ex]{.0ex}{3.0ex}\tt\small}p{2.5cm}>{\tt\small}p{3.5cm}>{\small}p{7.2cm}}
&&\\[-0.7cm]
\rm\small Option  & \rm\small Values    & \rm\small Description\\[0.1cm]\hline
&&\\[-0.45cm]}%
{\\[-0.1cm]\end{tabular*}\end{Sbox}\vspace{1ex} \fbox{\TheSbox}}

{\setlength{\gfitterboxwidth}{\textwidth}\addtolength{\gfitterboxwidth}{-2\fboxsep}%
\begin{Sbox}\begin{tabular*}{\gfitterboxwidth}{>{\rule[-0.5ex]{.0ex}{3.0ex}\tt\small}p{4.5cm}>{\tt\small}p{3.2cm}>{\small}p{5.5cm}}
&&\\[-0.7cm]
\rm\small Option  & \rm\small Values    & \rm\small Description\\[0.1cm]\hline
&&\\[-0.45cm]}%
{\\[-0.1cm]\end{tabular*}\end{Sbox}\vspace{1ex} \fbox{\TheSbox}}

{\setlength{\gfitterboxwidth}{\textwidth}\addtolength{\gfitterboxwidth}{-2\fboxsep}%
\begin{Sbox}\begin{tabular*}{\gfitterboxwidth}{>{\rule[-0.5ex]{.0ex}{3.0ex}\tt\small}p{4.0cm}>{\tt\small}p{3.0cm}>{\small}p{6.2cm}}
&&\\[-0.7cm]
\rm\small Option  & \rm\small Values    & \rm\small Description\\[0.1cm]\hline
&&\\[-0.45cm]}%
{\\[-0.1cm]\end{tabular*}\end{Sbox}\vspace{1ex} \fbox{\TheSbox}}

\newfloat{programs}{H}{loc}
\floatname{programs}{\sf Table}

\usepackage{tabularx}
{\setlength{\gfitterboxwidth}{\textwidth}\addtolength{\gfitterboxwidth}{-2\fboxsep}%
\begin{Sbox}\begin{tabular*}{\gfitterboxwidth}{>{\rule[-0.5ex]{.0ex}{3.0ex}\tt\small}p{4.1cm}>{\small}p{9.5cm}}
&\\[-0.7cm]
\rm\small Macro & \rm\small Description\\[0.1cm]\hline
&\\[-0.45cm]}%
{\\[-0.1cm]\end{tabular*}\end{Sbox}\vspace{1ex} \fbox{\TheSbox}}

\newcommand\Gfitter{Gfitter\xspace}

\newcommand{\qbar}  {\ensuremath{\overline q}\xspace}

\newcommand{\cbar}  {\ensuremath{\overline c}\xspace}

\newcommand{\bbar}  {\ensuremath{\overline b}\xspace}

\newcommand{\fbar} {\ensuremath{\overline f}\xspace}
\mathchardef\Upsilon="7107
\def\Y#1S{\ensuremath{\Upsilon{(#1S)}}\xspace}

\newcommand{\mt}{\ensuremath{m_{t}}\xspace}
\newcommand{\as}{\ensuremath{\alpha_{\scriptscriptstyle S}}\xspace}
\newcommand{\aas}{\ensuremath{a_{\scriptscriptstyle S}}\xspace}

\newcommand{\asZ}{\ensuremath{\as(M_Z^2)}\xspace}
\newcommand{\MSb}{\ensuremath{\overline{\mathrm{MS}}}\xspace}

\newcommand{\NNNLO}{\ensuremath{\rm 3NLO}\xspace}

\newcommand{\ass}{\ensuremath{\as(s)}\xspace}

\renewcommand\l{\ell}

\newcommand{\tw}{\ensuremath{\theta_{\scriptscriptstyle W}}\xspace}
\newcommand{\sw}{\ensuremath{s^2_{\scriptscriptstyle W}}\xspace}
\newcommand{\cw}{\ensuremath{c^2_{\scriptscriptstyle W}}\xspace}
\newcommand{\tzw}{\ensuremath{\theta_{\scriptscriptstyle W}}\xspace}

\newcommand{\Afbz}[1]{{\ensuremath{A_{\rm\scriptscriptstyle FB}^{0,#1}}}\xspace}
\newcommand{\gv}[1]{{\ensuremath{g_{{\scriptscriptstyle V},#1}}}\xspace}
\newcommand{\ga}[1]{{\ensuremath{g_{{\scriptscriptstyle A},#1}}}\xspace}

\newcommand{\gvs}[1]{{\ensuremath{g^2_{{\scriptscriptstyle V},#1}}}\xspace}
\newcommand{\gas}[1]{{\ensuremath{g^2_{{\scriptscriptstyle A},#1}}}\xspace}
\newcommand{\gls}[1]{{\ensuremath{g^2_{{\scriptscriptstyle L},#1}}}\xspace}
\newcommand{\grs}[1]{{\ensuremath{g^2_{{\scriptscriptstyle R},#1}}}\xspace}
\newcommand{\gvz}[1]{{\ensuremath{g_{{\scriptscriptstyle V},#1}^{(0)}}}\xspace}
\newcommand{\gaz}[1]{{\ensuremath{g_{{\scriptscriptstyle A},#1}^{(0)}}}\xspace}
\newcommand{\glz}[1]{{\ensuremath{g_{{\scriptscriptstyle L},#1}^{(0)}}}\xspace}
\newcommand{\grz}[1]{{\ensuremath{g_{{\scriptscriptstyle R},#1}^{(0)}}}\xspace}
\newcommand{\glrz}[1]{{\ensuremath{g_{{\scriptscriptstyle L(R)},#1}^{(0)}}}\xspace}

\newcommand{\GF}{{\ensuremath{G_{\scriptscriptstyle F}}}\xspace}
\newcommand{\GFe}[1]{{\ensuremath{G_{\scriptscriptstyle F}^{#1}}}\xspace}

\newcommand{\NCf}[1]{{\ensuremath{N_{\scriptscriptstyle C}^{#1}}}\xspace}

\newcommand{\Kbar    }{\kern 0.2em\overline{\kern -0.2em K}{}\xspace}

\newcommand{\Kz      }{\ensuremath{K^0}\xspace}
\newcommand{\Kzb     }{\ensuremath{\Kbar^0}\xspace}
\newcommand{\KzKzb   }{\ensuremath{\Kz \kern -0.16em \Kzb}\xspace}
\newcommand{\Kp      }{\ensuremath{K^+}\xspace}
\newcommand{\Km      }{\ensuremath{K^-}\xspace}

\newcommand{\KpKm    }{\ensuremath{\Kp \kern -0.16em \Km}\xspace}

\newcommand\Dbar    {\kern 0.18em\overline{\kern -0.18em D}{}\xspace}

\newcommand\Bbar    {\kern 0.18em\overline{\kern -0.18em B}{}\xspace}

\newcommand\Bz      {\ensuremath{B^0}\xspace}

\newcommand\Bzb     {\ensuremath{\Bbar^0}\xspace}
\newcommand\Bu      {\ensuremath{B^+}\xspace}
\newcommand\Bub     {\ensuremath{B^-}\xspace}

\newcommand\BpBm    {\ensuremath{\Bu {\kern -0.16em \Bub}}\xspace}
\newcommand\Bs      {\ensuremath{B^0_{s}}\xspace}
\newcommand\Bsb     {\ensuremath{\Bbar^0_{s}}\xspace}

\newcommand\BzBzb   {\ensuremath{\Bz {\kern -0.16em \Bzb}}\xspace}
\newcommand\BszBszb {\ensuremath{\Bs {\kern -0.16em \Bsb}}\xspace}

\newcommand\invab{\ensuremath{\;{\rm ab}^{-1}}\xspace}
\newcommand\invfb{\ensuremath{\;{\rm fb}^{-1}}\xspace}

\newcommand\deltatheo{\ensuremath{\delta_{\rm th}}\xspace}
\newcommand\Dim{{\rm dim}\xspace}
\newcommand\Rfit{{\em R}fit\xspace}
\newcommand{\CP}{\ensuremath{C\!P}\xspace}
\newcommand{\ft}{\footnotesize}
\newcommand{\multic}{\multicolumn}

\newcommand{\Tau}{\ensuremath{\tau}\xspace}

\newcommand{\Order}{\ensuremath{{\cal O}}\xspace}
\newcommand{\BR}{\ensuremath{{\mathcal B}}\xspace}

\newcommand{\ee}{\ensuremath{e^+e^-}\xspace}

\newcommand{\tev}{\ensuremath{\mathrm{\;Te\kern -0.1em V}}\xspace}
\newcommand{\gev}{\ensuremath{\mathrm{\;Ge\kern -0.1em V}}\xspace}
\newcommand{\mev}{\ensuremath{\mathrm{\;Me\kern -0.1em V}}\xspace}
\newcommand{\kev}{\ensuremath{\mathrm{\;ke\kern -0.1em V}}\xspace}
\newcommand{\ev}{\ensuremath{\mathrm{\,e\kern -0.1em V}}\xspace}
\newcommand{\gevc}{\ensuremath{{\mathrm{\,Ge\kern -0.1em V\!/}c}}\xspace}
\newcommand{\mevc}{\ensuremath{{\mathrm{\,Me\kern -0.1em V\!/}c}}\xspace}
\newcommand{\gevcc}{\ensuremath{{\mathrm{\,Ge\kern -0.1em V\!/}c^2}}\xspace}
\newcommand{\mevcc}{\ensuremath{{\mathrm{\,Me\kern -0.1em V\!/}c^2}}\xspace}

\newcommand{\bei}{\begin{itemize}}
\newcommand{\eei}{\end{itemize}}
\newcommand{\beq}{\begin{equation}}
\newcommand{\eeq}{\end{equation}}
\newcommand{\beqn}{\begin{eqnarray}}
\newcommand{\eeqn}{\end{eqnarray}}
\newcommand{\beqns}{\begin{eqnarray*}}
\newcommand{\eeqns}{\end{eqnarray*}}
\newcommand{\bitm}{\begin{itemize}}
\newcommand{\eitm}{\end{itemize}}

\renewcommand{\arraystretch}{1.25}

\newcommand{\alphaMZ}{\ensuremath{\alpha(M_Z^2)}\xspace}
\newcommand{\dalphaHadMZ}{\ensuremath{\Delta\alpha_{\rm had}^{(5)}(M_Z^2)}\xspace}
\newcommand{\dalphaHads}{\ensuremath{\Delta\alpha_{\rm had}^{(5)}(s)}\xspace}
\newcommand{\dalphaTops}{\ensuremath{\Delta\alpha_{\rm top}(s)}\xspace}
\newcommand{\dalphaTopMZ}{\ensuremath{\Delta\alpha_{\rm top}(M_Z^2)}\xspace}
\newcommand{\dalphaLeps}{\ensuremath{\Delta\alpha_{\rm lep}(s)}\xspace}

\newcommand\ie{{\it i.e.}\xspace} 
\newcommand\eg{{\it e.g.}\xspace}

\newcommand\cf{{\em cf.}\xspace}

\newcommand\rs{\raisebox{1.5ex}[-1.5ex]}

\newcommand{\rar}{\rightarrow}
\def\@citex[#1]#2{\if@filesw\immediate\write\@auxout{\string\citation{#2}}\fi
  \@tempcnta\z@\@tempcntb\m@ne\def\@citea{}\@cite{\@for\@citeb:=#2\do
    {\@ifundefined
       {b@\@citeb}{\@citeo\@tempcntb\m@ne\@citea
        \def\@citea{,\penalty\@m\ }{\bf ?}\@warning
       {Citation `\@citeb' on page \thepage \space undefined}}%
    {\setbox\z@\hbox{\global\@tempcntc0\csname b@\@citeb\endcsname\relax}%
     \ifnum\@tempcntc=\z@ \@citeo\@tempcntb\m@ne
       \@citea\def\@citea{,\penalty\@m}
       \hbox{\csname b@\@citeb\endcsname}%
     \else
      \advance\@tempcntb\@ne
      \ifnum\@tempcntb=\@tempcntc
      \else\advance\@tempcntb\m@ne\@citeo
      \@tempcnta\@tempcntc\@tempcntb\@tempcntc\fi\fi}}\@citeo}{#1}}

\def\@citeo{\ifnum\@tempcnta>\@tempcntb\else\@citea
  \def\@citea{,\penalty\@m}%
  \ifnum\@tempcnta=\@tempcntb\the\@tempcnta\else
   {\advance\@tempcnta\@ne\ifnum\@tempcnta=\@tempcntb \else
\def\@citea{--}\fi
    \advance\@tempcnta\m@ne\the\@tempcnta\@citea\the\@tempcntb}\fi\fi}

\xspace

\newcommand\rhobar{\ensuremath{\overline \rho}\xspace}
\newcommand\etabar{\ensuremath{\overline \eta}\xspace}

\newcommand\CL{{\rm CL}}

\newcommand\tho{{\rm theo}}
\newcommand\experi{{\rm exp}}
\def\mod{{\rm mod}}
\newcommand\mini{{\rm min}}

\newcommand\xexp{x_{\experi}}
\newcommand\xthe{x_{\tho}}
\newcommand\ymod{y_{\mod}}
\newcommand\yguessmod{{\overline y}_{\mod}}
\newcommand\yhatmod{\hat y_{\mod}}

\newcommand\yNP{y_{\rm NP}}

\newcommand\Lik{{\cal L}}

\newcommand\ndof{n_{\rm dof}}

\newcommand\Likexp{{\cal L}_{\experi}}
\newcommand\Likthe{{\cal L}_{\tho}}

\def\a{a}
\newcommand\Mu{\mu}
\newcommand\Muhat{{\hat\mu}}

\newcommand\ChiMinGlob{\ensuremath{\chi^2_{\mini ;\yhatmod}}\xspace}
\newcommand\ChiMin{\ensuremath{\chi^2_{\mini}}\xspace}
\newcommand\DeltaChi{\ensuremath{\Delta\chi^2}\xspace}
\newcommand\Prob{{\cal P}}
\newcommand\ProbCERN{{\rm Prob}}

\newcommand\Nmod{N_{\mod}}

\newcommand\Nexp{N_{\experi}}

\newcommand\minchitwo{\ensuremath{\chi^2_{\rm min}}\xspace}

\newcommand{\Dalphahad}{\dalphaHadMZ}

\def\Re{\ensuremath{{\rm Re}}\xspace}

\newcommand{\drhov}{\delta\rho}
\newcommand{\dkapv}{\delta\kappa}

\newcommand{\drhovb}{\delta{\hat{\rho}}}

\newcommand{\seffsf}[1]{\sin\!^2\theta^{#1}_{{\rm eff}}}

\newcommand{\rZ}[1]{\rho^{#1}_{Z}}

\newcommand{\kZ}[1]{\kappa^{#1}_{Z}}

\newcommand{\sinfeff}{\sin\!^2\theta^f_{{\rm eff}}}
\newcommand{\sinleff}{\seffsf{\ell}}
\newcommand{\sinbeff}{\sin\!^2\theta^b_{{\rm eff}}}

\newcommand{\mzs}{M_Z^2}
\newcommand{\zb}{Z}
\newcommand{\mws}{M_W^2}
\newcommand{\wb}{W}

\newcommand{\mts }{m^2_t}
\newcommand{\sss}[1]{\scriptscriptstyle{#1}}
\newcommand{\mq }{\ensuremath{\overline{m}_q  }\xspace}
\newcommand{\mqP}{\ensuremath{\overline{m}_q^\prime}}
\newcommand{\mqS}{\ensuremath{\overline{m}^2_q}\xspace}

\newcommand{\mc }{\ensuremath{\overline{m}_c}\xspace}
\newcommand{\mcS}{\ensuremath{\overline{m}^2_c}\xspace}
\newcommand{\mcQ}{\ensuremath{\overline{m}^4_c}\xspace}
\newcommand{\mb }{\ensuremath{\overline{m}_b  }\xspace}
\newcommand{\mbS}{\ensuremath{\overline{m}^2_b}\xspace}
\newcommand{\mbQ}{\ensuremath{\overline{m}^4_b}\xspace}
\newcommand{\mqQ}{\ensuremath{\overline{m}^4_q}\xspace}
\newcommand{\mqX}{\ensuremath{\overline{m}^6_q}\xspace}

\newcommand{\mqpQ}{\ensuremath{\overline{m}^{\prime 4}_q}\xspace}
\newcommand{\ztwo}{\zeta(2)}
\newcommand{\ztri}{\zeta(3)}
\newcommand{\zfor}{\zeta(4)}
\newcommand{\zfiv}{\zeta(5)}
\newcommand{\nf}{n_f}
\newcommand{\stws}{s^2_{\sss{W}}}
\newcommand{\stwf}{s^4_{\sss{W}}}
\newcommand{\ctws}{c^2_{\sss{W}}}
\newcommand{\ctwf}{c^4_{\sss{W}}}
\newcommand{\ctwsix}{c^6_{\sss{W}}}
\newcommand{\delrho}[1]{{\Delta \rho}^{#1}}
\newcommand{\zmz}{\Sigma^{\prime}_{_{\zb\zb}}(\mzs)}

\newcommand{\Pzg}{\Pi_{Z\gamma}}

\newcommand{\btoxsg}{\ensuremath{B \rar X_s \gamma}\xspace}
\newcommand{\btotaunu}{\ensuremath{B \rar \tau \nu}\xspace}
\newcommand{\btomunu}{\ensuremath{B \rar \mu \nu}\xspace}
\newcommand{\btodtaunu}{\ensuremath{B \rar D \tau \nu}\xspace}
\newcommand{\btodenu}{\ensuremath{B \rar D e \nu}\xspace}
\newcommand{\ktomunu}{\ensuremath{K \rar \mu \nu}\xspace}
\newcommand{\pitomunu}{\ensuremath{\pi \rar \mu \nu}\xspace}

\newcommand{\Ctaunp}{\ensuremath{C^{\tau}_{\rm NP}}\xspace}

\newcommand{\MHp}{\ensuremath{M_{H^{\pm}}}\xspace}
\newcommand{\tanb}{\ensuremath{\tan\!\beta}\xspace}
\newcommand{\cotb}{\ensuremath{\cot\!\beta}\xspace}

\makeindex
\begin{document}
%
%
{\small
\color{mygray}
\begin{flushright}
{\sf\em arXiv:0811.0009} \\
{\sf\em CERN-OPEN-2008-024} \\
{\sf\em DESY-08-160} \\
{\sf\em November 3, 2008} \\
\def\UrlFont{\sf\em}
\url{http://cern.ch/gfitter} 
\end{flushright}
}
\def\UrlFont{\rm}

\vspace{1.3cm}

{\sf\LARGE\bfseries
  Revisiting the Global Electroweak Fit of the \\[0.2cm]
  Standard Model and Beyond with Gfitter
}

\vspace{1.0cm}

{\Large \em 
  The Gfitter Group \\[0.2cm]
}
{\Large
  H.~Fl\"acher$^{a}$, M.~Goebel$^{b,c}$, J.~Haller$^{c}$, 
  A.~Hoecker$^{a}$, K.~M\"onig$^{b}$, J.~Stelzer$^{b}$
}

\vspace{0.5cm}

{\normalsize
  $^{a}$CERN, Geneva, Switzerland \\
  $^{b}$DESY, Hamburg and Zeuthen, Germany \\ 
  $^{c}$Institut f\"ur Experimentalphysik, Universit\"at Hamburg, Hamburg, Germany

\vspace{1.0cm}

\begin{details}
{\sf\bfseries Abstract} ---
The global fit of the Standard Model to electroweak precision data, routinely performed
by the LEP electroweak working group and others, demonstrated impressively the predictive
power of electroweak unification and quantum loop corrections. We have revisited 
this fit in view of $(i)$ the development of the new generic fitting package, {\em \Gfitter}, 
allowing flexible and efficient model testing in high-energy physics, $(ii)$ the 
insertion of constraints from direct Higgs searches at LEP and the Tevatron, and $(iii)$ 
a more thorough statistical interpretation of the results. \Gfitter is a modular fitting 
toolkit, which features predictive theoretical models as independent plugins, and a 
statistical analysis of the fit results using toy Monte Carlo techniques. The state-of-the-art 
electroweak Standard Model is fully implemented, as well as generic extensions to it. 
Theoretical uncertainties are explicitly included in the fit through scale parameters 
varying within given error ranges. 

This paper introduces the \Gfitter project, and presents 
state-of-the-art results for the global electroweak fit in the Standard Model (SM), and for a 
model with an extended Higgs sector (2HDM). Numerical and graphical results for fits with and without 
including the constraints from the direct Higgs searches at LEP and Tevatron are given. 
Perspectives for future colliders are analysed and discussed.

In the SM fit including the direct Higgs searches, we 
find $M_H\ =\ 116.4^{\,+18.3}_{\,-1.3}\gev$, and the $2\sigma$ 
and $3\sigma$ allowed regions $[114,\,145]\gev$ and $[[113,\,168]\,{\rm and}\,[180,\,225]]\gev$, respectively. 
For the strong coupling strength at fourth perturbative order we obtain 
$\alpha_S(M_Z^2)=0.1193^{\,+0.0028}_{\,-0.0027} (\rm exp) \pm 0.0001 ({\rm theo})$.
Finally, for the mass of 
the top quark, excluding the direct measurements, we find $\mt=178.2^{\,+9.8}_{\,-4.2}\gev$.
In the 2HDM we exclude a charged-Higgs mass below $240\gev$ 
at 95\% confidence level. This limit increases towards larger \tanb, \eg, 
$\MHp<780\gev$ is excluded for $\tanb=70$.
\end{details}

\vfill

\thispagestyle{empty}
\newpage

%
%
\pagenumbering{roman}

\setcounter{tocdepth}{3}
\addtolength{\parskip}{-0.40\baselineskip}
{\footnotesize\sl
\tableofcontents
}
\addtolength{\parskip}{+0.34\baselineskip}

\newpage

\pagenumbering{arabic}
%
%
\section{Introduction}
\label{sec:introduction}

Precision measurements allow us to probe physics at much higher energy scales than the 
masses of the particles directly involved in experimental reactions by exploiting 
contributions from quantum loops. These tests do not only 
require accurate and well understood experimental data but also theoretical predictions 
with controlled uncertainties that match the experimental precision. Prominent examples 
are the LEP precision measurements, which were used in conjunction with the 
Standard Model (SM) to predict via multidimensional parameter fits the mass of the top 
quark~\cite{Alexander:1991vi}, prior to its discovery at the Tevatron~\cite{Abachi:1995iq,Abe:1995hr}. 
Later, when combined with the measured top mass, the same approach led to the prediction 
of a light Higgs boson~\cite{:1994qa}. Other examples are fits to constrain parameters of 
Supersymmetric or extended Higgs models, using as inputs the anomalous magnetic 
moment of the muon, results on neutral-meson mixing, \CP violation, rare loop-induced decays 
of $B$ and $K$ mesons, and the relic matter density of the universe determined from 
fits of cosmological models to data.

Several theoretical libraries within and beyond the SM have been developed in the past,
which, tied to a multi-parameter minimisation program, allowed to constrain the unbound 
parameters of the SM~\cite{Arbuzov:2005ma,Bardin:1999yd,Montagna:1998kp,Erler:2000cr}.
However, most of these programs are relatively old, were implemented in outdated programming 
languages, and are difficult to maintain in line with the theoretical and experimental progress. 
It is unsatisfactory to rely on them during the forthcoming era of the Large Hadron Collider (LHC)
and the preparations for future linear collider projects. Improved measurements of 
important input observables are expected and new observables from discoveries may augment 
the available constraints. None of the previous programs were modular 
enough to easily allow the theoretical predictions to be extended to models beyond the SM,
and they are usually tied to a particular minimisation package. 

These considerations led to the development of the generic fitting package {\em Gfitter}~\cite{web}, 
designed to provide a framework for model testing in high-energy physics. Gfitter is implemented 
in C++ and relies on ROOT~\cite{Brun:1997pa} functionality. Theoretical models are inserted
as plugin packages, which may be hierarchically organised. Tools for the handling of the data, 
the fitting, and statistical analyses such as toy Monte Carlo sampling are provided by a core 
package, where theoretical errors, correlations, and inter-parameter dependencies are consistently 
dealt with. The use of dynamic parameter caching avoids the recalculation of unchanged
results between fit steps, and thus significantly reduces the amount of computing 
time required for a fit. 

The first theoretical framework implemented in \Gfitter has been the SM predictions for the
electroweak precision observables measured by the LEP, SLC, and the Tevatron experiments.
State-of-the-art calculations have been used, and -- wherever possible -- the results have 
been cross-checked against the ZFITTER package~\cite{Arbuzov:2005ma}. For the $W$ mass 
and the effective weak mixing angle, which exhibit the strongest constraints on the Higgs 
mass through radiative corrections, the full second order corrections are 
available~\cite{Awramik:2003rn,Awramik:2004ge,Awramik:2006uz}. Furthermore, the corrections of order 
${\cal O}(\alpha \as^2)$ and the leading three-loop corrections in an expansion of the 
top-mass-squared ($\mt^2$) are included. 
The full three-loop corrections are known in the large $M_H$ limit,
however they turn out to be negligibly small \cite{Boughezal:2004ef,Boughezal:2005eb}.

The calculations of the partial and total widths of the $Z$ and of the total width of 
the $W$ boson have been integrated from the ZFITTER 
package~\cite{Arbuzov:2005ma,Bardin:1999yd} into the Gfitter subpackage GSM 
and are co-authored by both groups~\cite{zfittergfitter}.\footnote
{
  Usage of the Gfitter subpackage GSM should include a citation of the 
  ZFITTER package~\cite{Arbuzov:2005ma,Bardin:1999yd}.\label{fn}
} It includes up to two-loop electroweak
corrections~\cite{Akhundov:1985fc,Arbuzov:2005ma,Bardin:1986fi,Barbieri:1992dq,Fleischer:1993ub,Bardin:1999yd,Degrassi:1994tf,Degrassi:1995mc,Degrassi:1996mg,Degrassi:1999jd,Bardin:1997xq,Bardin:1999ak} 
and all known QCD corrections~\cite{Arbuzov:2005ma,Bardin:1999yd,Kniehl:1989yc}.
Among the new developments included in the SM library is 
the fourth-order (\NNNLO) perturbative calculation of the massless QCD Adler 
function~\cite{Baikov:2008jh}, contributing to the vector and axial-vector radiator functions 
in the prediction of the $Z$ hadronic width (and other observables). It allows to fit the 
strong coupling constant with unique theoretical accuracy~\cite{Baikov:2008jh,Davier:2008sk}. 

Among the experimental precision data used are the $Z$ mass, measured with relative 
precisions of $2 \cdot 10^{-5}$, the hadronic pole cross section at the $Z$ mass and 
the leptonic decay width ratio of the $Z$ with $10^{-3}$ relative precision. The effective 
weak mixing angle $\sinleff$ is known from the LEP experiments and SLD to a relative precision 
of $7 \cdot 10^{-4}$.  The $W$ mass has been measured at LEP and the Tevatron to an overall 
relative precision of $3 \cdot 10^{-4}$. The mass of the top quark occurs quadratically 
in loop corrections of many observables. A precision measurement (currently $7\cdot10^{-3}$) 
is mandatory. Also required is the precise knowledge of the electromagnetic and weak coupling 
strengths at the appropriate scales. Energy-dependent photon vacuum polarisation contributions 
modify the QED fine structure constant, which at the $Z$-mass scale has been evaluated to a 
relative precision of $8 \cdot 10^{-3}$. The Fermi constant, parametrising the weak coupling
strength, is known to $10^{-5}$ relative precision.

We perform global fits in two versions: the {\em standard (``blue-band'') fit} makes
use of all the available information except for the direct Higgs searches performed at LEP 
and the Tevatron; the {\it complete fit} uses also the constraints from the direct Higgs 
searches. Results in this paper are commonly derived for both types of fits.

Several improvements are expected from the LHC~\cite{atlastdr,CMSTDR}. The uncertainty on the
$W$-boson and the top-quark masses should shrink to $1.8\cdot10^{-4}$ and $5.8\cdot10^{-3}$ respectively. 
In addition, the Higgs boson should be discovered leaving the SM without an unmeasured parameter (excluding
here the massive neutrino sector, requiring at least nine additional parameters, which are 
however irrelevant for the results discussed in this paper). The primary focus of the global 
SM fit would then move from parameter estimation to the analysis of the goodness-of-fit with 
the goal to uncover inconsistencies between the model and the data, indicating the presence 
of new physics. Because the Higgs-boson mass enters only logarithmically in the loop 
corrections, a precision measurement is not required for this purpose. 
Dramatic improvements on SM observables are expected from the 
ILC~\cite{Djouadi:2007ik}. The top and Higgs masses may be measured to a relative 
precision of about $1\cdot 10^{-3}$, corresponding to absolute uncertainties of $0.2\gev$ and 
$50\mev$, respectively. Running at lower energy with polarised beams, the $W$ mass could be 
determined to better than $7\cdot10^{-5}$ relative accuracy, and the weak mixing angle to a relative 
precision of $5 \cdot 10^{-5}$. Moreover, new precision measurements would enter the 
fit, namely the two-fermion cross section at higher energies and the triple gauge couplings 
of the electroweak gauge bosons, which are sensitive to models beyond the SM.
Most importantly, however, both machines are directly sensitive to new phenomena
and thus either provide additional constraints on fits of new physics models or -- if the 
searches are successful -- may completely alter our view of the physics at the terascale. 
The SM will then require extensions, the new parameters of which must be determined by a 
global fit, whose goodness must also be probed. 
To study the impact of the expected experimental improvements on the SM parameter
determination, we perform fits under the assumption of various prospective setups (LHC, ILC, 
and ILC with GigaZ option). 

As an example for a study beyond the SM we investigate models with an extended Higgs sector 
of two doublets (2HDM). We constrain the mass of the charged Higgs and the ratio of the 
vacuum expectation values of the two Higgs doublets using current measurements 
of observables from the $B$ and $K$ physics sectors and the most recent theoretical 2HDM 
predictions.

The paper is organised as follows. A disquisition of statistical considerations required
for the interpretation of the fit results is given in Section~\ref{subsec:statistics}. It
is followed in Section~\ref{sec:gfitter} by an introduction to the \Gfitter project and 
toolkit. The calculation of electroweak precision observables, the results of the global 
fit, and its perspectives are described in Section~\ref{sec:smfit}. 
Section~\ref{sec:2hdmfit} discusses results obtained for the Two Higgs Doublet Model. 
Finally, a collection of formulae used in the theoretical libraries of \Gfitter is given 
in the appendix. We have chosen to give rather exhaustive information here for the purpose 
of clarity and reproducibility of the results presented.

%
%
\section{The Statistical Analysis}
\label{subsec:statistics}

The fitting tasks are performed with the Gfitter toolkit described in Section~\ref{sec:gfitter}.
It features the minimisation of a test statistics and its interpretation 
using frequentist statistics. Confidence intervals and p-values are obtained 
with the use of toy Monte Carlo (MC) simulation or probabilistic 
approximations where mandatory due to resource limitations. This section
introduces the three statistical analyses performed in the paper: $(i)$ determination of 
SM parameters, $(ii)$ probing the overall goodness of the SM, and $(iii)$ probing 
SM extensions and determining its parameters. The SM part is represented by the global
fit at the electroweak scale (Section~\ref{sec:smfit}), while as example for 
beyond SM physics we analyse an extension of the Higgs sector to two scalar 
doublets (Section~\ref{sec:2hdmfit}). The statistical treatment of all three 
analyses relies on a likelihood function formed to measure the agreement between
data and theory. The statistical discussion below follows
in many aspects Refs.~\cite{Charles:2004jd,Hocker:2001xe} with additional input 
from~\cite{Charles:SOS,Demortier:CDF} and other statistical literature. 

\subsection{Model Parameters}
\label{subsec:ModelParameters}

We consider an analysis involving a set of $\Nexp$ measurements $(\xexp)_{i=1..\Nexp}$, 
described by a corresponding set of theoretical expressions $(\xthe)_{i=1..\Nexp}$.  
The theoretical expressions are functions of a set of $\Nmod$ model parameters 
$(\ymod)_{j=1..\Nmod}$.  Their precise definition is irrelevant for the present discussion 
besides the fact that:
\bei
\item a subset of $(\ymod)$ may be unconstrained parameters of the
      theory (\eg, the Higgs mass in the SM, if the results from the direct searches are 
      not used);

\item another subset of $(\ymod)$ are theoretical parameters for which prior 
      knowledge from measurements or calculations is available and 
      used (\eg, the $Z$-boson mass and the hadronic vacuum polarisation 
      contribution to the running electromagnetic coupling strength); 

\item the remaining $(\ymod)$ parametrise theoretical
      uncertainties, which are based on hard-to-quantify
      educated guesswork (\eg, higher order QCD corrections to a truncated 
      perturbative series).

\eei
It may occur that $\xexp$ or $\ymod$ parameters have statistical {\em and} theoretical
errors, requiring a proper treatment for both of these. In the following we use 
the shorthand notations $\ymod$ ($\xexp$, $\xthe$) to label both, sets of and individual 
parameters (measurements, theoretical expressions). 

\subsection{Likelihood Function}
\label{subsec:LikelihoodFunction}

We adopt a least-squares like notation and define the test statistics
\beq
\label{eq:chi2Function}
   \chi^2(\ymod)\equiv-2\ln\Lik(\ymod)\,,
\eeq
where the likelihood function, $\Lik$, is the product of two contributions
\beq
\label{eq:likFunction}
   \Lik(\ymod)=\Likexp(\xthe(\ymod)-\xexp)\cdot \Likthe(\ymod)~.
\eeq
The experimental likelihood, $\Likexp$, measures the agreement between 
$\xthe$ and $\xexp$, while the theoretical likelihood, $\Likthe$, 
expresses prior knowledge of some of the $\ymod$ parameters. In most cases 
$\Likexp$ incorporates well-behaved statistical errors as well as (mostly)
non-statistical experimental systematic uncertainties. In some instances it may also include 
theoretical uncertainties and/or specific treatments that may account for 
inconsistent measurements. On the contrary, $\Likthe$ relies on educated guesswork, 
akin to experimental systematic errors, but in most cases  
less well defined. The impact of (mostly strong interactions related) theoretical 
uncertainties and their treatment on the analysis may be strong, as it is the case 
for the global CKM fit~\cite{Charles:2004jd,Hocker:2001xe}. The statistical treatment 
\Rfit~\cite{Hocker:2001xe,Charles:2004jd} (described below) is designed to deal with the problem of  
theoretical errors in a clear-cut and conservative manner. Evidently though,
an ill-defined problem cannot be treated rigorously, and results that 
strongly depend on theory uncertainties must be interpreted with care. 
For the present analysis, by virtue of the large electroweak mass scale so that 
QCD is in the perturbative regime, purely theoretical errors are small and controlled, 
so that the fit results are well behaved. Increasing experimental precision may 
alter this picture in the future. 

\subsubsection*{\normalsize The Experimental Likelihood}

The experimental component of the likelihood is given by the product
\beq
\label{eq:likExp}
   \Likexp(\xthe(\ymod)-\xexp)=\prod_{i,j=1}^{\Nexp}\Likexp(i,j)\,,
\eeq
where the $\Nexp$ individual likelihood components $\Likexp(i,j)$ account 
for observables that may be independent or not. The model predictions of 
the observables depend on a subset of the $\ymod$ parameters, and are used to 
constrain those. Ideally, all likelihood components are independent 
(\ie $\Likexp(i,j)=0$ for $i\ne j$)  Gaussian functions, each with a standard deviation 
estimating the experimental statistical uncertainty.\footnote
{
   \label{ftn:errors}
   The fitting procedure described in Section~\ref{subsec:parameterDetermination}
   uses $\chi^2$ minimisation to obtain the best match
   between a test hypothesis, represented by a certain parameter set, and the data.
   This requires the use of {\it expected} experimental errors corresponding to the 
   test hypothesis in the experimental likelihood, rather than the {\it measured} 
   experimental errors. However, the expected experimental errors are usually not 
   available for all possible test hypotheses, and the measured experimental errors 
   are used instead. This may be a
   reasonable approximation for test values in close vicinity of the measured experimental 
   results. Nonetheless, one should expect that for regions that are strongly disfavoured 
   by the likelihood estimator the statistical analysis is less precise, so that large 
   deviations in terms of ``sigmas'' must be interpreted with care. We shall revisit this 
   point in Section~\ref{subsec:sminputData} when including results from the direct 
   searches for the Higgs boson in the fit.   
} 
In practise however, one has to deal with correlated measurements and with additional 
experimental and theoretical systematic uncertainties. In accordance with the approach 
adopted by most published analyses, experimental systematic errors are assumed to express Gaussian 
standard deviations, so that different systematic errors can be added in quadrature.\footnote
{
   This introduces a Bayesian flavour to the statistical analysis. 
} 
Theoretical errors are treated according to the \Rfit scheme described below.

\subsubsection*{\normalsize The Theoretical Likelihood}
\label{sec:TheoreticalLikelihood}

The theoretical component of the likelihood is given by the product
\beq
\label{eq:likThe}
        \Likthe(\ymod)=\prod_{i=1}^{\Nmod} \Likthe(i)~.
\eeq
The individual components $\Likthe(i)$ can be constant everywhere in case of no {\em a-priori}
information, be bound, or may express a probabilistic function when such information 
is reliably available. Ideally, one should incorporate in $\Likexp$ measurements 
(or equivalent determinations such as Lattice gauge theory, provided well-controlled 
theoretical assumptions are made)
from which constraints on the $\ymod$ parameters can be derived. If such constraints are 
not available, or if a component has been explicitly introduced to parametrise theoretical  
uncertainty, the $\Likthe(i)$ components must be incorporated by hand in 
Eq.~(\ref{eq:likThe}). They are statistically ill-defined and can hardly be treated 
as probability density functions. 

In the {\em range fit} approach, \Rfit, it is proposed 
that the theoretical likelihoods 
$\Likthe(i)$ do not contribute to the $\chi^2$ of the fit when the corresponding $\ymod$ 
parameters take values within allowed ranges denoted $[\ymod]$. Usually these ranges are 
identified with the intervals $[\yguessmod-\sigma_{\rm theo}\ ,\yguessmod+\sigma_{\rm theo}]$, 
where $\yguessmod$ is a best-guess value, and $\sigma_{\rm theo}$ is the theoretical 
systematic error assigned to $\ymod$. Hence all {\em allowed} $\ymod$ values are 
treated on equal footing, irrespective of how close they are to the edges of 
the allowed range. Instances where even only one of the $\ymod$ parameters lies outside 
its nominal range are not considered. This is the unique assumption made in the \Rfit
scheme: $\ymod$ parameters for which {\em a-priori} information exists are bound to remain 
within {\em predefined} allowed ranges. The \Rfit scheme departs from a perfect 
frequentist analysis only because the allowed ranges $[\ymod]$ do not always extend to 
the whole physical space.\footnote
{
  Some $\ymod$ parameters do not have any {\em a-priori} information and are hence fully
  unbound in the fit.
}
This minimal assumption, is nevertheless a strong constraint: all the results obtained 
should be understood as valid only if all the assumed allowed ranges contain the true 
values of their $\ymod$ parameters. Because there is in general no guarantee for it being
the case, a certain arbitrariness of the results remains and must be kept in mind.\footnote
{
   If a theoretical parameter is bound to an allowed range, and if this range is 
   narrower than what the fit would yield as constraint for the parameter if let free
   to float, the best fit value of this (bound) parameter usually occurs on the edge 
   of the allowed range. A modification of this range will thus have immediate 
   consequences for the central values of the fit.   
}
Although in general range errors do not need to be of theoretical origin, but could as 
well parametrise hard-to-assess experimental systematics, or set physical 
boundaries, we will collectively employ the term ``theoretical (or theory) errors'' 
to specify range errors throughout this paper. 

\subsection{Parameter Estimation}
\label{subsec:parameterDetermination}

When estimating model parameters one is not interested in the quality of the agreement 
between data and the theory as a whole. Rather, taking for granted that the theory is 
correct, one is only interested in the quality of the agreement between data and various 
realisations (models) of the theory, specified by distinct sets of $\ymod$ values. In the 
following we denote $\ChiMinGlob$ the absolute minimum value of the $\chi^2$ function of 
Eq.~(\ref{eq:chi2Function}), obtained when letting all the $\ymod$ parameters free to 
vary within their respective bounds, with a fit converging at the solution 
$\yhatmod$.\footnote
{
  The application of the \Rfit scheme in presence of theoretical uncertainties
  may lead to a non-unique $\{\yhatmod\}$ solution space. 
}
One now attempts to estimate confidence intervals for the complete $\ymod$ set. This 
implies the use of the offset-corrected test statistics
\beq
\label{eq:deltachi2}
        \Delta\chi^2(\ymod)=\chi^2(\ymod)-\ChiMinGlob\,,
\eeq
where $\chi^2(\ymod)$ is the $\chi^2$ for a given set of model parameters $\ymod$. 
Equation~(\ref{eq:deltachi2}) represents the logarithm of a profile likelihood.
The minimum value $\Delta\chi^2(\yhatmod)$ is zero, by construction. This ensures that, 
consistent with the assumption that the model is correct, exclusion confidence levels\footnote
{
   Throughout this paper the term {\em confidence level} denotes 1 minus the p-value of a given 
   $\Delta\chi^2$ (or $\chi^2$) test statistics, and is hence a measure of the exclusion probability 
   of a hypothesis. This is not to be confounded with a {\em confidence interval}, which expresses
   an inclusion probability. 
} 
(CL) equal to zero are obtained when exploring the $\ymod$ space. 

In general, the $\ymod$ parameters in Eq.~(\ref{eq:deltachi2}) are divided into relevant 
and irrelevant ones. The relevant parameters (denoted $\a$) are scanned for 
estimation purposes, whereas the irrelevant ones (the {\em nuisance} parameters $\Mu$) 
are adjusted such that $\Delta\chi^2(\a,\Mu)$ is at a minimum for $\Mu=\Muhat$. Since 
in frequentist statistics one cannot determine probabilities for certain $\a$ values to 
be true, one must derive exclusion CLs. The goal is therefore to set exclusion CLs in the 
$\a$ space irrespective of the $\Mu$ values. 

A necessary condition is that the confidence interval (CI) constructed from the $\Delta\chi^2(\ymod)$ 
test statistics provides sufficient coverage, that is, the CI for a parameter under consideration 
covers the true parameter value with a frequency of at least the $\CL$ values at the CI boundaries
if the measurement were repeated many times. For a Gaussian problem, the test statistics follows a 
$\chi^2$ distribution~\cite{wilks:1938} and one finds 
\beq
\label{eq:clprob}
   1-\CL(\a,\Muhat)=\ProbCERN(\Delta\chi^2(\a,\Muhat),\Dim[\a])\,,
\eeq
where $\Dim[a]$ is the dimension of the $\a$ space, which is the number of 
degrees of freedom\footnote
{
   \label{ftn:dofdiscussion}
   Note that the effective number of degrees of freedom may not always be equal 
   to the dimension of the $\a$ space. For example, if $\Dim[a]=2$ but   
   a single observable ${\cal O}=f(\a)$ is scanned in $\a$, only one of the two
   dimensions of $\a$ is independent, while the other can be derived via ${\cal O}$
   so that the effective $\Dim[a]$ to be used here is one~\cite{Hocker:2001xe}.   
   Similarly, the available observables may only constrain
   one of the two dimensions of $\a$. Again, the effective dimension to be used
   in Eq.~(\ref{eq:clprob}) would be one. Intermediate cases, mixing strong and 
   weak constraints in different dimensions of $\a$ may lead to an ill-posed
   situation, which can only be resolved by means of a full toy MC analysis.
   Such an analysis is performed at some instances in this paper (see in particular
   Section~\ref{sec:2hdm_combinedFit} for the two-dimensional case).
} 
of the offset-corrected $\Delta\chi^2$. Here the probability density distribution of 
$\Delta\chi^2$ is independent of $\mu$. In a non-Gaussian case the CI for 
$\a$ must be evaluated with toy MC simulation for any possible set of true $\Mu$ 
values using, \eg, a Neyman construction~\cite{Neyman} with likelihood-ratio 
ordering~\cite{Feldman:1997qc,Conrad:2002kn}.\footnote
{
   An ordering scheme is required because the construction of a Neyman CL belt 
   is not unique. It depends on the definition of the test statistics used. 
} 
One may then choose for each $a$ the set of $\mu$ that gives the smallest $\CL(a)$. This 
``supremum'' approach~\cite{Charles:SOS} (also described in Ref.~\cite{Demortier:CDF} with 
however a somewhat different meaning) provides the most conservative result, which 
however overcovers in general. (Note also that the approach depends on the ordering algorithm 
used~\cite{Punzi}). It may  lead to the paradoxical situation that $\mu$ values
excluded by the data may be chosen as the true set to determine $\CL(a)$. As a modification 
to this scheme, one could only consider $\mu$ values that are within predefined 
$\Delta\chi^2(\a,\mu)$ bounds, thus guaranteeing a minimum compatibility with the 
data~\cite{Berger:1994,Silvapulle:1996}.
A vast literature on this topic is available (see PhyStat conference proceedings 
and, \eg, Ref.~\cite{Demortier:CDF}), mostly attempting to prescribe a limitation of the 
$\Mu$ space while maintaining good coverage properties.\footnote
{
   We recall here the reserve expressed in Footnote~\ref{ftn:errors} on page~\pageref{ftn:errors}
   affecting the accuracy of any approach: the dependence of the measured errors on the outcome 
   of the observables (determined by $a$ and $\mu$) -- if significant -- must be taken into account.
}
We point out that the naive ``plugin'' approach that consists of using the set of $\Muhat$ 
that minimises $\Delta\chi^2(\a,\Muhat)$ in the fit to estimate the true 
$\Mu$ is incorrect in general (it is trivially correct 
if the problem is strictly Gaussian, as then the $\Delta\chi^2$ distribution is $\mu$-independent).
It may lead to serious undercoverage if the $\Delta\chi^2(\a,\Mu)$ frequency distribution 
is strongly dependent on $\Mu$ (\cf the analysis of the CKM phase $\gamma$~\cite{Charles:SOS}). 

As a shortcut to avoid the technically challenging full Neyman construction in presence
of nuisance parameters, one may choose a Gaussian interpretation of the profile 
likelihood $\Lik(\a,\Muhat)$ versus $\a$, which corresponds to a MINOS~\cite{James:1975dr} 
parameter scan. Simple tests suggest satisfying coverage properties of the profile 
likelihood (see, \eg, \cite{Reid:2003vb,Rolke:2004mj,cranmer-2005}). Mainly because 
of its simplicity this assumption will be adopted for most (though not all) of the results 
presented in this paper.

\subsection{Probing the Standard Model}
\label{sec:probingTheSM}

By construction, the parameter estimation via the offset-corrected $\Delta\chi^2$
is unable to detect whether the SM fails to 
describe the data. This is because Eq.~(\ref{eq:deltachi2}) wipes out the information 
contained in $\ChiMinGlob$. This value is a test statistics for the best possible 
agreement between data and theory. The agreement can be quantified by the p-value 
$\Prob(\ChiMin\ge\ChiMinGlob|\rm SM)$, which is the tail probability to observe a 
test statistics value as large as or larger than $\ChiMinGlob$, if the SM is the 
theory underlying the data. It hence quantifies the probability of wrongly rejecting 
the SM hypothesis. In a Gaussian case, $\ChiMinGlob$ can be readily 
turned into a p-value via $\ProbCERN(\ChiMinGlob,\ndof)$.\footnote
{
   The corresponding ROOT function is TMath::Prob(...).
} 
In presence of non-Gaussian effects, a toy MC simulation must be performed. Again, a 
full frequentist analysis requires the scan of all possible (or ``likely'') true nuisance 
parameters, followed by toy MC studies to derive the corresponding p-values. Chosen is the
set of true $\yhatmod$ that maximises $\Prob(\ChiMinGlob|\rm SM)$, where here exact 
coverage is guaranteed by construction (note that in this phase no explicit parameter 
determination is performed so that all $\ymod$ are nuisance parameters).

Such a {\em goodness-of-fit} test may not be the most 
sensitive manner to uncover physics beyond the SM (BSM). If the number of 
degrees of freedom is large in the global fit, and if observables that are sensitive 
to the BSM physics are mixed with insensitive ones, the fluctuations in the latter 
observables dilute the information contained in the global p-value (or deficiencies
in the SM description may fake presence of new physics). It is therefore 
mandatory to also probe specific BSM scenarios.\footnote
{
   This problem is similar to those occurring in goodness-of-fit (GoF) tests in experimental 
   maximum-likelihood analyses. If, for instance, the data sample with respect to which 
   a likelihood analysis 
   is performed is dominated by background events with a small but significant signal 
   excess a successful global GoF test would only reveal agreement with the background 
   model and say little about the signal. Similarly, a small p-value for the null 
   hypothesis may reflect problems in the background description rather than an excess
   of signal events. A possible remedy here would be to restrict the GoF test to signal-like 
   events, or more specifically, to test the GoF in all likelihood bins independently.
}

\subsection{Probing New Physics}
\label{sec:probingNewPhysics}

If the above analysis establishes that the SM cannot accommodate the data, that is, the p-value 
is smaller than some critical value, the next step is to probe the BSM physics revealed by the 
observed discrepancy. The goal is akin to the determination of the SM parameters: it is to 
measure new sets of physical parameters $\yNP$ that complement the $\ymod$ SM parameters.
The treatment is identical to the one of Section~\ref{subsec:parameterDetermination}, using 
$\a=\{\yNP\}$. Even if the SM cannot be said to be in significant disagreement with the data, 
the estimation of $\yNP$ remains interesting because the most sensitive observables, and the 
precision to be aimed at for their determination can only be derived by this type of 
analysis. Moreover, the specific analysis might be able to faster detect the first signs 
of a discrepancy between data and the SM if the theoretical extension used in the 
analysis turns out to be the right one. 


\section{The Gfitter Package}
\label{sec:gfitter}

The generic fitting package \Gfitter comprises a statistical framework 
for model testing and parameter estimation problems. It is specifically designed 
to provide a modular environment for complex fitting tasks, such as the global 
SM fit to electroweak precision data, and fits beyond the SM.
\Gfitter is also a convenient framework for averaging problems, ranging from 
simple weighted means using or not correlated input data, to more involved 
problems with non-Gaussian PDFs and/or common systematic errors, requiring or 
not consistent rescaling due to parameter interdependencies.

\subsubsection*{Software}

The \Gfitter package~\cite{GfitterSW} consists of abstracted
object-oriented code in C++, relying on ROOT
functionality~\cite{Brun:1997pa}. The core fitting code and the
physics content are organised in separate packages, each physics model
package can be invoked as a plugin to the framework. The user interfaces
\Gfitter through data cards in XML format, where all the input data and
driving options are defined. The fits are run alternatively as ROOT
macros or executables, interactively or in a batch system.

\subsubsection*{\Gfitter Parameters and Theories}

\Gfitter defines only a single data container, denoted {\em parameter}, which can have three 
distinct manifestations according to its use case:
\bei

\item[(A)] Measurements $\xexp$ that are predicted by the model (\eg, $W$ mass in the SM):
      parameters of this type are not varied in the fit, but contribute to 
      the log-likelihood function through comparison between the model prediction and the
      corresponding measurement.
    
\item[(B)] Model parameters $\ymod$ that are not predicted by the theory but for which a direct 
      measurement exists (\eg, top mass in the SM):
      parameters of this type are varied in the fit, and they contribute to 
      the log-likelihood function through comparison between the fit parameter value
      and the corresponding measurement.
    
\item[(C)] Model Parameters $\ymod$ that are not predicted by the theory and for which
      no direct measurement exist (\eg, Higgs mass in the SM), or which parametrise theoretical 
      uncertainties according to the \Rfit prescription (\cf Section~\ref{subsec:LikelihoodFunction}): 
      parameters of this type are varied freely in the fit within bounds (if exist), 
      and they do not contribute themselves to the log-likelihood function.

\eei
A parameter is uniquely defined via a name (and optionally an alias to allow the user
to declare several correlated measurements of the same parameter, and to design theoretical 
predictions in a polymorph class hierarchy) in the data card, and stored in a global parameter 
container. These parameters are objects (of the {\em GParameter} class) that cannot be destroyed 
nor be recreated. 
Upon creation of a parameter, \Gfitter searches automatically in the physics libraries for a 
corresponding theory (an object of the {\em GTheory} class), identified through the name of 
the parameter. If a theory is found, the corresponding class object is instantiated\footnote
{
   A GTheory can depend on auxiliary theory objects (derived from GTheory) that are used to 
   outsource complex computation tasks. Caching of results from repetitive calculations also 
   benefits from outsourcing. 
} 
and the parameter is categorised as of type (A); if no theory is found, it is of type (B) 
or (C) depending on the presence of a measurement in the data card.\footnote 
{
  Measurement results can be given as central value and Gaussian (possibly asymmetric) and/or theoretical 
  errors, or as a user-defined log-likelihood function encoded in ROOT objects (\eg histograms, 
  graphs or functions).
}
The categorisation of parameters is 
performed automatically by \Gfitter maintaining full transparency for the user. 

\subsubsection*{Parameter Errors, Ranges, Correlations and Rescaling}}
\label{sec:rescaling}

\Gfitter distinguishes three types of errors: normal errors following a Gaussian distribution
describing statistical and experimental systematic errors, a user-defined log-likelihood functions 
including statistical and systematic uncertainties, and allowed ranges describing physical limits
or hard-to-assess systematic errors (mostly of theoretical origin). All errors can be asymmetric
with respect to the central values given. All parameters may have combinations of Gaussian 
and range errors (but only a single user-defined likelihood function). Parameters of type (A) 
and (B) do not contribute to the log-likelihood functions if the theory prediction or floating
parameter value is compatible with the central value of the parameter within the ranges of 
the theoretical errors attributed to the parameter (\cf Section~\ref{subsec:ModelParameters} concerning
the implications of the term ``theoretical error''). Only beyond 
these ranges, a Gaussian parabolic contribution to the log-likelihood function occurs. For 
example, the combined log-likelihood function of a parameter with central value $x_0$, 
positive (negative) Gaussian error $\sigma_{\rm Gauss}^{+}$ ($\sigma_{\rm Gauss}^{-}$), and  
positive (negative) theoretical error $\sigma_{\rm theo}^{+}$ ($\sigma_{\rm theo}^{-}$), for 
a given set of $\ymod$ parameters and theoretical prediction $f(\ymod)$ reads\footnote
{
   In the log-likelihood definition of Eq.~(\ref{eq:logLikelihood}), the central value 
   $x_0$ corresponds to the value with the largest likelihood, which is not necessarily 
   equal to the arithmetic average in case of asymmetric errors. 
}
\beq
\label{eq:logLikelihood}
   -2\log\Lik(\ymod) = \left\{\begin{array}{cl}
       0\;, & \mbox{if: $-\sigma_{\rm theo}^{-} \le f(\ymod) - x_0 \le \sigma_{\rm theo}^{+}$}\;, \\ 
       \left(\frac{f(\ymod) - \left(x_0 + \sigma_{\rm theo}^{+}\right)}{\sigma_{\rm Gauss}^{+}}\right)^{\!\!2}\;, &
           \mbox{if: $f(\ymod) - x_0 > \sigma_{\rm theo}^{+}$}\;, \\[0.4cm]
       \left(\frac{f(\ymod) - \left(x_0 - \sigma_{\rm theo}^{-}\right)}{\sigma_{\rm Gauss}^{-}}\right)^{\!\!2}\;, &
           \mbox{if: $x_0 - f(\ymod) > \sigma_{\rm theo}^{-}$}\;. 
       \end{array}
       \right.
\eeq
Parameters of type (C) vary freely within the ranges set by the theoretical errors if 
available, or are unbound otherwise. 

Parameters can have correlation coefficients identified and set in the data card via the parameter 
names (and alias if any). These correlations are taken into account in the log-likelihood test 
statistics as well as for the creation of toy MC experiments. 

It is possible to introduce dependencies among parameters, which can be used to parametrise correlations 
due to common systematic errors, or to rescale parameter values and errors with newly available results
for parameters on which other parameters depend. For example, in the global SM fit the 
experimental value used of the parameter \dalphaHadMZ depends on \asZ. The value 
for \asZ used when evaluating \dalphaHadMZ
may have been updated in the meantime, or may be updated in each fit step, which leads to 
a (not necessarily linear) shift of \dalphaHadMZ and also to a reduced systematic error (for details
see Footnote~\ref{ftn:alpha} on page~\pageref{ftn:alpha}). 
The rescaling mechanism of \Gfitter allows to automatically account for arbitrary functional 
interdependencies between an arbitrary number of parameters.

\subsubsection*{Caching}

An important feature of \Gfitter is the possibility to cache computation results between
fit steps. Each parameter holds pointers to the theory objects that depend on it, and the 
theories keep track of all auxiliary theory objects they depend on. Upon computation 
of the log-likelihood function in a new fit step, only those theories (or part of theories) 
that depend on modified parameters (with respect to the previous fit step) are recomputed.
More importantly, time intensive calculations performed by auxiliary theories that are shared
among several theories are made only once per fit step. The gain in CPU time of this caching 
mechanism is substantial, and can reach orders of magnitudes in many-parameter fitting problems.

\subsubsection*{Fitting}

The parameter fitting is transparent with respect to the fitter implementation, which by 
default uses TMinuit~\cite{James:1975dr}, but which is extensible via the driving card to the 
more involved global minima finders Genetic Algorithm and Simulated Annealing, implemented 
in the ROOT package TMVA~\cite{Hocker:2007ht}.

\subsubsection*{Parameter Scans and Contours}

\Gfitter offers the possibility to study the behaviour of the log-likelihood test 
statistics as a function of one or two parameters by one- or two-dimensional scans,
respectively. If a parameter is of type (A), penalty contributions are added to the 
log-likelihood test statistics forcing the fit to yield the parameter value under study. 
In addition, two-dimensional contour regions of the test statistics can be computed 
using the corresponding TMinuit functionality.

\subsubsection*{Toy Monte Carlo Analyses}

\Gfitter offers the possibility to perform toy Monte Carlo (MC) analyses repeating 
the minimisation step for input parameter values that are randomly generated around 
expectation values according to specified errors and correlations.
For each MC  experiment the fit results are recorded allowing
a statistical analysis, \eg, the determination of a p-value and an overall 
goodness-of-fit probability. All parameter scans
can be optionally performed that way, as opposed to using a Gaussian approximation
to estimate the p-value for a given scan point (manifestation of true values). 
%
%
\section{The Standard Model Fit to Electroweak Precision Data}
\label{sec:smfit}

In recent particle physics history, coined by the success of the electroweak unification 
and Quantum Chromodynamics (QCD), fits to experimental precision data have substantially 
contributed to our knowledge of the Standard Model (SM). The first application of global fits to 
electroweak data has been performed by the LEP Electroweak Working Group~\cite{LEPEWWG} 
in the last decade of the 20$^{\rm th}$ century, unifying LEP and SLD precision data. The primary
results of these fits were a prediction of the top-quark mass (today's fit precision $\simeq9\,\gev$)
prior to its discovery, an accurate and theoretically well controlled determination of the strong 
coupling constant at the $Z$-mass scale (today available at the \NNNLO level~\cite{Baikov:2008jh}),
and a logarithmic constraint on the Higgs mass establishing that the SM Higgs must be light.
Other areas related to particle physics where global fits are performed are neutrino
oscillation~\cite{Goodman:url}, leading to constraints on mixing parameters and mass 
hierarchies, flavour physics, with constraints on the parameters of the quark-flavour 
mixing (CKM) matrix and related quantities~\cite{Charles:2004jd,Bona:2005vz}, 
and cosmology~\cite{lambda:url}, leading to a large number of phenomenological results 
such as the universe's curvature, 
the relic matter and energy density, neutrino masses and the age of the universe. Global fits 
also exist for models beyond the SM such as Supersymmetry~\cite{Lafaye:2004cn,Bechtle:2004pc}
with however yet insufficient high-energy data for successfully constraining the parameters
of even a minimal model so that simplifications are in order. 

We emphasise that the goal of such fits is twofold (\cf Section~\ref{subsec:statistics}): $(i)$ 
the determination of the free model parameters, and $(ii)$ a goodness-of-fit test measuring the 
agreement between model and data after fit convergence. This latter goal can be only 
achieved if the model is overconstrained by the available measurements. The situation 
is particularly favourable in the CKM sector, where the primary 
goal of experiments and phenomenological analysis has been moved from CKM parameter determination 
to the detection of new physics via inconsistencies in the CKM phase determination. The 
relatively young field of neutrino oscillation measurements on the contrary does not yet 
provide significant overconstraints of the neutrino flavour mixing matrix. 

In the following we revisit the global electroweak fit at the $Z$-mass scale using the \Gfitter
package. We recall the relevant observables, their SM predictions, perform fits under various 
conditions, and discuss the results.

\subsection{Formalism and Observables}
\label{sec:FormalismAndObservables}

The formal analysis of this section is placed within the framework of the SM. The electroweak 
fit focuses on the parameters directly related to the $Z$ and $W$ boson properties, and to radiative
corrections to these, providing the sensitivity to heavy particles like the top quark and 
the Higgs boson. The floating parameters of the fit are the Higgs and $Z$-boson masses, the $c$, 
$b$, and $t$-quark masses, as well as the electromagnetic and strong coupling 
strengths at the $Z$ pole. Most of these parameters are also directly constrained by measurements
included in the fit. 

We have put emphasis on the completeness of the information given in this 
paper, with a large part of the relevant formulae quoted in the main text and the appendices. 
Readers seeking for a more pedagogical introduction are referred to the many excellent reviews 
on this and related topics (see, \eg, 
Refs.~\cite{Bardin:1999ak,Bardin:1997xq,:2005ema,Erler:2004nh}). Section~\ref{sec:SMrelations}
provides a formal introduction of tree-level relations, and quantum loop corrections sensitive
to particles heavier than the $Z$. The observables used in the global fit and their SM
predictions are summarised in Section~\ref{sec:SMobservables} and Section~\ref{sec:SMtheoobservables}
respectively. Theoretical uncertainties are discussed in Section~\ref{sec:SMtheoerrors}.

\subsubsection{Standard Model Tree-Level Relations and Radiative Corrections} 
\label{sec:SMrelations}

The tree-level vector and axial-vector couplings occurring in the $Z$ boson to 
fermion-antifermion vertex $i\fbar\gamma_\mu(\gvz{f}+\gvz{f}\gamma_5)f Z_\mu$ are 
given by\footnote
{
   Throughout this paper the superscript '$(0)$' is used to label tree-level 
   quantities.
}
\begin{align}
   \gvz{f}\ \equiv\ \glz{f} + \grz{f} &\ =\   I^{f}_{3} - 2Q^{f} \sin^{2}\tzw\,, \\
   \gaz{f}\ \equiv\ \glz{f} - \grz{f} &\ =\   I^{f}_{3}\,, 
   \label{eq:treecouplings2}
\end{align}
where $\glrz{f}$ are the left-handed (right-handed) fermion couplings, 
and $Q^f$ and $I^{f}_{3}$ are respectively the charge and the third component of the weak isospin. 
In the (minimal) SM, containing only one Higgs doublet, the weak mixing 
angle is defined by 
\beq
   \sin^{2}\tzw\ =\  1-\frac{M_{W}^2}{M_{Z}^2}\,.
\label{eq:massrelation}
\eeq
Electroweak radiative corrections modify these relations, leading to an effective weak mixing 
angle and effective couplings
\begin{align}
\label{eq:Zcouplings_sinfeff}
   \sinfeff & \ =\   \kZ{f} \sin^{2}\tzw\,, \\
\label{eq:Zcouplings_rZ}
   \gv{f}   & \ =\   \sqrt{\rZ{f}} \left( I^{f}_{3} - 2Q^{f} \sinfeff \right)\,, \\
\label{eq:Zcouplings_kZ}
   \ga{f}   & \ =\   \sqrt{\rZ{f}}  I^{f}_{3}\,, 
\end{align} 
where $\kZ{f}$ and $\rZ{f}$ are form factors absorbing the radiative corrections. They 
are given in Eqs.~(\ref{eq:rho}) and (\ref{eq:kappa}) of Appendix~\ref{app:ewformfactors}.
Due to non-zero absorptive parts in the self-energy and vertex correction diagrams, the effective 
couplings and the form factors are complex quantities. The {\em observable} effective mixing 
angle is given by the real parts of the couplings
\beq
\label{eq:gvgaratio}
   \frac{\Re(\gv{f})}{\Re(\ga{f})}\ =\  1-4|Q_f|\sinfeff\,.
\eeq
Electroweak unification leads to a relation between weak and electromagnetic couplings, 
which at tree level reads
\beq
\label{eq:GF}
     \GF \ =\  \frac{\pi \alpha}{\sqrt{2} (M_{W}^{(0)})^2 \bigg(1-\frac{(M^{(0)}_{W})^2}{M_{Z}^2}\bigg)}\,.
\eeq
Radiative corrections are parametrised by multiplying the r.h.s. of Eq.~(\ref{eq:GF}) with 
the form factor $(1-\Delta r)^{-1}$. Using Eq.~(\ref{eq:massrelation}) and resolving for $M_W$ gives
\beq
\label{eq:MWdeltar}
    M_{W}^2 \ =\  \frac{M_{Z}^2}{2}
              \left(1+\sqrt{1-\frac{\sqrt{8}\,\pi\alpha(1-\Delta r)}{\GF{M_{Z}^2}}}\right)\,.
\eeq
The form factors $\rZ{f}$, $\kZ{f}$ and $\Delta r$ depend nearly quadratically on $\mt$ and logarithmically 
on $M_H$. They have been calculated including two-loop corrections in the on-shell renormalisation 
scheme (OMS)~\cite{Passarino:1978jh,Sirlin:1980nh,Aoki:1982ed}, except for $b$ quarks where an 
approximate expression, including the full one-loop correction and the known leading 
two-loop terms $\propto \mt^4$, is provided. The 
relevant formulae used in this analysis are summarised in Appendix~\ref{app:ewformfactors}. 
Since $\Delta r$ also depends on $M_W$ an iterative method is needed to solve 
Eq.~(\ref{eq:MWdeltar}). The calculation of $M_W$ has been performed including the 
complete one-loop correction, two-loop and three-loop QCD corrections of order 
$\Order(\alpha \as)$ and $\Order(\alpha \alpha^{2}_{s})$,
fermionic and bosonic two-loop electroweak corrections of order $\Order(\alpha^{2})$, and the 
leading $\Order(\GFe{2}\alpha_S\mt^4)$ and $\Order(\GFe{3}\mt^6)$ three-loop 
contributions~\cite{Awramik:2003rn,Awramik:2004ge,Awramik:2006uz}.
Four-loop QCD corrections have been calculated 
for the $\rho$-parameter~\cite{Chetyrkin:2006bj,Schroder:2005db,Boughezal:2006xk}. Since they 
affect the $W$ mass by 2\mev only, they have been neglected in this work. 

For the SM prediction of $M_W$ we use the 
parametrised formula~\cite{Awramik:2003rn}
\begin{align}
\label{eq:fitformula2}
M_{W} \ =\  M_W^{\rm ini} &- c_1 \, \mathrm{dH} - c_2 \, \mathrm{dH}^2 
       + c_3 \, \mathrm{dH}^4 + c_4 (\mathrm{dh} - 1)
       - c_5 \, \mathrm{d}\alpha + c_6 \, \mathrm{dt} \nonumber\\
       &- c_7 \, \mathrm{dt}^2 
       - c_8 \, \mathrm{dH} \, \mathrm{dt} 
       + c_9 \, \mathrm{dh} \, \mathrm{dt} - c_{10} \, \mathrm{d}\as
       + c_{11} \, \mathrm{dZ} \,,
\end{align}
with
\begin{align}
\mathrm{dH} &\ =\  \ln\left(\frac{M_H}{100 \gev}\right)\,, &
\mathrm{dh} &\ =\  \left(\frac{M_H}{100 \gev}\right)^{\!\!2}\,,  &
\mathrm{dt} &\ =\  \left(\frac{\mt}{174.3 \gev}\right)^{\!2} - 1\,, \nonumber\\
\mathrm{dZ} &\ =\  \frac{M_Z}{91.1875 \gev} -1\,, \quad &
\mathrm{d}\alpha &\ =\  \frac{\Delta\alpha(M_Z^2)}{0.05907} - 1\,, &
\mathrm{d}\as &\ =\  \frac{\asZ}{0.119} - 1\; ,\nonumber
\end{align}
where here and below all masses are in units of GeV, and where \mt is the top-quark pole mass,
$M_Z$ and $M_H$ are the $Z$ and Higgs boson masses, $\Delta\alpha(M_Z^2)$ is the sum of the leptonic
and hadronic contributions to the running QED coupling strength at $M_Z^2$ (\cf Appendix~\ref{app:alpha}), 
\asZ is the running strong coupling constant at $M_Z^2$ (\cf Appendix~\ref{app:alphas}), and 
where the coefficients $M_W^{\rm ini}, c_1, \ldots, c_{11}$ read
\begin{align}
M_{W}^{\rm ini} &\ =\  80.3799 \gev, &\quad  c_1 &\ =\  0.05429 \gev,  
   &\quad c_2 &\ =\  0.008939 \gev ,\nonumber\\
c_3 &\ =\  0.0000890 \gev,  &\quad  c_4 &\ =\  0.000161 \gev, 
   &\quad c_5 &\ =\  1.070 \gev ,  \nonumber\\
c_6 &\ =\  0.5256 \gev,     &\quad  c_7 &\ =\  0.0678 \gev,  
   &\quad  c_8 &\ =\  0.00179 \gev ,  \nonumber\\
c_9 &\ =\  0.0000659 \gev,  &\quad  c_{10}&\ =\  0.0737 \gev,
   &\quad c_{11}&\ =\  114.9 \gev . \nonumber
\end{align}
The parametrisation reproduces the full result for $M_{W}$ to better than 0.5\mev over 
the range $10\, \gev < M_H < 1\, \tev$, if all parameters are within their expected (year 
2003) $2\sigma$ intervals~\cite{Awramik:2003rn}. 

The effective weak mixing angle of charged and neutral leptons and
light quarks has been computed~\cite{Awramik:2004ge,Awramik:2006uz}
with the full electroweak and QCD one-loop and two-loop corrections,
and the leading three-loop corrections of orders
$\Order(\GFe{2}\alpha_S\mt^4)$ and $\Order(\GFe{3}\mt^6)$. The
corresponding parametrisation formula for charged leptons reads
\begin{align}
\label{eq:sinfeff1}
   \sinleff \ =\  s_0 &+ d_1 L_H + d_2  L_H^2 + d_3  L_H^4 + d_4  (\Delta_H^2 -1) + d_5  \Delta_\alpha  \nonumber\\
            & + d_6  \Delta_t + d_7  \Delta_t^2 
              + d_8  \Delta_t  (\Delta_H -1)
              + d_9  \Delta_{\as} + d_{10} \Delta_Z\,,
\end{align}
with
\begin{align}
   L_H          &\ =\  \ln\left(\frac{M_H}{100 \gev}\right), &
   \Delta_H     &\ =\  \frac{M_H}{100 \gev}, &
   \Delta_\alpha &\ =\  \frac{\Delta \alpha(M_Z)}{0.05907}-1\,, \nonumber \\
   \Delta_t     &\ =\  \left(\frac{\mt}{178.0 \gev}\right)^{\!2} -1\,, &
   \Delta_{\as}  &\ =\  \frac{\asZ}{0.117}-1\,, &
   \Delta_Z     &\ =\  \frac{M_Z}{91.1876 \gev} -1\,, \nonumber
\end{align}
and the numerical values
\begin{align}
   s_0 &\ =\  0.2312527, &\quad d_1 &\ =\   4.729 \cdot 10^{-4}, &\quad d_2 &\ =\  2.07 \cdot  10^{-5},  \nonumber\\  
   d_3 &\ =\   3.85 \cdot  10^{-6}, &\quad  d_4  &\ =\  -1.85 \cdot 10^{-6}, &\quad d_5 &\ =\  0.0207,   \nonumber\\ 
   d_6 &\ =\  -0.002851, &\quad d_7 &\ =\  1.82 \cdot 10^{-4}, &\quad d_8 &\ =\   -9.74 \cdot 10^{-6},  \nonumber\\ 
   d_9 &\ =\  3.98 \cdot 10^{-4}, &\quad d_{10} &\ =\  -0.655. && \nonumber
\end{align}
Equation~(\ref{eq:sinfeff1}) reproduces the full expression with maximum (average) deviation
of $4.5\cdot10^{-6}$ ($1.2\cdot 10^{-6}$), if the Higgs-boson mass lies within 
$10\,\gev < M_H < 1 \,\tev$, and if all parameters are within their expected (year 2003) 
$2\sigma$ intervals~\cite{Awramik:2006uz}. 

The prediction of the effective weak mixing angle for the remaining light fermions 
($u,d,s,c$ quarks and neutrinos) differs slightly from the prediction for charged leptons. 
Again a parametrisation formula is provided~\cite{Awramik:2006uz}, which is used in this analysis. 
For bottom quarks, new diagrams with additional top-quark propagators enter the calculation
and the $b$ quark specific two-loop vertex corrections do not exist.\footnote
{
   After completion of this work the two-loop electroweak fermionic corrections to $\sinbeff$
   have been published\cite{Awramik:2008gi}. They will be included in future updates of this analysis. 
} 
Instead we use Eq.~(\ref{eq:Zcouplings_sinfeff}) and the calculation of 
$\kappa_Z^b$ (\cf Appendix~\ref{app:ewformfactors}), which includes the full one-loop correction 
and the known leading two-loop terms $\propto \mt^4$.

\subsubsection{Summary of Electroweak Observables}
\label{sec:SMobservables}

The following classes of observables are used in the fit. 

\setlength{\tabcolsep}{0.0pc}
\begin{tabularx}{\textwidth}{Xp{10.9cm}}

{\em Z resonance parameters}: & $Z$ mass and width, and total $\ee\to Z\to{\rm hadron}$ 
      production cross section (\ie, corrected for photon exchange contributions). \\[0.15cm]
{\em Partial Z cross sections}: & Ratios of leptonic to hadronic, and heavy-flavour
      hadronic to total hadronic cross sections. \\[0.15cm]
{\em Neutral current couplings}: & Effective weak mixing angle, and left-right and forward-backward 
      asymmetries for universal leptons and heavy quarks.\footnote
      {
         Left-right and forward-backward asymmetries have been also measured for strange 
         quarks, with however insufficient precision to be included here.
      } \\
{\em W boson parameters}: & $W$ mass and width. \\[0.15cm]
{\em Higgs boson parameters}: & Higgs mass. \\[0.15cm]
{\em Additional input parameters}: & Heavy-flavour ($c,b,t$) quark masses (masses of lighter
      quarks and leptons are fixed to their world averages), QED and QCD coupling strengths at the $Z$-mass scale. 
\end{tabularx}

\subsubsection{Theoretical Predictions of Electroweak Observables}
\label{sec:SMtheoobservables}

Parity violation in neutral current reactions $\ee\to f\fbar$ resulting from the different 
left and right-handed $Z$-boson couplings to fermions leads to fermion polarisation in the 
initial and final states and thus to observable asymmetry effects. They can be conveniently 
expressed by the asymmetry parameters
\beq
\label{eq:asymmetry}
   A_f  \ =\  \frac{\gls{f} - \grs{f}}{\gls{f} + \grs{f}}
        \ =\   2 \frac { \gv{f}/\ga{f}} { 1 + (\gv{f}/\ga{f})^2}\,, 
\eeq
where only the real parts of the couplings are considered as the asymmetries refer to pure $Z$ 
exchange. For instance, the forward-backward asymmetry
$\Afbz{f}=(\sigma^{0}_{F,f} - \sigma^{0}_{B,f})/(\sigma^{0}_{F,f} + \sigma^{0}_{B,f})$, 
where the superscript '0' indicates that the observed values have been corrected for radiative 
effects and photon exchange, can be determined from the asymmetry parameters~(\ref{eq:asymmetry}) 
as follows
\beq
\label{eq:AFB0f}
   \Afbz{f} \ =\  \frac{3}{4} A_e A_f\,.
\eeq
The $A_f$ are obtained from Eqs.~(\ref{eq:asymmetry}) and~(\ref{eq:gvgaratio}) 
using $\sinfeff$ from the procedure described in the previous section.

Unlike the asymmetry parameters, the partial decay width $\Gamma_f\ = \ \Gamma(Z\to f\fbar)$
is defined inclusively, \ie, it contains all real and virtual corrections such that the 
imaginary parts of the couplings must be taken into account. One thus has
\beq
\label{eq:partialZWidth}
   \Gamma_f\ = \ 4\,\NCf{f}\Gamma_0|\rZ{f}|(I_3^f)^2
                 \left(\left|\frac{\gvs{f}}{\gas{f}}\right| R_V^f(M_Z^2) + R_A^f(M_Z^2)\right)\,,
\eeq
where $\NCf{\ell(q)}=1(3)$ is the colour factor, $R_V^f(M_Z^2)$ and $R_A^f(M_Z^2)$ are 
{\em radiator functions} (defined further below), and $\Gamma_0$ is given by
\beq
   \Gamma_0 \ =\  \frac{\GF M_Z^3}{24\sqrt{2}\pi}\,.
\eeq
The $\sinfeff$ term entering through the ratio of coupling constants in Eq.~(\ref{eq:partialZWidth}) 
is modified by the real-valued contribution $I_f^2$ resulting from the product of two imaginary parts of 
polarisation operators~\cite{Bardin:1999yd}
\beq
   \sinfeff \rightarrow \sinfeff + I_f^2\,,
\eeq
where
\beq
   I_f^2 \ =\ \alpha^2(M_Z^2) \frac{35}{18} 
              \left(1 - \frac{8}{3}\Re(\kappa_Z^f) \sin^2\tw \right)\,.
\eeq
The full expression for the partial leptonic width reads~\cite{Bardin:1999yd}
\begin{align}
\label{eq:Gammal}
   \Gamma_\ell &\ =\  \Gamma_0 \big|\rho_Z^\ell\big| \sqrt{1-\frac{4 m_\ell^2}{M_Z^2}}
      \left[ \left(  1+\frac{2 m_\ell^2}{M_Z^2}\right) \left( \left|\frac {\gv{\ell}}{\ga{\ell}} \right|^2 + 1 \right)
      -\frac{6 m_\ell^2}{M_Z^2}\right] \cdot
      \left( 1+\frac{3}{4}\frac{\alpha(M_Z^2)}{\pi}Q_\ell^2 \right)\,,
\end{align}
which includes effects from QED final state radiation. 
The partial widths for $q\qbar$ final states, $\Gamma_q$, involve radiator functions 
that describe the final state QED and QCD vector and axial-vector corrections for quarkonic decay 
modes. Furthermore, they contain ${\rm QED}\otimes{\rm QCD}$ and finite quark-mass corrections. For 
the massless perturbative QCD correction, the most recent fourth-order result
is used~\cite{Baikov:2008jh}. Explicit formulae for the radiator functions are given in
Appendix~\ref{app:radfunctions}. The influence of non-factorisable ${\rm EW}\otimes{\rm QCD}$ 
corrections, $\Delta_{\rm EW/QCD}$, that must be added to the width~(\ref{eq:partialZWidth})
for quark final states is small (less than $10^{-3} $). They are assumed to be 
constant~\cite{Czarnecki:1996ei,Harlander:1997zb}, and take the values
\beq
   \Delta_{\rm EW/QCD}  \ =\  \left\{
   \begin{array}{l}
      -0.113\,\mev  \mbox{~for $u$ and $c$ quarks,} \\
      -0.160\,\mev  \mbox{~for $d$ and $s$ quarks,} \\
      -0.040\,\mev  \mbox{~for the $b$ quark.}
      \end{array}
   \right.
\label{eq:deltaEWQCD}
\eeq
The total $Z$ width for three light neutrino generations obeys the sum
\beq
   \Gamma_Z \ =\  \Gamma_e + \Gamma_{\mu} + \Gamma_{\tau} + 3\Gamma_{\nu} + \Gamma_{\rm had}\,, 
\eeq
where $\Gamma_{\rm had}=\Gamma_u + \Gamma_d + \Gamma_c + \Gamma_s + \Gamma_b$ 
is the total hadronic $Z$ width. From these the {\em improved} tree-level total hadronic 
cross-section at the $Z$ pole is given by
\beq
   \sigma^0_{\rm had} \ =\  \frac{12\pi}{M_Z^2} \frac{\Gamma_e \Gamma_{\rm had}}{\Gamma_Z^2}\,.
\eeq
To reduce systematic uncertainties, the LEP experiments have determined the partial-$Z$-width 
ratios $R^0_\ell =\Gamma_{\rm had}/\Gamma_\ell$ and $R^0_q=\Gamma_q/\Gamma_{\rm had}$, which are
used in the fit.

The computation of the $W$ boson width is similar to that of the $Z$ boson, but it is only
known to one electroweak loop. The expression adopted in this analysis can be found 
in~\cite{Bardin:1986fi}. An improved, gauge-independent formulation exists~\cite{Kniehl:2000rb},
but the difference with respect to the gauge-dependent result is small (0.01\%) compared to the 
current experimental error (3\%).

The value of the QED coupling constant at the $Z$ pole is obtained using three-loop results for
the leptonic contribution, and the most recent evaluation of the hadronic vacuum polarisation 
contribution for the five quarks lighter than $M_Z$. Perturbative QCD is used for the small 
top-quark contribution. The relevant formulae and references are given in Appendix~\ref{app:alpha}. 

The evaluation of the running QCD coupling constant uses the known four-loop expansion of the QCD 
$\beta$-function, including three-loop matching at the quark-flavour thresholds (\cf 
Appendix~\ref{app:alphas}). The running of the $b$ and $c$ quark masses is obtained 
from the corresponding four-loop $\gamma$-function (\cf Appendix~\ref{app:runningmasses}).
All running QCD quantities are evaluated in the modified minimal subtraction renormalisation
scheme (\MSb).


\subsubsection{Theoretical Uncertainties}
\label{sec:SMtheoerrors}

Within the \Rfit scheme, theoretical errors based on educated guesswork are introduced via
bound theoretical scale parameters in the fit, thus providing a consistent numerical 
treatment. For example, the effect from a truncated perturbative series is included by adding a 
{\em deviation parameter}, \deltatheo, describing the varying perturbative prediction as a function of 
the contribution from the unknown terms. Leaving the deviation parameter floating within estimated 
ranges allows the fit to adjust it when scanning a parameter, such that the likelihood estimator 
is increased (thus improving the fit compatibility). 

The uncertainties in the form factors $\rZ{f}$ and $\kZ{f}$ are estimated using different
renormalisation schemes, and the maximum variations found are assigned as theoretical errors.
A detailed numerical study has been performed in~\cite{goebel} leading to the following 
real-valued relative theoretical errors
\begin{align}
   \deltatheo  \rZ{f}/|1-\rZ{f}| & \ \approx\ 5 \cdot10^{-3}\,, \\
   \deltatheo  \kZ{f}/|1-\kZ{f}| & \ \approx\ 5 \cdot10^{-4}\,, 
\end{align}
which vary somewhat depending on the fermion flavour. The corresponding absolute theoretical 
errors are around $2 \cdot10^{-5}$ for both $\deltatheo \rZ{f}$ and $\deltatheo  \kZ{f}$ and are 
treated as fully correlated in the fit. These errors, albeit included, have a negligible effect 
on the fit results. 

More important are theoretical uncertainties affecting directly the $M_W$ and $\sinleff$ predictions.
They arise from three 
dominant sources of unknown higher-order corrections~\cite{Awramik:2003rn,Awramik:2006uz}:
$(i)$ $\Order(\alpha^2\as)$ terms beyond the known contribution of $\Order(\GFe{2} \as \mt^4)$, 
$(ii)$ $\Order(\alpha^3)$ electroweak three-loop corrections, and  
$(iii)$ $\Order(\as^3)$ QCD terms. The quadratic sums of the above corrections amount to
\begin{align}
   \deltatheo M_W      & \ \approx\  4 \; \mev\,, \\
   \deltatheo \sinleff & \ \approx\  4.7 \cdot 10^{-5}\,,
   \label{eq:Uncer}    
\end{align}
which are the theoretical ranges used in the fit. 
The empirical $W$ mass parametrisation~(\ref{eq:fitformula2}) is only valid for a
relatively light Higgs boson, $M_H \lesssim 300\; \gev$, for which the error introduced by 
the approximation is expected to be negligible~\cite{Awramik:2003rn}. For larger Higgs masses, 
the total theoretical error used is linearly increased up to $\deltatheo M_W=6\,\mev$ at $M_H = 1\; \tev$, 
which is a coarse estimate along the theoretical uncertainties given in~\cite{Awramik:2003rn}. 

Theoretical uncertainties affecting the top mass from non-perturbative colour-reconnection effects 
in the fragmentation process~\cite{Skands:2007zg,Wicke:2008iz} and due to ambiguities in the 
top-mass definition~\cite{Hoang:2008yj,Hoang:2008xm} have been recently estimated to approximately
0.5\gev each. The systematic error due to shower effects may be larger~\cite{Skands:2007zg}. 
Especially the colour-reconnection and shower uncertainties, estimated by means of a toy model, 
need to be verified with experimental data and should be included in the top-mass result published 
by the experiments. Both errors have been neglected for the present analysis.

Other theoretical uncertainties are introduced via the evolution of the QED and QCD couplings
and quark masses, and are discussed in Appendices~\ref{app:alpha} and~\ref{app:QCD}.

\subsection{Global Standard Model Analysis}

The last two decades have been proliferous in providing precision experimental data at the 
electroweak scale. Driven by measurements at LEP, SLC and the Tevatron, and significant 
theoretical progress, many phenomenological analyses have been performed, of which we re-examine
below the global SM fit. The primary goal of this re-analysis is $(i)$ to validate the new 
fitting toolkit \Gfitter and its SM library with respect to earlier 
results~\cite{Arbuzov:2005ma,Bardin:1999yd,Montagna:1998kp,Erler:2000cr}, $(ii)$
to include the results from the direct Higgs searches at LEP and the Tevatron in the global fit, 
$(iii)$ to revisit the impact of theoretical uncertainties on the results, and $(iv)$ 
to perform more complete statistical tests.

\subsubsection{Floating Fit Parameters}

The SM parameters relevant for the global electroweak analysis are the coupling constants of 
the electromagnetic ($\alpha$), weak (\GF) and strong interactions ($\as$), and the 
masses of the elementary bosons ($M_{\gamma}$, $M_Z$, $M_W$, $M_H$) and fermions ($m_f$ with 
$f=e,\mu,\tau,\nu_e,\nu_{\mu},\nu_{\tau},u,c,t,d,s,b,$), where neutrinos are taken to be massless. 
The fit simplifies with electroweak unification resulting in a massless photon and a relation
between the $W$ mass and the electromagnetic coupling $\alpha$, the $Z$ mass, 
and the weak coupling \GF, according to Eq.~(\ref{eq:GF}). Further simplification of the fit 
arises from fixing parameters with insignificant uncertainties compared to the sensitivity of 
the fit.
\bei
\item Compared to $M_Z$ the masses of leptons and light quarks are 
      small and/or sufficiently well known so that their uncertainties are negligible 
      in the fit. They are fixed to their world average values~\cite{pdg}. 
      Only the masses of the heavy quarks,\footnote
      {
         In the analysis and throughout this paper we use the 
         \MSb renormalised masses of the $c$ and $b$ quarks, $\mc(\mc)$ and $\mb(\mb)$, at 
         their proper scales. In the following they are denoted with $\mc$ and $\mb$ respectively.
      }
      $\mc$, $\mb$ and $\mt$, are floating in the fit while being constrained to their 
      experimental values. The top mass uncertainty has the strongest impact on the fit.

\item The weak coupling constant \GF 
      has been accurately determined through the measurement of the $\mu$ lifetime, giving
      $\GF=1.16637(1)\cdot 10^{-5}\gev^{-2}$~\cite{pdg}. The parameter is fixed
      in the fit. 

\item The leptonic and top-quark vacuum polarisation contributions to the running of the 
      electromagnetic coupling are precisely known or small. Only the hadronic contribution
      for the five lighter quarks, $\Dalphahad$, adds significant uncertainties and replaces 
      the electromagnetic 
      coupling $\alpha(M_Z^2)$ as floating parameter in the fit (\cf Appendix~\ref{app:alpha}).
\eei
With the \Rfit treatment of theoretical uncertainties four deviation parameters are introduced in the 
fit. They vary freely within their corresponding error ranges (\cf Section~\ref{sec:SMtheoerrors}).
The theoretical uncertainties in the predictions of $M_W$ and $\sinleff$ are parametrised by
$\deltatheo M_W$ and $\deltatheo \sinleff$. The form factors $\kZ{f}$ and $\rZ{f}$ 
have theoretical errors $\deltatheo \kZ{f}$ and $\deltatheo \rZ{f}$, which are treated as fully 
correlated in the fit.

In summary, the floating parameters in the global electroweak fit are 
the coupling parameters $\Dalphahad$ and $\asZ$, the masses  $M_Z$, $\mc$, $\mb$, 
$\mt$ and $M_H$, and four theoretical error parameters.

\subsubsection{Input Data}
\label{subsec:sminputData}

\begin{table}
\setlength{\tabcolsep}{0.0pc}
{\small
\begin{tabular*}{\textwidth}{@{\extracolsep{\fill}}lccccc} 
\hline\noalign{\smallskip}
& & Free & \multic{2}{c}{Results from global EW fits:} & \multic{1}{c}{\em Complete fit w/o}   \\[-0.1cm]
\rs{Parameter} & \rs{Input value} & in fit & \multic{1}{c}{\em Standard fit} & \multic{1}{c}{\em Complete fit} & \multic{1}{c}{\em exp. input in line} \\
\noalign{\smallskip}\hline\noalign{\smallskip}
$M_{Z}$ {\ft [GeV]} &  $91.1875\pm0.0021$ & yes &  $91.1874\pm0.0021$ &  $91.1877\pm0.0021$ &  $91.2001^{\,+0.0174}_{\,-0.0178}$\\
$\Gamma_{Z}$ {\ft [GeV]} &  $2.4952\pm0.0023$ & -- &  $2.4959\pm0.0015$ &  $2.4955\pm0.0015$ &  $2.4950\pm0.0017$\\
$\sigma_{\rm had}^{0}$ {\ft [nb]} &  $41.540\pm0.037$ & -- &  $41.477\pm0.014$ &  $41.477\pm0.014$ &  $41.468\pm0.015$\\
$R^{0}_{\l}$ &  $20.767\pm0.025$ & -- &  $20.743\pm0.018$ &  $20.742\pm0.018$ &  $20.717^{\,+0.029}_{\,-0.025}$\\
$A_{\rm FB}^{0,\l}$ &  $0.0171\pm0.0010$ & -- &  $0.01638\pm0.0002$ &  $0.01610\pm0.9839$ &  $0.01616\pm0.0002$\\
 $A_\ell$ $^{(\star)}$  & $0.1499\pm0.0018$ & --  & $0.1478^{+0.0011}_{-0.0010}$ & $0.1471^{+0.0008}_{-0.0009}$ & --\\
$A_{c}$ &  $0.670\pm0.027$ & -- &  $0.6682^{\,+0.00046}_{\,-0.00045}$ &  $0.6680^{\,+0.00032}_{\,-0.00046}$ &  $0.6680^{\,+0.00032}_{\,-0.00047}$\\
$A_{b}$ &  $0.923\pm0.020$ & -- &  $0.93470^{\,+0.00011}_{\,-0.00012}$ &  $0.93464^{\,+0.00008}_{\,-0.00013}$ &  $0.93464^{\,+0.00008}_{\,-0.00011}$\\
$A_{\rm FB}^{0,c}$ &  $0.0707\pm0.0035$ & -- &  $0.0741\pm0.0006$ &  $0.0737^{\,+0.0004}_{\,-0.0005}$ &  $0.0737^{\,+0.0004}_{\,-0.0005}$\\
$A_{\rm FB}^{0,b}$ &  $0.0992\pm0.0016$ & -- &  $0.1036\pm0.0007$ &  $0.1031^{\,+0.0007}_{\,-0.0006}$ &  $0.1036\pm0.0005$\\
$R^{0}_{c}$ &  $0.1721\pm0.0030$ & -- &  $0.17224\pm0.00006$ &  $0.17224\pm0.00006$ &  $0.17225\pm0.00006$\\
$R^{0}_{b}$ &  $0.21629\pm0.00066$ & -- &  $0.21581^{\,+0.00005}_{\,-0.00007}$ &  $0.21580\pm0.00006$ &  $0.21580\pm0.00006$\\
$\sinleff(Q_{\rm FB})$ &  $0.2324\pm0.0012$ & -- &  $0.23143\pm0.00013$ &  $0.23151^{\,+0.00012}_{\,-0.00010}$ &  $0.23149^{\,+0.00013}_{\,-0.00009}$\\
\noalign{\smallskip}\hline\noalign{\smallskip}
$M_{H}$ {\ft [GeV]} $^{(\circ)}$ & Likelihood ratios & yes & $ 80^{+ 30[+ 75]}_{- 23[- 41]}$ & $116.4^{+ 18.3[+ 28.4]}_{-\;\,1.3[-\;\,2.2]}$ & $ 80^{+ 30[+ 75]}_{- 23[- 41]}$\\
\noalign{\smallskip}\hline\noalign{\smallskip}
$M_{W}$ {\ft [GeV]} &  $80.399\pm0.025$ & -- &  $80.382^{\,+0.014}_{\,-0.016}$ &  $80.364\pm0.010$ &  $80.359^{\,+0.010}_{\,-0.021}$\\
$\Gamma_{W}$ {\ft [GeV]} &  $2.098\pm0.048$ & -- &  $2.092^{\,+0.001}_{\,-0.002}$ &  $2.091\pm0.001$ &  $2.091^{\,+0.001}_{\,-0.002}$\\
\noalign{\smallskip}\hline\noalign{\smallskip}
$\mc$ {\ft [GeV]} &  $1.25\pm0.09$ & yes &  $1.25\pm0.09$ &  $1.25\pm0.09$ & -- \\
$\mb$ {\ft [GeV]} &  $4.20\pm0.07$ & yes &  $4.20\pm0.07$ &  $4.20\pm0.07$ & -- \\
$m_{t}$ {\ft [GeV]} &  $172.4\pm1.2$ & yes &  $172.5\pm1.2$ &  $172.9\pm1.2$ &  $178.2^{\,+9.8}_{\,-4.2}$\\
$\dalphaHadMZ$ $^{(\dag\bigtriangleup)}$ &  $2768\pm  22$ & yes & $2772\pm  22$ & $2767^{+  19}_{-  24}$ & $2722^{+  62}_{-  53}$\\
$\alpha_{s}(M_{Z}^{2})$ & -- & yes &  $0.1192^{\,+0.0028}_{\,-0.0027}$ &  $0.1193^{\,+0.0028}_{\,-0.0027}$ &  $0.1193^{\,+0.0028}_{\,-0.0027}$\\
\noalign{\smallskip}\hline\noalign{\smallskip}
$\deltatheo M_W$ {\ft [MeV]}  & $[-4,4]_{\rm theo}$ & yes  & $4$ & $4$ & -- \\
$\deltatheo \sinleff$ $^{(\dag)}$  & $[-4.7,4.7]_{\rm theo}$ & yes  & $4.7$ & $-1.3$ & -- \\
$\deltatheo \rZ{f}$ $^{(\dag)}$ &  $[-2,2]_{\rm theo}$ & yes  & $2$ & $2$ & -- \\
$\deltatheo \kZ{f}$ $^{(\dag)}$ &  $[-2,2]_{\rm theo}$ & yes  & $2$ & $2$ & -- \\
\noalign{\smallskip}\hline
\noalign{\smallskip}
\end{tabular*}
{\ft
$^{(\star)}$Average of LEP ($A_\ell=0.1465\pm0.0033$) and SLD ($A_\ell=0.1513\pm0.0021$) measurements.
The {\em complete fit} w/o the LEP (SLD) measurement gives $A_\ell=$ $0.1472^{\,+0.0008}_{\,-0.0011}$ 
($A_\ell=$ $0.1463\pm0.0008$ 
).
$^{(\circ)}$In brackets the $2\sigma$ errors. 
$^{(\dag)}$In units of $10^{-5}$.
$^{(\bigtriangleup)}$Rescaled due to $\alpha_s$ dependency.
}}
\caption[.]{Input values and fit results for parameters of the global electroweak fit. The first and 
         second columns list respectively the observables/parameters used in the fit, and their 
         experimental values or phenomenological estimates (see text for references). 
         The subscript ``theo'' labels theoretical error ranges. 
         The third column indicates whether a parameter is floating in the fit.
         The fourth (fifth) column quotes the results of the {\em standard} ({\em complete}) {\em fit} 
         not including (including) the constraints from the direct Higgs searches at LEP and
         Tevatron in the fit. 
         In case of floating parameters the fit results are directly given,
         while for observables, the central values and errors are obtained by individual profile 
         likelihood scans. The errors are derived from the $\Delta\chi^2$ profile
         using a Gaussian approximation.
         The last column gives the fit results for each parameter without using the corresponding 
         experimental constraint in the fit (indirect determination). 
}         
\label{tab:results}
\end{table}

A summary of the input data used in the fit is given in the second column of Table~\ref{tab:results},
and discussed below.

\bei

\item The mass and width of the $Z$ boson, the hadronic pole cross section $\sigma^0_{\rm had}$, the 
partial widths ratio $R^0_\ell$, and the forward-backward asymmetries for leptons $\Afbz{\ell}$, 
have been determined by fits to the $Z$ lineshape measured precisely at LEP (see~\cite{:2005ema}
and references therein). Measurements of the \Tau polarisation at LEP~\cite{:2005ema} and the
left-right asymmetry at SLC~\cite{:2005ema} have been used to determine the lepton 
asymmetry parameter $A_\ell$. The corresponding $c$ and $b$-quark asymmetry parameters $A_{c(b)}$, 
the forward-backward asymmetries $\Afbz{c(b)}$, and the widths ratios $R^0_c$ and $R^0_b$, 
have been measured at LEP and SLC~\cite{:2005ema}. In addition, the forward-backward charge asymmetry 
($Q_{\rm FB}$) measurement in inclusive hadronic events at LEP was used to directly determine 
the effective leptonic weak mixing angle $\sinleff$~\cite{:2005ema}.
The log-likelihood function used in the fit includes the linear correlation coefficients among the 
$Z$-lineshape and heavy-flavour observables given in Table~\ref{tab:Corr}. 
\begin{table}
\setlength{\tabcolsep}{0.0pc}
{\small
\begin{tabular*}{0.43\textwidth}{@{\extracolsep{\fill}}lrrrrr} 
    \hline\noalign{\smallskip}
     & $M_Z$ & $\Gamma_Z$ & $\sigma^0_{\rm had}$ & $R^0_\ell$ & $\Afbz{\ell}$ \\
    \noalign{\smallskip}\hline\noalign{\smallskip}
    $M_Z$          & 1    &$-0.02$ &$-0.05$ & 0.03 & 0.06 \\
    $\Gamma_Z$     &      & 1     &$-0.30$ & 0.00 & 0.00 \\
    $\sigma^0_{\rm had}$  &      &       & 1     & 0.18 & 0.01 \\
    $R^0_\ell$              &      &       &       & 1     &$-0.06$ \\
    $\Afbz{\ell}$      &      &       &       &       & 1     \\
    \noalign{\smallskip}\hline
\end{tabular*}
\hspace{0.58cm}
\setlength{\tabcolsep}{0.0pc}
\begin{tabular*}{0.51\textwidth}{@{\extracolsep{\fill}}lrrrrrr} 
    \hline\noalign{\smallskip}
     & $\Afbz{c}$ & $\Afbz{b}$ & $A_{c}$ & $A_{b}$ & $R^0_{c}$ & $R^0_{b}$ \\
    \noalign{\smallskip}\hline\noalign{\smallskip}
    $\Afbz{c}$ & 1   & 0.15 & 0.04 &$-0.02$ &$-0.06$ & 0.07 \\  
    $\Afbz{b}$ &     & 1    & 0.01 & 0.06 & 0.04 &$-0.10$ \\ 
    $A_{c}$          &     &      & 1    & 0.11 &$-0.06$ & 0.04 \\
    $A_{b}$          &     &      &      & 1    & 0.04 &$-0.08$ \\
    $R^0_{c}$        &     &      &      &      & 1    &$-0.18$ \\
    \noalign{\smallskip}\hline
\end{tabular*}
}
\caption[]{Correlation matrices for observables determined by the $Z$ lineshape fit (left),
         and by heavy flavour analyses at the $Z$ pole (right)~\cite{:2005ema}.}
\label{tab:Corr}
\end{table}

\item For the running quark masses $\mc$ and $\mb$, the world average values derived in~\cite{pdg} 
are used. The combined top-quark mass is taken from the Tevatron Electroweak Working 
Group~\cite{:2008vn}.

\item For the five-quark hadronic contribution to \alphaMZ, the most recent phenomenological result 
is used~\cite{Hagiwara:2006jt} (see also the discussion in~\cite{pdg_g2review}). Its dependence 
on \asZ requires a proper rescaling in the fit (\cf Section~\ref{sec:gfitter}).\footnote
{
   \label{ftn:alpha}
   In~\cite{Hagiwara:2006jt} the light-quark hadronic contribution to \alphaMZ
   was found to be $\dalphaHadMZ=0.02768\pm 0.00022\pm 0.00002$, where the second
   error singles out the uncertainty from the strong coupling constant for which 
   $\asZ=0.118\pm0.003$ was used. Linear rescaling leads to the modified central value
   $\dalphaHadMZ=0.02768+0.00002\cdot(\asZ_{\rm fit}-0.118)/0.003$. Since $\asZ$ is a free
   fit parameter and has no uncertainty in a certain fit step the error on 
   $\dalphaHadMZ$ used in the log-likelihood function does no longer include
   the contribution from \asZ, but the corresponding variation is included in the rescaling
   of the central value only.
}

\item The LEP and Tevatron results for the $W$ mass and width are respectively
$M_W=(80.376\pm 0.033)\gev$, $\Gamma_W=(2.196\pm0.083)\gev$~\cite{Alcaraz:2006mx}, 
and $M_W=(80.432\pm 0.039)\gev$, $\Gamma_W=(2.056\pm 0.062)\gev$~\cite{:2008ut, Abazov:2003sv}.
Their weighted averages~\cite{:2008ut}, quoted without the correlation coefficient between 
mass and width, are used in the fit (\cf Table~\ref{tab:results}). 
Since a modest correlation has insignificant impact on the fit results\footnote
{
   A correlation of 0.2 between $W$ mass and width was reported for the Tevatron Run-I 
   results~\cite{:2005ema}. Assuming the same correlation for the 
   LEP and Tevatron combined values of $W$ mass and width leads to an increase of the
   $\chi^2_{\rm min}$ of the {\em standard fit} ({\em complete fit}) by 0.09 (0.23). In 
   the {\em complete fit} the central value of the Higgs mass estimate is unchanged
   (only the $+1\sigma$ bound slightly reduces by $0.6\gev$), whereas a downward shift of $1.1\gev$
   of the central value is observed for the {\em standard fit}. In both fits the changes 
   in the other parameters are negligible.
} 
it is ignored in the following. 
 
\item The direct searches for the SM Higgs boson at LEP~\cite{Barate:2003sz} and at the 
Tevatron~\cite{Group:2008ds,Bernardi:2008ee} use as test statistics the negative logarithm
of a likelihood ratio, $-2\ln\!Q$, of the SM Higgs signal plus background ($\rm s+b$) to the 
background-only ($\rm b$) hypotheses. This choice guarantees $-2\ln\!Q=0$ when there
is no experimental sensitivity to a Higgs signal. The corresponding one-sided 
confidence levels $\CL_{\rm s+b}$ and $\CL_{\rm b}$ describe the 
probabilities of upward fluctuations of the test statistics in presence
and absence of a signal ($1-\CL_{\rm b}$ is thus the probability of a false discovery).
They are derived using toy MC experiments.\footnote
{
   \label{ftn:lnQ}
   For a counting experiment with $N$ observed events and $N_s$ ($N_b\gg N_s$) expected
   signal (background) events, one has 
   $-\ln\!Q=N_s-N\ln(N_s/N_b+1)\simeq N_s(1-N/N_b)$, leading to 
   small $-\ln\!Q$ values for large $N$ (signal-like) and large $-\ln\!Q$ values for small 
   $N$ (background-like). For sufficiently large $N_s+N_b$, the test statistics $-\ln\!Q$  
   has a symmetric Gaussian probability density function. 
}

In the modified frequentist approach~\cite{read, Junk:1999kv,Read:2002hq}, a hypothesis 
is considered excluded at 95\%~CL if the ratio $\CL_{\rm s}=\CL_{\rm s+b}/\CL_{\rm b}$ is 
equal or lower than 0.05. The corresponding exclusion confidence levels defined by 
Eq.~(\ref{eq:clprob}) are given by $1-\CL_{\rm s}$ and $1-\CL_{\rm s+b}$, 
respectively. The use of $\CL_{\rm s}$ leads to a more conservative 
limit~\cite{Barate:2003sz} than the (usual) approach based on $\CL_{\rm s+b}$.\footnote
{
   Assuming a simple counting experiment with a true number of 100 background and 30 
   signal events, the one-sided probability $\CL_{\rm s+b}$ to fluctuate to equal or less 
   than 111 observed event is 0.05. The corresponding value $\CL_{\rm s}=0.05$ (which 
   does not represent a probability) is however already reached between 105 and 106 events. 
}
Using this method the combination of LEP searches~\cite{Barate:2003sz} has set the lower limit
$M_H>114.4\gev$ at 95\%~CL. For the Tevatron combination~\cite{Group:2008ds,Bernardi:2008ee}, 
ratios of the 95\% CL cross 
section limits to the SM Higgs boson production cross section as a function of the 
Higgs mass are derived, exhibiting a minimum of 1.0 at $M_H=170\gev$. 
The LEP Higgs Working Group provided the observed and expected 
$-2\ln\!Q$ curves for the $\rm s+b$ and $\rm b$ hypotheses, 
and the corresponding values of the aforementioned confidence levels up to $M_H=120\gev$. 
The Tevatron New Phenomena and Higgs Working Group (TEVNPH) made the same information available for 
10 discrete data points in the mass range $155\gev\le M_H\le 200\gev$ based on preliminary 
searches using data samples of up to $3\invfb$ integrated luminosity~\cite{Bernardi:2008ee}.
For the mass range $110\gev\le M_H\le 200\gev$, Tevatron results based on 
$2.4\invfb$ are provided for $-2\ln\!Q$~\cite{Group:2008ds}, however not for the 
corresponding confidence levels.

To include the direct Higgs searches in the {\em complete} SM fit we 
interpret the $-2\ln\!Q$ results for a given Higgs mass hypothesis
as measurements and derive a log-likelihood estimator quantifying the deviation of 
the data from the corresponding SM Higgs expectation. For this purpose we transform 
the one-sided $\CL_{\rm s+b}$ into two-sided confidence levels\footnote
{
   The experiments integrate only the tail towards larger $-2\ln\!Q$ values of the probability 
   density function to compute $\CL_{\rm s+b}$ (corresponding to a counting experiment with to 
   too few observed events with respect to the $\rm s+b$ hypothesis), which is later used 
   to derive $\CL_{\rm s}$ in the modified frequentist approach. They thus quantify 
   Higgs-like (not necessarily SM Higgs) enhancements in the data.
   In the global SM fit, however, one is interested in 
   the compatibility between the SM hypothesis and the experimental data as a whole, 
   and must hence account for any deviation, including the tail towards smaller
   $-2\ln\!Q$ values (corresponding to a counting experiment with too many Higgs candidates
   with respect to the $\rm s+b$ hypothesis where, $\rm s$ labels the {\em SM Higgs} signal). 
} 
using $\CL^{\rm 2\mbox{-}sided}_{\rm s+b}= 2 \CL_{\rm s+b}$ for 
$\CL_{\rm s+b}\le0.5$ and $\CL^{\rm 2\mbox{-}sided}_{\rm s+b}= 2(1-\CL_{\rm s+b})$ for $\CL_{\rm s+b}>0.5$.
The contribution to the $\chi^2$ estimator of the fit is then obtained via
$\delta\chi^2=2\cdot[{\rm Erf}^{-1}(1-\CL^{\rm 2\mbox{-}sided}_{\rm s+b})]^2$, where ${\rm Erf}^{-1}$ 
is the inverse error function,\footnote
{
   The use of ${\rm Erf}^{-1}$ provides a consistent error interpretation when 
   (re)translating the $\chi^2$ estimator into a confidence level via
   ${\rm CL} = 1-{\rm Prob}(\chi^2,1)={\rm Erf}(\sqrt{\chi^2/2})$.
}
and where the underlying probability density function has been assumed to be symmetric 
(\cf Footnote~\ref{ftn:lnQ} on page~\pageref{ftn:lnQ}). 

For the complete mass range available for the LEP searches ($M_H\le 120\gev$), and for the high 
mass region of the Tevatron searches ($155\gev\le M_H\le 200\gev$), we employ the $\CL_{\rm s+b}$ 
values determined by the experiments. For the low-mass Tevatron results ($110\gev\le M_H\le 150\gev$), 
where the $\CL_{\rm s+b}$ values are not provided, they are estimated from the measured $-2\ln\!Q$ 
values that are compared with those expected for the $\rm s+b$ hypothesis, and using the errors 
derived by the experiments for the b hypothesis. We have tested this approximation in the high-mass 
region, where the experimental values of $\CL_{\rm s+b}$ from the Tevatron are provided, and found 
a systematic overestimation of the contribution to our $\chi^2$ test statistics of about 30\%, with 
small dependence on the Higgs mass. We thus rescale the test statistics in the mass region 
where the $\CL_{\rm s+b}$ approximation is used (\ie $110\gev\le M_H\le 150\gev$) by the
correction factor 0.77.\footnote
{
   The correction factor reduces the value of the $\chi^2$ test statistics. 
   As described in Footnote~\ref{ftn:rescale}, its application has little impact on the 
   fit results.
} 
Once made available by the TEVNPH Working Group, this approximation will be replaced 
by the published $\CL_{\rm s+b}$ values.

\begin{figure}[!t]
  \centering
   \epsfig{file=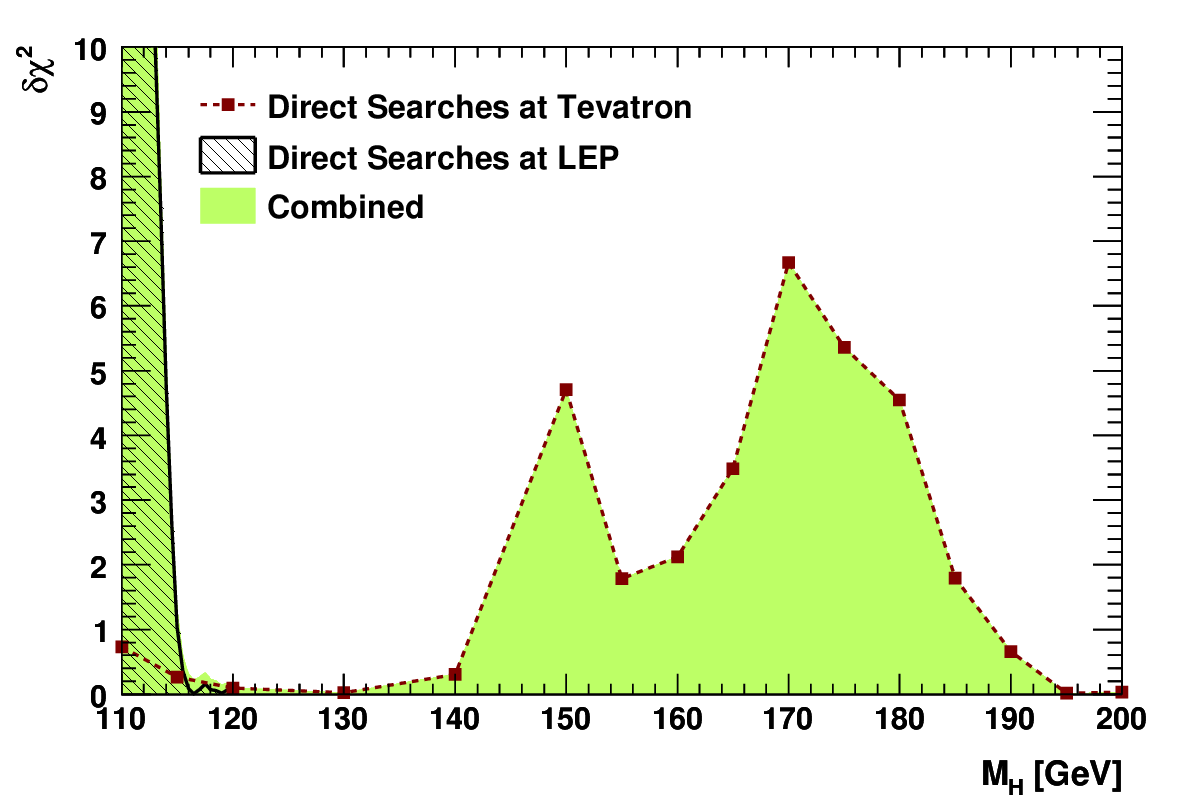, scale=\defaultFigureScale}
   \vspace{-0.0cm}
   \caption[]{The contribution to the $\chi^2$ estimator versus $M_H$ derived from the experimental information 
            on direct Higgs boson searches made available by the LEP Higgs Boson and the Tevatron 
            New Phenomena and Higgs Boson Working Groups~\cite{Barate:2003sz,Group:2008ds,Bernardi:2008ee}. 
            The solid dots indicate the Tevatron measurements. Following the original 
            figure they have been interpolated by straight lines for the purpose of presentation
            and in the fit. See text for a description of the method applied. } 
   \label{fig:directHiggsSearches}
\end{figure}

Our method follows the spirit of a global SM fit and takes advantage from downward 
fluctuations of the background in the sensitive region to obtain a more restrictive 
limit on the SM Higgs production as is obtained with the modified frequentist approach.
The resulting $\chi^2$ curves versus $M_H$ are shown in Fig.~\ref{fig:directHiggsSearches}. The 
low-mass exclusion is dominated by the LEP searches, while the information above $120\gev$ is 
contributed by the Tevatron experiments. Following the original figure, the Tevatron 
measurements have been interpolated by straight lines for the purpose of presentation
and in the fit which deals with continuous $M_H$ values. 

\eei

Constraints on the weak mixing angle can also be derived from atomic parity violation 
measurements in caesium, thallium, lead and bismuth. For heavy atoms one determines the 
{\em weak charge}, $Q_W\approx Z(1-4\sin^2\tw)-N$. Because the present experimental 
accuracy of 0.6\% (3.2\%) for $Q_W$ from Cs~\cite{Wood:1997zq,Guena:2004sq} 
(Tl~\cite{PhysRevLett.74.2654,Vetter:1995vf}) is still an order of magnitude away from a 
competitive constraint on $\sin\!^2\tw$, we do not include it into the fit. 
(Including it would reduce the error on the fitted Higgs mass by 0.2\,\gev). Due to the 
same reason we do not include the parity violation left-right asymmetry measurement using 
fixed target polarised M\o ller scattering at low 
$Q^2=0.026\,\gev^2$~\cite{Anthony:2005pm}.\footnote
{
   The main success of this measurement is to have established the running of the 
   weak coupling strength at the $6.4\sigma$ level.
}

The NuTeV Collaboration measured ratios of neutral and charged current cross sections in
neutrino-nucleon scattering at an average $Q^2\simeq 20\gev^2$ using both muon neutrino 
and muon anti-neutrino beams~\cite{Zeller:2001hh}. The results derived for the effective 
weak couplings are not included in this analysis because of unclear theoretical 
uncertainties from QCD effects such as next-to-leading order corrections 
and nuclear effects of the bound nucleon parton distribution functions~\cite{Eskola:2006ux}
(for reviews see, \eg, Refs.~\cite{Davidson:2001ji, McFarland:2003jw}).

Although a large number of precision results for $\as$ at various scales are available, 
including recent \NNNLO determinations at the \Tau-mass 
scale~\cite{Baikov:2008jh,Davier:2008sk,Beneke:2008ad,Maltman:2008nf}, we do not include 
these in the fit, because -- owing to the weak correlation between $\asZ$ and $M_H$ (\cf 
Table~\ref{tab:cormat}) -- the gain in precision on the latter quantity is insignificant.\footnote
{
   Including the constraint $\asZ=0.1212\pm0.0011$~\cite{Davier:2008sk} into the fit moves
   the central value of $M_H$ by $+0.6\mev$, and provides no reduction in the error.
}
Leaving $\asZ$ free provides thus an independent and theoretically robust determination of 
the strong coupling at the $Z$-mass scale.

The anomaly of the magnetic moment of the muon $(g-2)_\mu$  has been measured very accurately to 
a relative precision of $5\cdot10^{-7}$. Because of the small muon mass the interesting weak 
corrections only set in at a similar size, and this observable is thus not included in the analysis. 
However, the sensitivity of $(g-2)_\mu$ to physics beyond the SM (expected to couple to the 
lepton mass-squared) is similar to that of the other observables.

\subsubsection{Fit Results}
\label{sec:fitResults}

\begin{figure}[p]
   \centerline{\epsfig{file=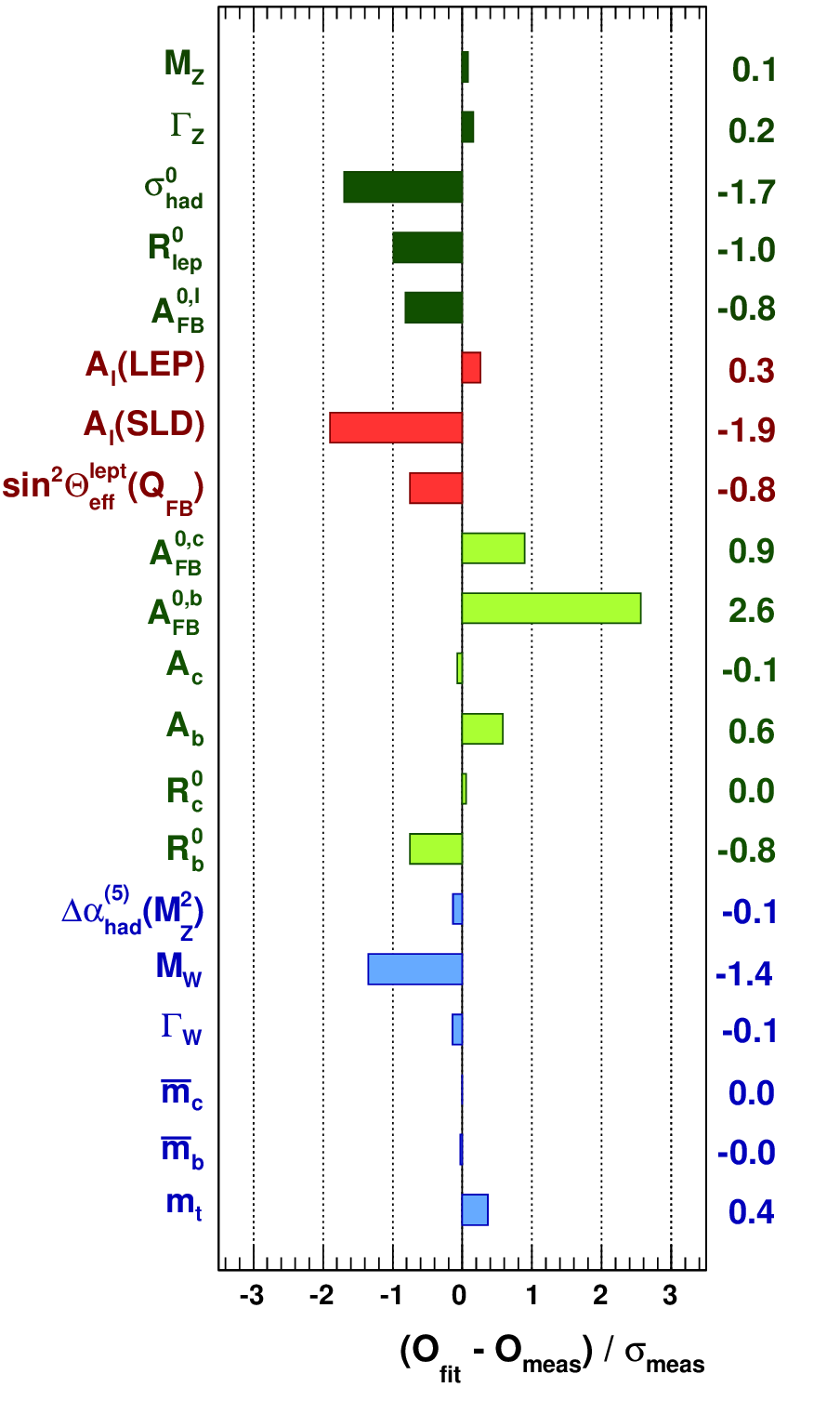, scale=0.53}
               \epsfig{file=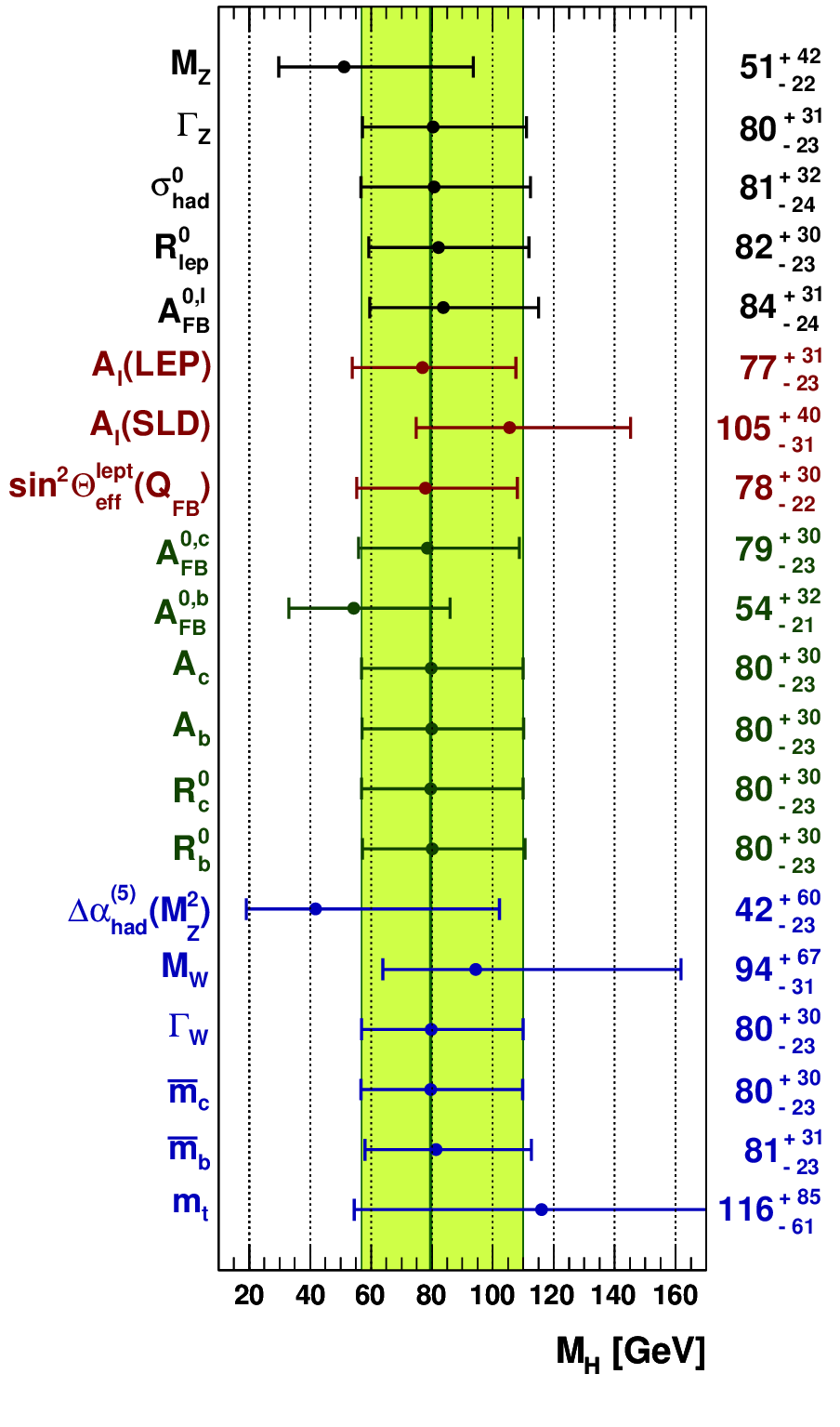, scale=0.53}}
   \vspace{-0.35cm}
   \caption{Comparing fit results with direct measurements: pull values for the 
            {\em complete fit} (left), and results for $M_H$ from the {\em standard fit} 
            excluding the respective measurements from the fit (right). }
   \label{fig:pullplot}
  \vspace{0.9cm}
  \centering
   \epsfig{file=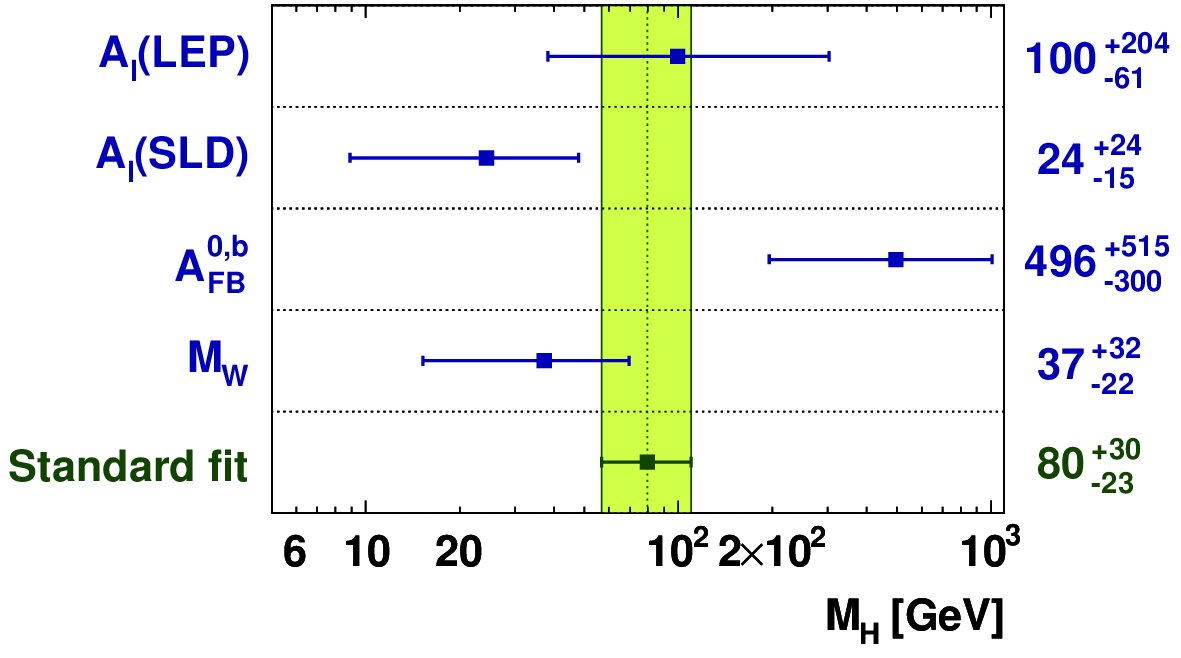, scale=0.45}
   \vspace{-0.1cm}
   \caption{Determination of $M_H$ excluding all the sensitive observables from the {\em standard fit},
            except for the one given. The results shown are not independent.
            The information in this figure is complementary to the one in the 
            right hand plot of Fig.~\ref{fig:pullplot}. 
            } 
   \label{fig:mainObservables}
\end{figure}

All fits discussed in this section minimise the test statistics $\chi^2(\ymod)$ defined in 
Eq.~(\ref{eq:chi2Function}). The $\chi^2$ function accounts for the deviations between the
observables given in Table~\ref{tab:results} and their SM predictions (including correlations).
Throughout this section we will discuss the results of two fits:
\bei

\item The {\em standard (``blue-band'') fit}, which includes all the observables listed 
      in Table~\ref{tab:results}, except for results from the direct Higgs searches.

\item The {\em complete fit} includes also the results from the direct searches for the Higgs 
      boson at LEP and the Tevatron using the method described in Section~\ref{subsec:sminputData}. 

\eei

The {\em standard} ({\em complete}) {\em fit} converges at the global minimum value 
$\ChiMin=16.4$ ($\ChiMin=18.0$) for 13 (14) degrees of freedom,
giving the naive p-value ${\rm Prob}(\ChiMin,13)=0.23$ (${\rm Prob}(\ChiMin,14)=0.21$).  
See Section~\ref{sec:probingSM} for a more accurate toy-MC-based determination of the 
p-value. The results for the parameters and observables of the two 
fits are given in columns four and five of Table~\ref{tab:results} together with their
one standard deviation ($\sigma$) intervals derived from the $\Delta\chi^2$ estimator 
using a Gaussian approximation.\footnote
{
   We have verified the Gaussian properties of the fit by sampling toy MC experiments. 
   The results are discussed in Section~\ref{sec:higgsprop}. In the following, unless otherwise
   stated, confidence levels and error ranges are derived using the Gaussian approximation 
   ${\rm Prob}(\DeltaChi,n_{\rm dof})$. 
}
We discuss in the following some of the outstanding findings and features of the fits.

\subsubsection*{Direct and Indirect Determination of Observables, Pulls}

To test the sensitivity of the SM fit to the various input observables, we consecutively 
disabled each of the observables in the fit and performed a log-likelihood scan of the disabled
observable. The corresponding results and the $1\sigma$ intervals are listed in the last column of 
Table~\ref{tab:results}. Comparing the errors obtained in these indirect determinations with the available 
measurements reveals their importance for the fit. For example, the measurement of $M_Z$ is a 
crucial ingredient, albeit the available accuracy is not required. The indirect and direct 
determinations of $M_W$ are of similar precision, such that an improved measurement would 
immediately impact the fit. The same is true for the asymmetry $A_\l$. On the other hand, 
due to an insufficient precision the heavy quark asymmetries $A_c$ and $A_b$ do not 
significantly impact the fit (the fit outperforms the measurements by almost two orders 
of magnitude in precision). 

For further illustration, the pull 
values obtained from the difference between the fit result and the measurement divided by 
the total experimental error (not including the fit error) are shown for the {\em complete fit}
in the left hand plot of Fig.~\ref{fig:pullplot} (the {\em standard fit} pulls are very similar). 
They reflect the known tension between the leptonic 
and hadronic asymmetries, though it is noticeable that no single pull value exceeds $3\sigma$. 
The pulls of the $c$ and $b$ quark masses are very small indicating that variations of 
these masses within their respective error estimates has negligible impact on the fit.
The same observation applies to $M_Z$ and \dalphaHadMZ (and to a lesser extent even to $\mt$). 
Thus, without significant impact on the goodness-of-fit fit these parameters could have been 
fixed.\footnote
{
   Fixing $\mc$, $\mb$, $\mt$, $M_Z$ and \dalphaHadMZ in the fit leads to only an insignificant 
   increase of 0.03 in the overall \ChiMin, reflecting the little sensitivity of 
   the fit to these parameters varying within the ranges of their (comparably small) measurement 
   errors. Of course, this does not prevent $M_H$ to strongly depend on the $\mt$ and \dalphaHadMZ 
   input values.
}

\begin{table}[!t]
\setlength{\tabcolsep}{0.0pc}
{\normalsize
\centering
\begin{tabular*}{\textwidth}{@{\extracolsep{\fill}}lrrrrrrr} 
    \hline\noalign{\smallskip}
      Parameter & $\ln M_{H}$ & \dalphaHadMZ & $M_{Z}$ & $\asZ $ & $\mt$ &  $\mc$ &  $\mb$ \\
    \noalign{\smallskip}\hline\noalign{\smallskip}
      $\ln M_{H}$   & 1    &  $-0.395$ &  0.113  &  0.041  & 0.309   &   $-0.001$ &  $-0.006$ \\
      \dalphaHadMZ   &      &  1     & $-0.006$ &   0.101 & $-0.007$ &  0.001 & 0.003 \\
      $M_{Z}       $  &      &        &  1     & $-0.019$ &  $-0.015$ &  $-0.000$ &  0.000 \\
      $\asZ $        &      &        &        &  1     &  0.021       &  0.011 & 0.043 \\
      $\mt       $ &      &        &        &        &  1          &     0.000 & $-0.003$ \\
       $\mc$         &      &        &        &        &             &  1     & 0.000 \\
    \noalign{\smallskip}\hline
\end{tabular*} 
}
\caption{Correlation coefficients between the free fit parameters in the {\em standard fit}. 
         The correlations with and between the varying theoretical error parameters \deltatheo 
         are negligible in all cases. The correlation between $M_H$ and the input parameter
         $M_W$ amounts to $-0.49$.}
\label{tab:cormat}
\end{table}

\subsubsection*{Correlations}

The correlation coefficients between the fit parameters of the {\em standard fit} are
given in Table~\ref{tab:cormat}. Significant are the correlations of $-0.40$ ($+0.31$) between
$\ln M_H$ and $\dalphaHadMZ$ ($\mt$). An excellent precision of these two latter quantities is 
hence of primary importance for the Higgs-mass constraint. The correlation between 
\dalphaHadMZ and \asZ is due to the dependence of the hadronic vacuum polarisation contribution 
on the strong coupling that is known to the fit (\cf comment in Footnote~\ref{ftn:alpha} on 
page~\pageref{ftn:alpha}). The correlation coefficients obtained with the {\em complete fit} 
are very similar.

\subsubsection*{Prediction of the Higgs Mass}

The primary target of the electroweak fit is the prediction of the Higgs mass. The main results
are discussed in this paragraph, while more detailed aspects concerning the statistical 
properties of the Higgs mass prediction are presented in Section~\ref{sec:higgsprop}.
The {\em complete fit} represents the most accurate estimation of $M_H$ considering all 
available data. We find
\beq 
\label{eq:MHresult1}
   M_H\ =\ 116.4^{\,+18.3}_{\,-1.3}\gev
\eeq
where the error accounts for both experimental and theoretical uncertainties. The theory parameters 
\deltatheo lead to an uncertainty of $8\gev$ on $M_H$, which does however not yet significantly impact
the error in~(\ref{eq:MHresult1}) because of the spread among the input measurements that are
sensitive to $M_H$ (\cf Fig.~\ref{fig:mainObservables}).\footnote
{  \label{ftn:theoerr}
   This is a subtle feature of the \Rfit treatment that we shall illustrate by mean of a simple 
   example. Consider two identical uncorrelated measurements of an
   observable $A$: $1\pm1\pm1$ and $1\pm1\pm1$, where the first errors are statistical and 
   the second theoretical. The weighted average of these measurements gives
   $\langle A\rangle=1\pm0.7\pm1=1\pm1.7$, where for the last term statistical and theoretical 
   errors (likelihoods) have been combined. If the two measurements 
   only barely overlap within their theoretical errors, \eg, $1\pm1\pm1$ and $3\pm1\pm1$, 
   their weighted average gives $\langle A\rangle=2\pm1$. Finally, if the two measurements are
   incompatible, \eg, $1\pm1\pm1$ and $5\pm1\pm1$, one finds $\langle A\rangle=3\pm0.7$, \ie,
   the theoretical errors are only used to increase the global likelihood value of the average,
   without impacting the error. This latter situation occurs in the $M_H$ fits discussed 
   here (although the theoretical errors in these fits are attached to the theory predictions 
   rather than to the measurements, which however does not alter the conclusion).
} 
As seen in Fig.~\ref{fig:future} of Section~\ref{sec:prospects}, once the measurements
are (made) compatible, the theoretical errors become visible by the uniform
plateau around the \DeltaChi minimum, and also fully contribute to the fit error.
The $2\sigma$ and  $3\sigma$ allowed regions of $M_H$, including all errors, are  
$[114,\,145]\gev$ and $[[113,\,168]\,{\rm and}\,[180,\,225]]\gev$, respectively.\footnote
{\label{ftn:rescale}
   A fit in which the estimated $\CL_{\rm s+b}$ values of the Tevatron searches in the 
   mass region $110\gev\le M_H\le 150\gev$ are not rescaled with the correction factor 
   0.77 (\cf. Section~\ref{subsec:sminputData}) leads to a significant increase of the 
   \DeltaChi value only for $M_H= 150\gev$. At lower masses the $\chi^2$ contributions 
   of the direct searches at the Tevatron are small. The central value of $M_H$ as well as 
   the $1\sigma$ and $3\sigma$ allowed regions are unchanged; only the $2\sigma$ intervall 
   is slightly reduced to $[114,\,144]\gev$ without the correction factor.
}
The result for the {\em standard fit} without the direct Higgs searches is
\beq
\label{eq:MHresult2}
   M_H\ =\ 80^{\,+30}_{\,-23}\:\gev\;.
\eeq
and the $2\sigma$ and $3\sigma$ intervals are respectively $[39,\,155]\gev$ and 
$[26,\,209]\gev$. The $3\sigma$ upper limit is tighter than for the complete fit because of 
the increase of the best fit value of $M_H$ in the {\em complete fit}. 
The contributions from the various measurements to the central value and error of $M_H$ in 
the {\em standard fit} are given in the right hand plot of Fig.~\ref{fig:pullplot}, where 
all input measurements except for the ones listed in a given line are used in the fit.
It can be seen that, \eg, the measurements of $\mt$ and $M_W$ are essential for an accurate
estimation of the $M_H$. 

Figure~\ref{fig:mainObservables} gives the complementary information. Among the four observables 
providing the strongest constraint on $M_H$, namely $A_\ell$(LEP), $A_\ell$(SLD), $A_{\rm FB}^{0,b}$
and $M_{W}$, only the one indicated in a given row of the plot is included in the fit.\footnote
{
   The uncertainty in the $\ymod$ parameters that are correlated to $M_H$ (mainly \dalphaHadMZ
   and \mt) contributes to the errors shown in Fig.~\ref{fig:mainObservables}, and generates
   a correlations between the four $M_H$ values found. 
}
The compatibility among these measurements (\cf Fig.~\ref{fig:mainObservables}) can be estimated
by (for example) repeating the global fit where the least compatible of the measurements (here 
$A_{\rm FB}^{0,b}$) is removed, and by comparing the $\ChiMin$ estimator obtained in that fit to the 
one of the full fit (here the {\em standard fit}). To assign a probability to the observation, the 
$\Delta\ChiMin$ obtained this way must be gauged with toy MC experiments to take into account the 
``look-elsewhere'' effect introduced by the explicit selection of the pull outlier. We find that 
in $(1.4\pm0.1)\%$ (``$2.5\sigma$'') of the toy experiments, the $\Delta\ChiMin$ found exceeds the 
$\Delta\ChiMin=8.0$ observed in current data.

In spite of the significant anticorrelation between $M_H$ and \dalphaHadMZ, the present 
uncertainty in the latter quantity does not strongly impact the precision obtained for $M_H$.
Using the theory-driven, more precise phenomenological value 
$\dalphaHadMZ=(277.0\pm1.6)\cdot10^{-4}$~\cite{Davier:1998si}, we find for the {\em standard
fit} $M_H\ =\ 80^{\,+28}_{\,-22}\gev$. 
For comparison, with $\dalphaHadMZ=(275.8\pm3.5)\cdot10^{-4}$~\cite{Burkhardt:2005se}, 
we find $M_H\ =\ 83^{\,+34}_{\,-26}\gev$, reproducing the result form the LEP Electroweak 
Working Group~\cite{LEPEWWG}.

\subsubsection*{Prediction of the Top Mass}

\begin{figure}[t!]
   \centerline{\epsfig{file=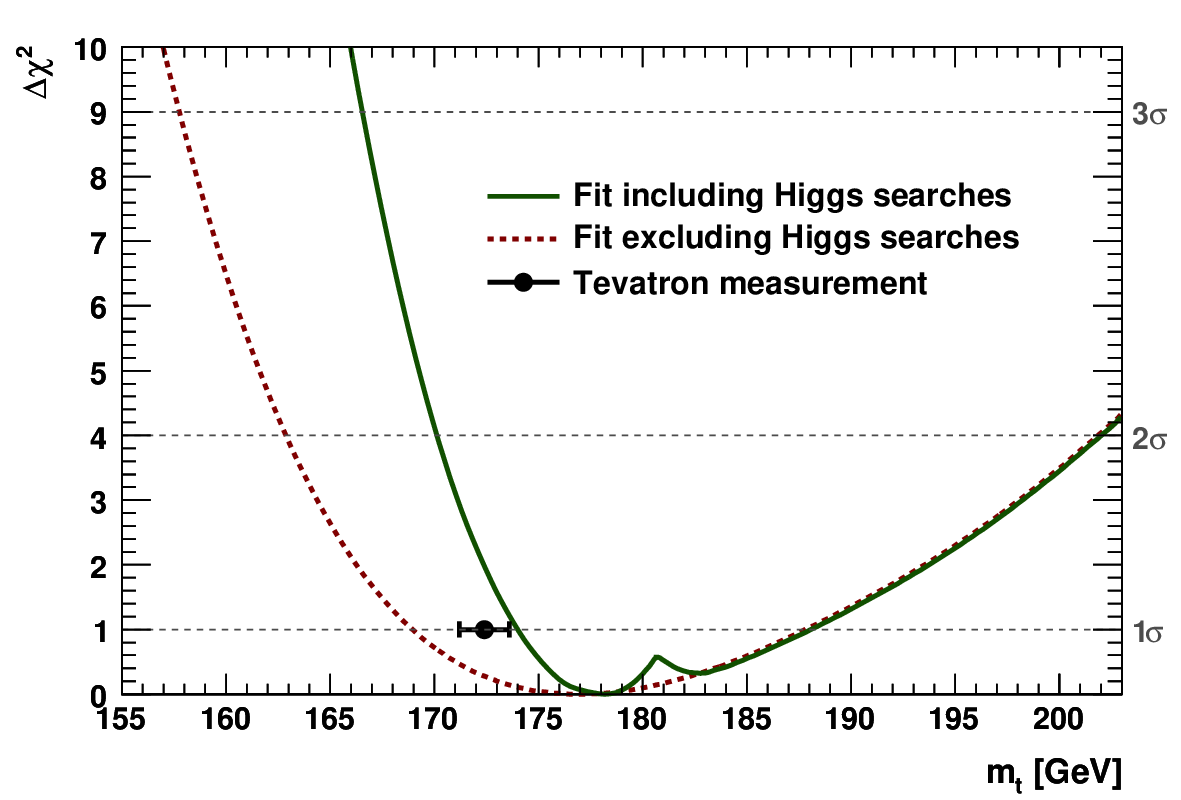, scale=\defaultFigureScale}}
  \vspace{-0.0cm}
   \caption{\DeltaChi versus $m_t$ for the {\em complete fit} (solid line) and the 
            {\em standard fit} (dashed), both excluding the direct $m_t$ measurement
            which is indicated by the dot with $1\sigma$ error bars.}
   \label{fig:topscan}
\end{figure}
Figure~\ref{fig:topscan} shows the $\DeltaChi=\chi^2-\ChiMin$ profile as a function of $m_t$ 
obtained for the {\em complete fit} (solid line) and the {\em standard fit} (dashed line), both
excluding the direct measurement of the top-quark mass from the fit. The one, two and three 
standard deviations from the minimum are indicated by the crossings with the corresponding 
horizontal lines. From the {\em complete fit} we find
\beq
     \mt\ = \ 178.2^{\,+9.8}_{\,-4.2} \:\gev\,,
\eeq
which, albeit less precise, agrees with the experimental number indicated in Fig.~\ref{fig:topscan} 
by the dot with $1\sigma$ error bars (\cf Table~\ref{tab:results}). 
The corresponding result for the {\em standard fit} is $\mt=177.0^{\,+10.8}_{\,-8.0}\gev$. 
The insertion of the direct (LEP) Higgs searches leads to a more restrictive constraint 
towards small top-quark masses. Because of the floating Higgs mass, and its positive 
correlation with $m_t$, the $\DeltaChi$ profile of the {\em standard fit} exhibits an 
asymmetry (the constraint is less restrictive towards {\em larger} $m_t$ values), which is 
opposite to the naive expectation from the dominantly quadratic $m_t$ dependence of the 
loop corrections. 

\subsubsection*{The Strong and Electromagnetic Couplings}

From the {\em complete fit} we find for the strong coupling at the $Z$-mass scale
\beq
\label{eq:alphasresult}
  \asZ=0.1193^{\,+0.0028}_{\,-0.0027} \pm 0.0001\,,
\eeq
where the first error is experimental (including also the propagated uncertainties from the 
errors in the $c$ and $b$ quark masses) and the second due to the truncation of the perturbative
QCD series. It includes variations of the renormalisation scale between 
$0.6\,M_Z<\mu<1.3\,M_Z$~\cite{Davier:2008sk}, 
of massless terms of order $\as^5(M_Z)$ and higher, and of quadratic massive terms of order and 
beyond $\as^4(M_Z)$ (\cf Appendix~\ref{app:radfunctions}).\footnote
{
   The uncertainty related to the ambiguity between the use of fixed-order perturbation theory 
   and the so-called contour-improved perturbation theory to solve the contour integration 
   of the complex Adler function has been found to be very small ($3\cdot 10^{-5}$) 
   at the $Z$-mass scale~\cite{Davier:2008sk}.
} 
Equation~(\ref{eq:alphasresult}) represents the theoretically most robust determination of $\as$
to date. It is in excellent agreement with the recent \NNNLO result from $\tau$ 
decays~\cite{Davier:2008sk,Baikov:2008jh}, $\asZ=0.1212\pm0.0005\pm0.0008\pm0.0005$, where
the errors are experimental (first) and theoretical (second and third), the latter error being
further subdivided into contributions from the prediction of the $\tau$ hadronic 
width (and spectral moments), and from the evolution to the $Z$-mass scale.\footnote
{
   Another analysis exploiting the $\tau$ hadronic width and its spectral functions, but using
   a different set of spectral moments than~\cite{Davier:2008sk}, finds 
   $\asZ=0.1187\pm0.0016$~\cite{Maltman:2008nf}.
   An analysis of the $\tau$ hadronic width relying on fixed-order perturbation theory finds 
   $\asZ=0.1180\pm0.0008$, where all errors have been added in quadrature~\cite{Beneke:2008ad}. 
} 
Because of their precision, and the almost two orders of magnitude scale difference,
the $\tau$ and $Z$-scale measurements of $\as$ represent the best current test of the 
asymptotic freedom property of QCD.

Finally, the fit result for \dalphaHadMZ without using the constraint from the phenomenological analysis 
in the fit (but including the constraint from the direct Higgs searches, \cf Table~\ref{tab:results})
precisely establishes a running QED coupling,\footnote
{
  This result is complementary (though more precise) to the LEP measurements of the scale dependence
  of $\alpha$ using, \eg, small and large-angle Bhabha scattering at low 
  energy~\cite{Abbiendi:2005rx,Acciarri:2000rx} and high energies~\cite{Achard:2005it}, 
  respectively, or cross section and asymmetry measurements at high energies~\cite{Abbiendi:2003dh}.
} 
and can be translated into the determination $\alpha^{-1}(M_Z)|_{\rm fit}=128.99\pm0.08$. The 
result is in agreement with the phenomenological value 
$\alpha^{-1}(M_Z)|_{\rm ph}=128.937 \pm 0.030$~\cite{Hagiwara:2006jt}.

\subsubsection{Properties of the Higgs-Mass Constraint}

\label{sec:higgsprop}

\begin{figure}[p]
   \centerline{\epsfig{file=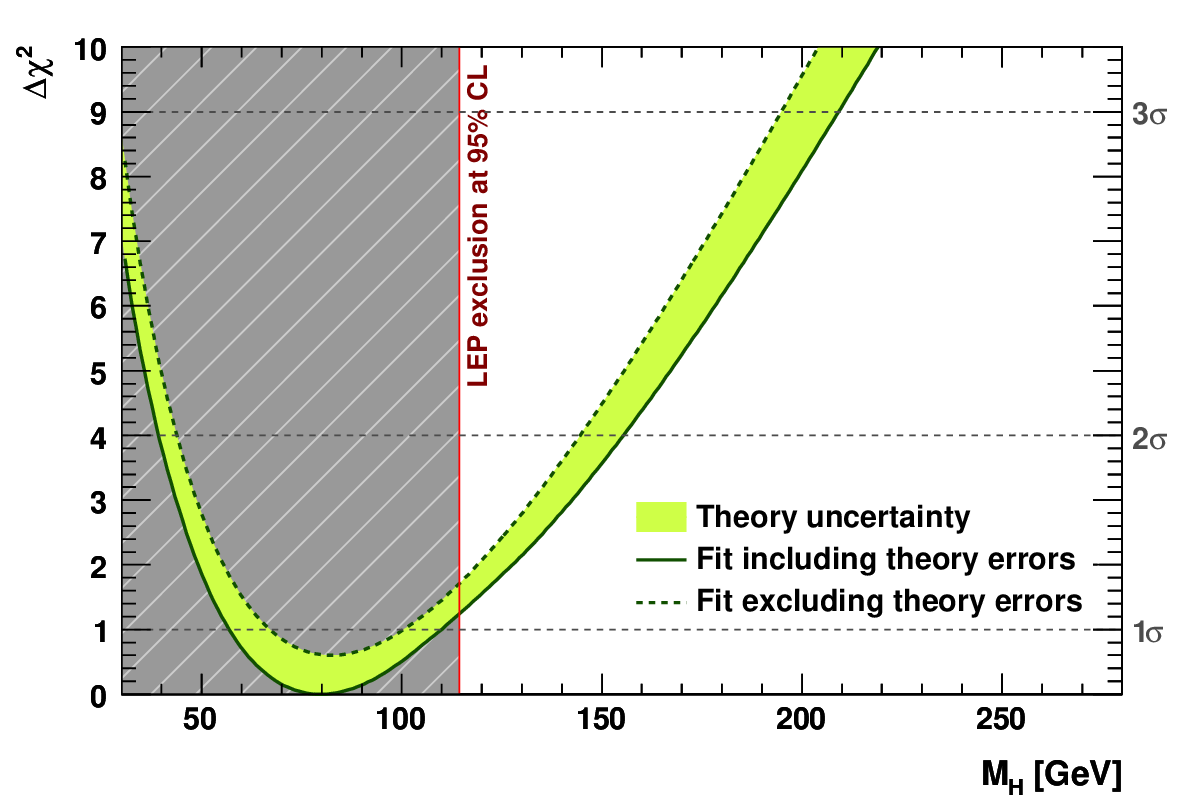, scale=\defaultFigureScale}}
  \vspace{0.3cm}
   \centerline{\epsfig{file=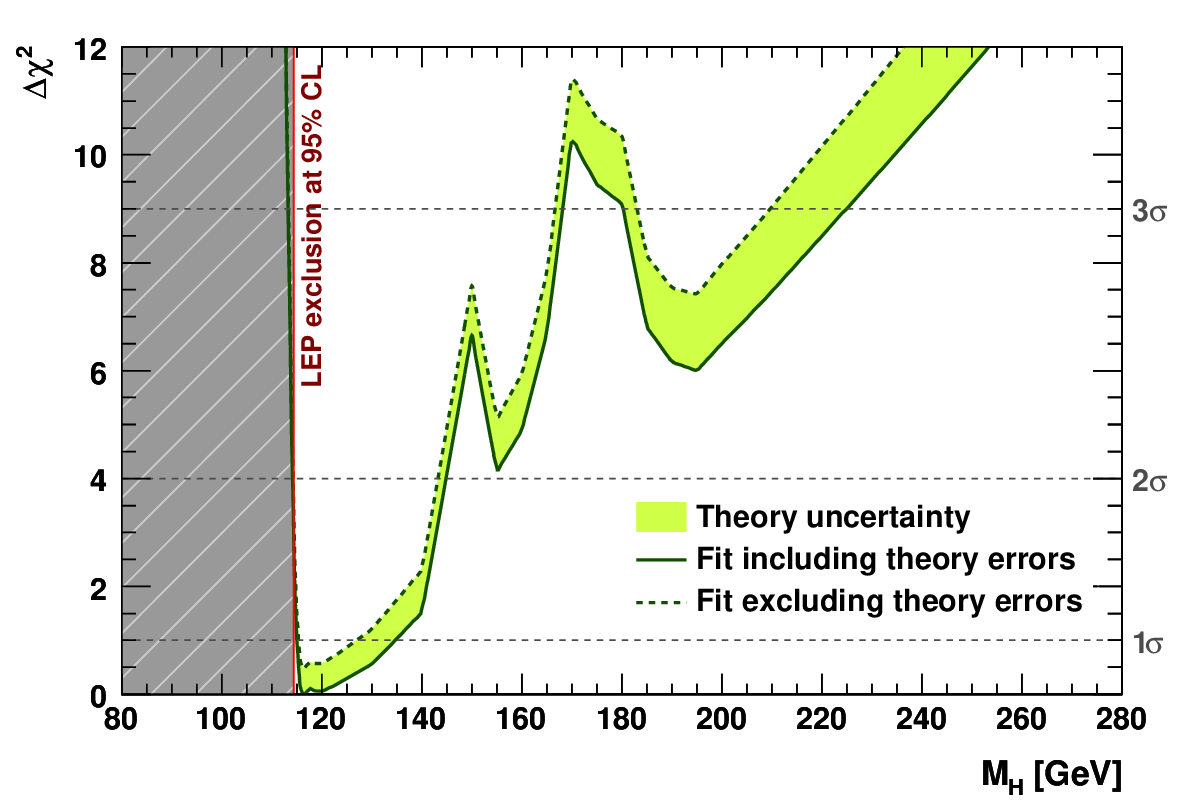, scale=\defaultFigureScale}}
  \vspace{ 0.2cm}
   \caption{\DeltaChi as a function of $M_H$ for the {\em standard fit} (top) and the 
            {\em complete fit} (bottom). The solid (dashed) lines give the results when including (ignoring) 
            theoretical errors. The minimum \ChiMin of the fit including theoretical errors is used for both 
            curves in each plot to obtain the offset-corrected \DeltaChi.}
   \label{fig:chi2lep}
\end{figure}
\begin{figure}[!t]
  \centering
   \epsfig{file=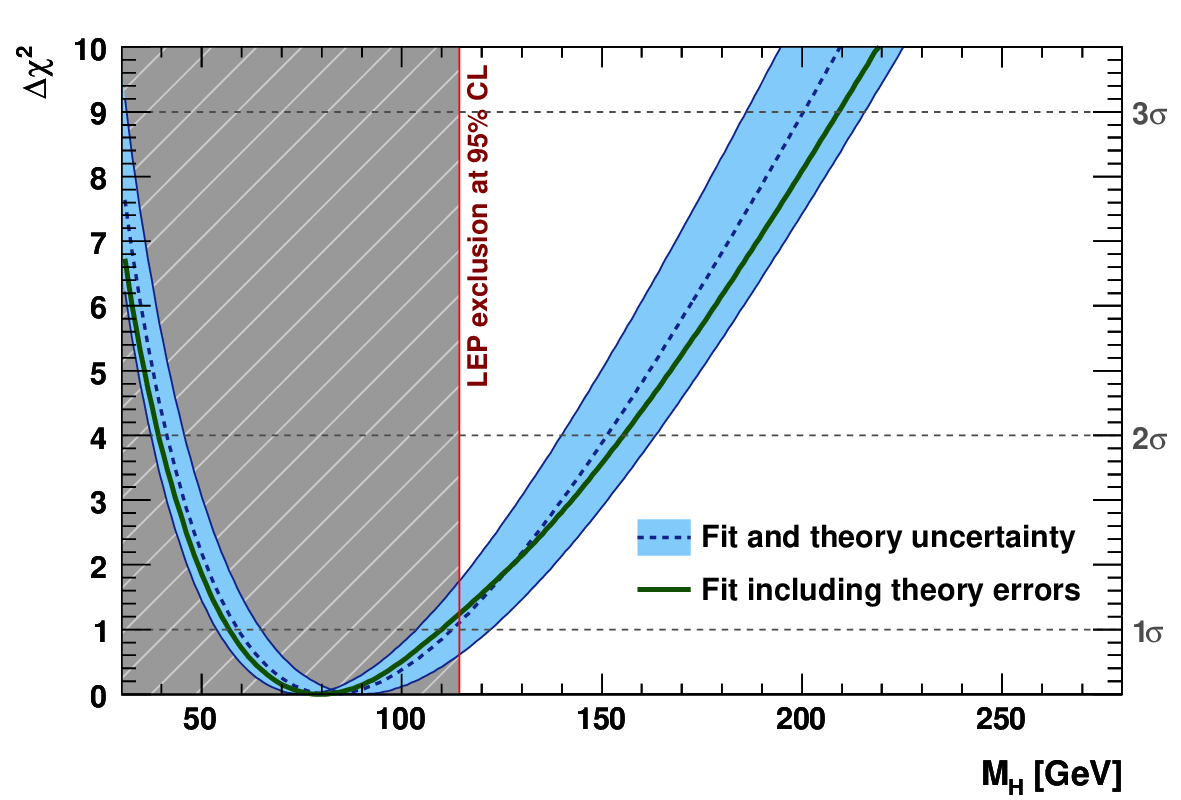, scale=\defaultFigureScale}
   \caption[]{\DeltaChi versus $M_H$ with an alternative treatment of theory 
     uncertainties~\cite{Alcaraz:2007ri}. Shown are the results of the {\em standard fit} 
     ignoring theoretical uncertainties (dotted line), the regions determined from the maximum 
     deviation in \DeltaChi achieved by shifting the SM predictions of all observables 
     according to 1 ``standard deviation'' of the various theory uncertainties (shaded band) 
     and for comparison the result of the {\em standard fit} (solid curve) in which 
     theoretical uncertainties are included in the $\chi^2$ calculation.}
   \label{fig:chi22}
\end{figure}

We proceed with studying the statistical properties of the constraints~(\ref{eq:MHresult1}) 
and~(\ref{eq:MHresult2}). Figure~\ref{fig:chi2lep} (top) shows the $\DeltaChi$ profile
versus $M_H$ obtained for the {\em standard fit} (outermost envelope). 
Also shown is the $95\%$ CL exclusion region obtained from the direct searches at 
LEP~\cite{Barate:2003sz}. It exceeds the best fit value of the {\em standard fit}. The \Rfit approach 
provides an inclusive treatment of all types of theoretical uncertainties considered in the fit. Fixing 
the $\deltatheo$ parameters at zero in the fit (which is equivalent to ignoring the corresponding 
theoretical uncertainties) results in a narrower log-likelihood curve, with a $+0.6$ larger global 
\ChiMin value, and a shift in $M_H$ at this minimum of $+2.4\gev$ with respect to 
the result of the {\em standard fit}.
The difference between the two envelopes obtained with freely varying and fixed $\deltatheo$ 
parameters is highlighted by the shaded band in Fig.~\ref{fig:chi2lep} (top).
\begin{details} 
   In previous electroweak fits~\cite{Alcaraz:2007ri} theoretical uncertainties were accounted for
   by independently shifting the SM prediction of each affected observable by the size of the 
   estimated theoretical uncertainty, and taking the maximum observed cumulative deviation in $M_H$
   as theoretical error. The error envelope obtained this way is shown in 
   Fig.~\ref{fig:chi22}. 
   The dotted curve in the middle of the shaded band is the result of a fit ignoring all
   theoretical uncertainties. The shaded band illustrates the maximum deviations of the 
   \DeltaChi curves obtained with shifted predictions. Including the systematic
   uncertainties in this way yields a $1\sigma$ interval of $[55,\,122]\gev$ and 95\% (99\%)~CL 
   upper limits of 162\gev (192\gev) respectively. For comparison the solid curve in Fig.~\ref{fig:chi22}
   shows the result of the {\em standard fit} using the \Rfit scheme.\footnote
   {
      The inclusion of the theory errors via freely varying parameters (\Rfit) leads to 
      a decrease in the global \ChiMin of the fit. Incompatibilities in the input observables
      (which may be due to statistical fluctuations) thus attenuate the numerical effect 
      of the theoretical errors on the fitted parameter (here $M_H$). See
      Footnote~\ref{ftn:theoerr} on page~\pageref{ftn:theoerr} for an illustration of this effect.
   } 
   More detailed studies of systematic theoretical uncertainties are reported in~\cite{goebel}.
\end{details}
The \DeltaChi curve versus $M_H$ for the {\em complete fit} is shown in Fig.~\ref{fig:chi2lep} (bottom). 
Again the shaded band indicates the difference between the two envelopes obtained with 
freely varying and fixed $\deltatheo$ parameters, both normalised to the same \ChiMin (from 
the fit with free \deltatheo parameters).
The inclusion of the direct Higgs search results from LEP leads to a strong rise 
of the \DeltaChi curve below $M_H=115\gev$. The data points from the direct Higgs 
searches at the Tevatron, available in the range $110\gev< M_H < 200\gev$ with linear 
interpolation between the points, increases the \DeltaChi estimator for Higgs masses above
$140\gev$ beyond that obtained from the {\em standard fit}.

\begin{figure}[!t]
   \vspace{+0.7cm}
  \centering
   \epsfig{file=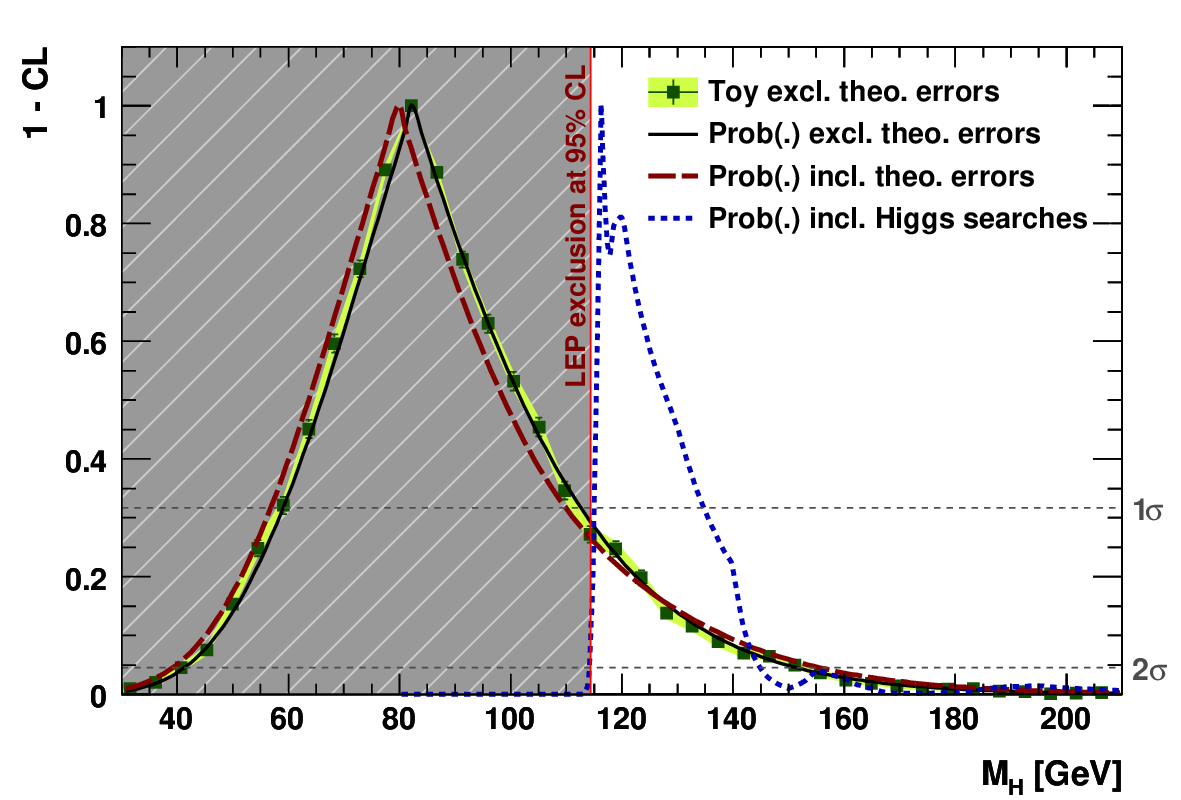, scale=\defaultFigureScale}
   \vspace{-0.0cm}
   \caption[.]{The $1-\CL$ function derived from the \DeltaChi estimator versus the $M_H$ hypothesis 
            (\cf Fig.~\ref{fig:chi2lep} (top) for \DeltaChi versus $M_H$) for the {\em standard fit}. Compared 
            are the Gaussian approximation ${\rm Prob}(\DeltaChi,1)$ for the {\em standard fit} with 
            (dashed/red line) and without theoretical errors (solid/black line), respectively, to an 
            evaluation based on toy MC simulation for which theoretical errors have been ignored.
            Also given is the result using ${\rm Prob}(.)$ for the {\em complete fit} (dotted/blue 
            line). }
   \label{fig:deltaChi2CL}
\end{figure}

We have studied the Gaussian (parabolic) properties of the \DeltaChi estimator to test whether 
the interpretation of the profile likelihood in terms of confidence levels can be simplified. 
Figure~\ref{fig:deltaChi2CL} 
gives the $1-\CL$ derived for \DeltaChi as a function of the $M_H$ hypothesis for various 
scenarios: Gaussian approximation ${\rm Prob}(\DeltaChi,1)$ of the {\em standard fit} including 
theory errors (dashed/red line), Gaussian approximation of the {\em standard fit} ignoring theory 
errors, \ie, fixing all \deltatheo parameters at zero (solid/black line), and an accurate evaluation 
using toy MC experiments ignoring theory errors (shaded/green area). Also shown is the {\em complete fit}
result with Gaussian approximation. The toy experiments are sampled using as underlying model 
the best fit parameters (and corresponding observables) obtained for each $M_H$ hypothesis.
As described in Section~\ref{sec:probingTheSM}, such a hypothesis is incomplete from a frequentist point 
of view because the true values of the nuisance parameters are unknown.\footnote
{  \label{ftn:stats}
   Examples from other particle physics areas, such as the determination of the CKM
   phase $\gamma$ via direct \CP violation measurements in $B$ decays involving charm, show that 
   this approximation can lead to severe undercoverage of the result~\cite{Charles:SOS}.
   As described in Section~\ref{subsec:statistics}, 
   the full treatment would require a numerical minimisation of the exclusion \CL with respect 
   to any true SM (nuisance) parameter set used to generate the toy MC samples 
   (\cf Refs.~\cite{Charles:SOS,Demortier:CDF}). More formally, this corresponds to solving 
   $\CL(M_H)={\rm min}_\mu\CL_\mu(M_H)$,
   where $\mu$ are the nuisance parameters of the fit and 
   $\CL_\mu(M_H)=\int_0^{\DeltaChi(M_H;{\rm data})}F(\DeltaChi|M_H,\mu)d\DeltaChi$, and where
   $F(\DeltaChi|M_H,\mu)$ is the probability density function of $\DeltaChi$ for true 
   $M_H$ and $\mu$ determined from toy MC simulation.
}
However, the persuasively Gaussian character of the fit makes us confident that our
assumption is justified in the present case (\cf the additional discussion and tests in 
Section~\ref{sec:probingSM}). The correlations given in Table~\ref{tab:Corr} are taken 
into account for the generation of the toy experiments. Theoretical errors
being of non-statistical origin have been excluded from this test, which aims at gauging the 
statistical properties of the test statistics. The curves in Fig.~\ref{fig:deltaChi2CL}
show agreement between the Gaussian approximation without theoretical errors, and the 
toy MC result. It proves that the fit is well behaved, and the \DeltaChi estimator can be 
interpreted as a true $\chi^2$ function. 

\begin{figure}[p]
   \centerline{\epsfig{file=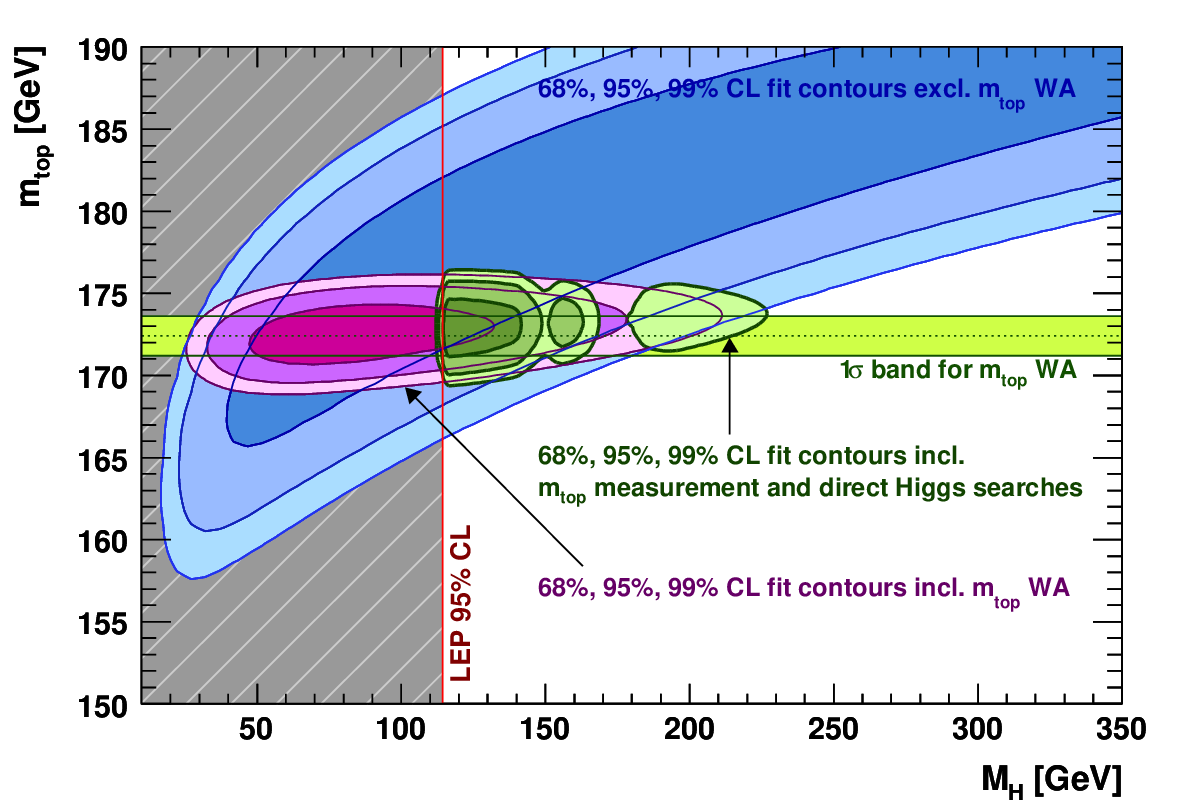, scale=\defaultFigureScale}}
   \vspace{0.3cm}
   \centerline{\epsfig{file=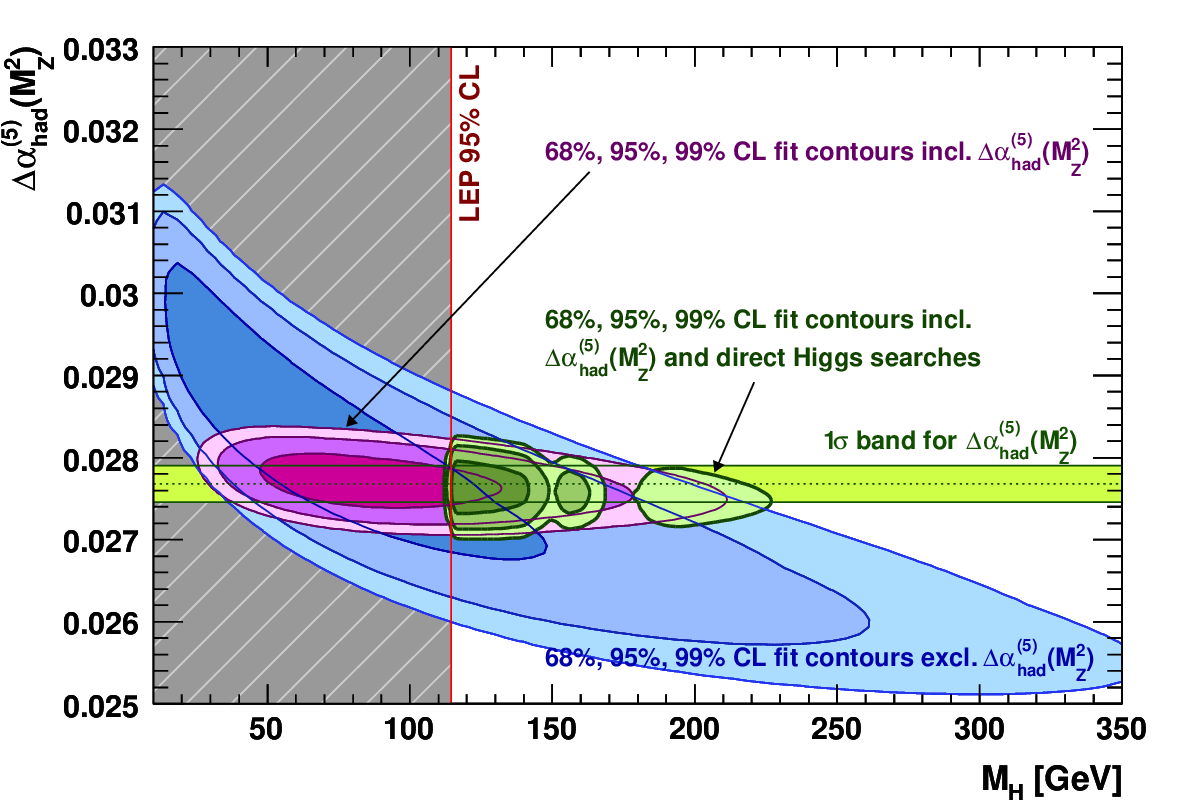, scale=\defaultFigureScale}}
   \vspace{0.2cm}
   \caption{Contours of 68\%, 95\% and 99\%~CL obtained from scans of fits with fixed
            variable pairs $\mt$ vs. $M_H$ (top) and \dalphaHadMZ vs. $M_H$ (bottom).
            The largest/blue (narrower/purple) allowed regions are the results of the {\em standard fit} 
            excluding (including) the measurements of $\mt$ (top) and \dalphaHadMZ (bottom). 
            The narrowest/green areas indicate the constraints obtained for the {\em complete fit} 
            including all the available data. The horizontal bands indicate the $1\sigma$
            regions of respectively the $\mt$ measurement and \dalphaHadMZ phenomenological 
            determination.}
   \label{fig:mtopmh}
\end{figure}

Figure~\ref{fig:mtopmh} shows the 68\%, 95\% and 99\% $\CL$ contours for the variable pairs 
$\mt$ vs. $M_H$ (top) and \dalphaHadMZ vs. $M_H$ (bottom), exhibiting the largest correlations 
in the fits. The contours are derived from the \DeltaChi values found in the profile scans using  
${\rm Prob}(\DeltaChi,2)$ (\cf discussion in Section~\ref{subsec:parameterDetermination}). 
Three sets of fits are shown in these plots: the largest/blue (narrower/purple) allowed 
regions are derived from the {\em standard fit} excluding (including) the measured values 
(indicated by shaded/light green horizontal bands) 
for respectively $\mt$ and \dalphaHadMZ in the fits. The correlations seen in these plots 
are approximately linear for $\ln M_H$ (\cf Table~\ref{tab:cormat}). The third set of fits, 
providing the narrowest constraints, uses the {\em complete fit}, \ie, including
in addition to all available measurements the direct Higgs searches. The structure of 
allowed areas reflects the presence of local minima in 
the bottom plot of Fig.~\ref{fig:chi2lep}. 

\begin{figure}[t]
  \centering
   \epsfig{file=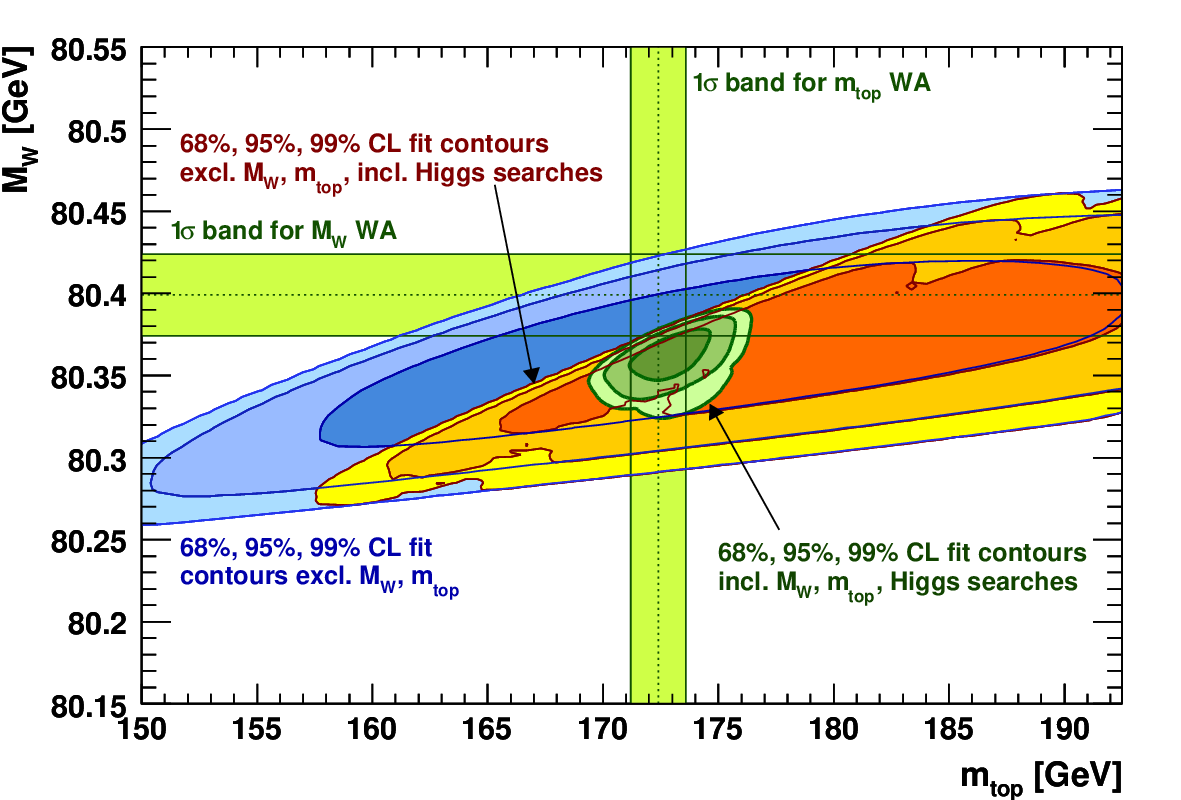, scale=\defaultFigureScale}
   \caption{Contours of 68\%, 95\% and 99\%~CL obtained from scans of fits with fixed
            variable pairs $M_W$ vs. $\mt$. The largest/blue allowed regions 
            are the results of the {\em standard} {\em fit} excluding
            the measurements of $M_W$ and $\mt$. The narrow/yellow 
            (narrowest/green) areas indicate the constraints obtained for the 
            {\em complete fit} excluding (including) the corresponding measurements. 
            The horizontal bands indicate the $1\sigma$ regions of the measurements 
            (world averages). }
   \label{fig:mwmt}
\end{figure}

Figure~\ref{fig:mwmt} compares the direct measurements of $M_W$ and $\mt$, shown by the 
shaded/green $1\sigma$ bands, with the 
68\%,  95\% and 99\%~CL constraints obtained with again three fit scenarios. The largest/blue
(narrowest/green) allowed regions are again the result of the {\em standard fit} ({\em complete fit}) 
excluding (including) the measured values of $M_W$ and $\mt$. The results of the {\em complete fit}
excluding the measured values are illustrated by the narrower/yellow allowed region. The
allowed regions of the indirect determination is significantly reduced with the insertion
of the direct Higgs searches. 
Good agreement is observed between $(i)$ indirect determination without (largest/blue area) and with 
(narrower/yellow area) the direct Higgs searches, and $(ii)$ the direct measurements (shaded/green bands). 

\subsubsection{Probing the Standard Model}
\label{sec:probingSM}

\begin{figure}[t]
  \centering
   \epsfig{file=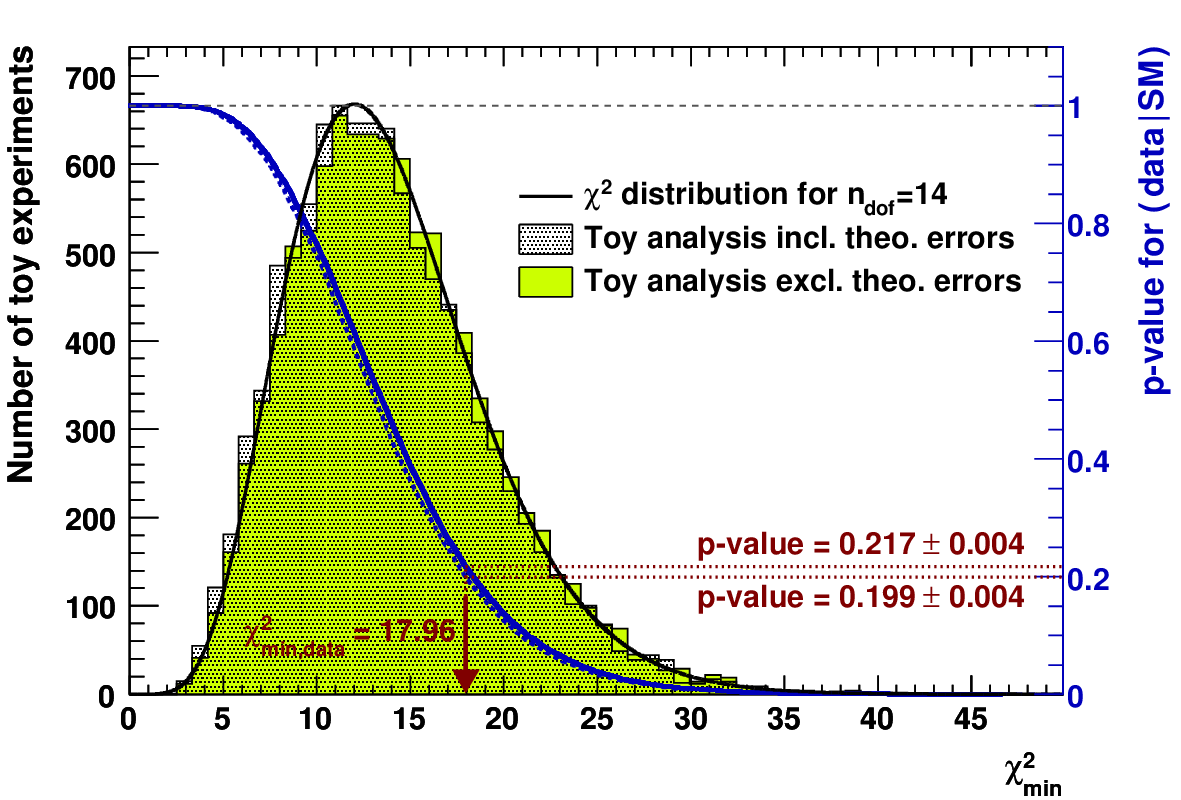, scale=\defaultFigureScale}
   \caption{Result of the MC toy analysis of the {\em complete fit}. Shown are the \ChiMin 
            distribution of a toy MC simulation (open histogram), the corresponding distribution 
            for a {\em complete fit} with fixed $\deltatheo$ parameters at zero (shaded/green histogram), 
            an ideal $\chi^2$ distribution assuming a Gaussian case with $n_{\rm dof}=14$ (black line) 
            and the p-value as a function of the \ChiMin of the global fit.}
   \label{fig:toyanalysis}
\end{figure}

We evaluate the p-value of the global SM fit following the prescription outlined in 
Section~\ref{sec:probingTheSM}. A toy MC sample with 10\,000 experiments has been 
generated using as true values for the SM parameters the outcomes of the global fit
(see the remarks below and in Section~\ref{sec:probingTheSM} and Footnote~\ref{ftn:stats} 
on page~\pageref{ftn:stats} about the limitation of this method).
For each toy simulation, the central values of all the observables used in the fit
are generated according to Gaussian distributions around their expected SM values 
(given the parameter settings) with standard deviations equal to the full experimental 
errors taking into account all correlations.\footnote
{
   Since only bounds on $M_H$ are available with no probability density information 
   given within these bounds, a random generation of $M_H$ toy measurements is not 
   possible. This experimental input is thus kept unchanged for all toy MC experiments. 
}
It is assumed that central values and errors are independent.
The \Rfit treatment of theoretical uncertainties allows the fit to adjust theoretical 
predictions and parameters at will within the given error ranges, and -- as opposed to 
measurements -- the theoretical parameters cannot be described by a probability density 
distribution and are thus not fluctuated in the toy MC. For each toy MC sample, the 
{\em complete fit} is performed (\ie, including the results 
from the direct Higgs searches) yielding the \ChiMin distribution shown
by the light shaded histogram in Fig.~\ref{fig:toyanalysis}. The distribution 
obtained when fixing the \deltatheo parameters at zero is shown by the 
dark shaded/green histogram. Including the theoretical uncertainties reduces the number 
of degrees of freedom in the data and hence shifts the distribution to lower values.
Overlaid is the $\chi^2$ function expected for Gaussian observables and 14 degrees of freedom. 
Fair agreement with the empirical toy MC distribution for fixed \deltatheo is observed.

The monotonously decreasing curves in Fig.~\ref{fig:toyanalysis} give the p-value
of the SM fit as a function of \ChiMin, obtained by integrating
the sampled normalised $\chi^2$ function between \ChiMin and infinity. The
value of the global SM fit is indicated by the arrow. Including theoretical errors in 
the fit gives
\beq
\label{eq:p-value}
    \mbox{p-value\,(data$|$SM)}\ = \ 0.22\pm0.01_{\,-0.02}\; ,
\eeq
where the first error is statistical, determined by the number of toy experiments performed,
and the second accounts for the shift resulting from fixed $\deltatheo$ parameters.
The probability of falsely rejecting the SM, expressed by the result~(\ref{eq:p-value}), is 
sufficient and no significant requirement for physics beyond the SM can be inferred from the fit.

To validate the $p_{\mu-{\rm best\;fit}}\approx{\rm min}_\mu p_\mu$ assumption used in the above
study, we have generated several true parameter sets ($\mu$) in the vicinity of 
the best fit result (varying parameters incoherently by $\pm1\sigma$ around their measurement
errors), and repeated the toy-MC based p-value evaluation for each of them. The $\chi^2$ probability
density distributions derived from these tests have been found to be compatible with each other, 
leading to similar p-values in all cases studied. It supports the robustness of 
the result~(\ref{eq:p-value}).

\begin{figure}[t]
  \centering
   \epsfig{file=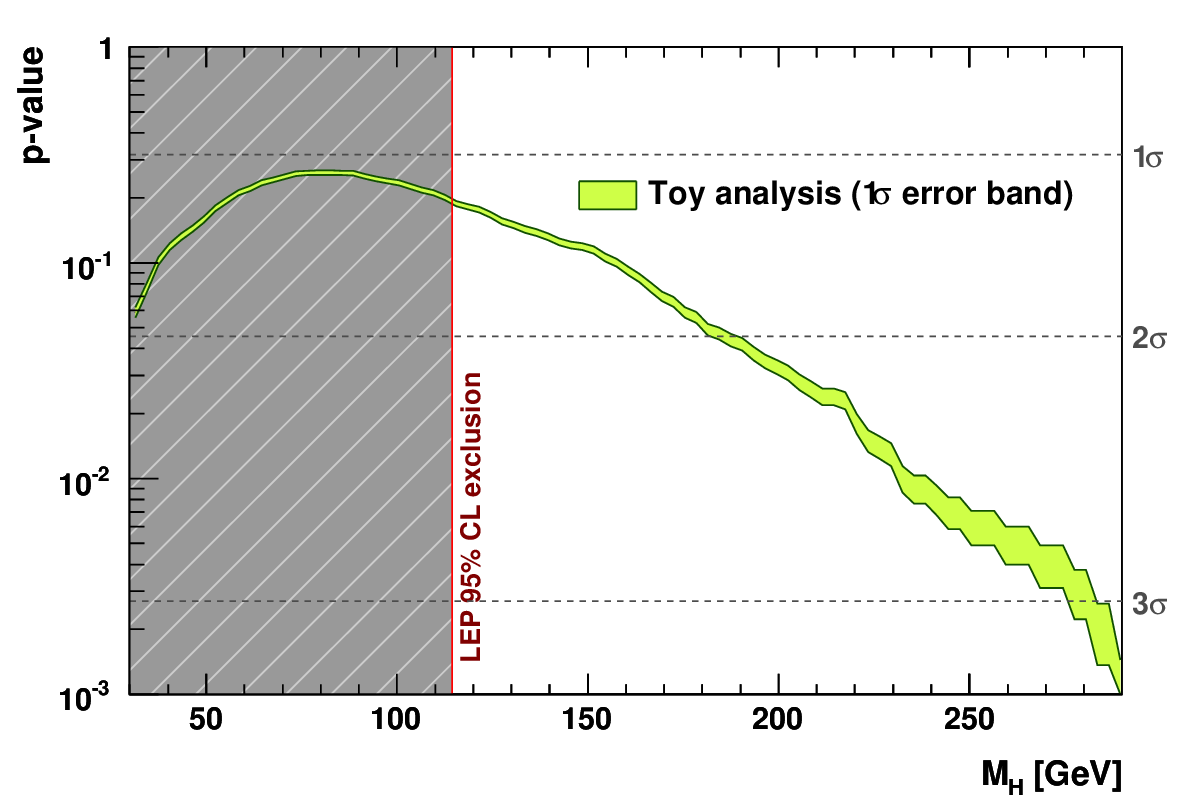, scale=\defaultFigureScale}
   \caption{P-value of the electroweak fit versus $M_H$ as obtained from toy MC simulation. 
            The error band represents the statistical error from the MC sampling.}
   \label{fig:pval}
\end{figure}

We have extended the above analysis by deriving p-values for the {\em standard fit} as a function
of the true Higgs mass. The results are shown in Fig.~\ref{fig:pval}. For values of $M_H$ around 
80\gev, corresponding to the \minchitwo of the {\em standard fit}, p-values of about 
0.25 are found.\footnote
{
   By fixing $M_H$ the number of degrees of freedom of the fit is increased compared to
   the {\em standard fit} resulting in a larger average \minchitwo and thus in a larger
   p-value.
}
With higher $M_H$ the p-value drops reaching the $2\sigma$ level at $M_H=190\gev$
and the $3\sigma$ level at $M_H=270\gev$.

\subsection{Prospects for the LHC and ILC}
\label{sec:prospects}

The next generation of particle colliders, namely LHC and ILC, have the potential to 
significantly increase the precision of most electroweak observables that are 
relevant to the fit. This will improve the predictive power of the fit, and -- in case of a Higgs 
discovery -- its sensitivity to physics beyond the SM by directly confronting theory and experiment, 
and by testing the overall goodness-of-fit of the SM.

At the LHC the masses of the $W$ boson and the top quark are expected to be
measured with precisions reaching $\sigma(M_W)=15\mev$ and
$\sigma(\mt)=1.0\gev$~\cite{atlastdr,CMSTDR,Haywood:1999qg,Borjanovic:2004ce},
respectively.\footnote
{ 
  CMS expects a systematic (statistical) precision of
  better than $20\mev$ ($10\mev$) for an integrated luminosity of
  $10\invfb$~\cite{Buge:2006dv,CMSTDR}. It uses a method based on solely the
  reconstruction of the charged lepton transverse momentum, which has reduced
  systematic uncertainties compared to reconstructing the transverse $W$ mass,
  with the downside of a smaller statistical yield. In an earlier study using
  the transverse-mass method, ATLAS finds a systematic (statistical)
  uncertainty of better than $25\mev$ ($2\mev$), for the same integrated
  luminosity~\cite{atlastdr}. Combining both, lepton channels and experiments,
  a final uncertainty of about $15\mev$ is anticipated
  in~\cite{Haywood:1999qg}, which is used here. A recent ATLAS study~\cite{Besson:2008zs}, 
  superseding their previous results, finds that uncertainties of $\sigma(M_W)\approx7\mev$ 
  may be achievable for each lepton channel (with similar uncertainties for both 
  aforementioned experimental approaches), by heavily relying on the calibration 
  of the lepton momenta and reconstruction efficiencies at the $Z$ pole. Using 
  this $\sigma(M_W)$ in the fit improves the $M_H$ determination for the LHC 
  prospective from $M_H=120^{+\,42}_{-\,33}$ to $M_H=120^{+\,31}_{-\,26}$ 
  (using the improved $\dalphaHadMZ$ error of $7\cdot 10^{-5}$ for both fits, \cf Table~\ref{tab:future}).
}  
At the ILC it is expected that the top mass can be measured to an experimental
precision of approximately $\sigma(\mt)=50\mev$ using a threshold scan and an
adapted mass definition~\cite{Djouadi:2007ik,Hoang:2000yr}. This should translates 
into an error of 100--200\mev on the \MSb-mass depending on the accuracy of the 
strong coupling constant~\cite{Djouadi:2007ik, Hoang:2000yr, Chetyrkin:1999qi}.  
More improvements are expected for a linear collider running with high 
luminosity and polarised beams at the $Z$ resonance (GigaZ). 
The $W$-mass can be measured to 6\mev from a scan of the $WW$ threshold~\cite{Djouadi:2007ik}. 
The effective weak mixing angle for leptons can be measured to a precision of $1.3\cdot10^{-5}$ 
from the left-right asymmetry, $A_{\rm LR}$~\cite{Djouadi:2007ik, Hawkings:1999ac}. At the same
time, the ratio of the $Z$ leptonic to hadronic partial decay widths, $R_\l^0$, can be
obtained to an absolute experimental precision of 0.004~\cite{Winter:2001}. 
These numbers do not include theoretical uncertainties since it is assumed that
substantial theoretical progress will be realised in the years left before these 
measurements are possible.

\begin{table}[t]
\setlength{\tabcolsep}{0.0pc}
{\normalsize
\begin{tabular*}{\textwidth}{@{\extracolsep{\fill}}lcccc} 
\hline\noalign{\smallskip}
 & \multic{4}{c}{Expected uncertainty} \\
  \rs{Quantity} & Present & LHC & ILC & GigaZ (ILC) \\
  \noalign{\smallskip}\hline\noalign{\smallskip}
  $M_W ~ [\mev]$              & 25   & 15 & 15   & 6   \\
  $\mt ~ [\gev]$              & 1.2  & 1.0 & 0.2 & 0.1 \\
  $\sinleff ~ [10^{-5}]$      & 17   & 17  & 17   & 1.3 \\
  $R^{0}_{\l} ~ [10^{-2}]$     & 2.5  & 2.5 & 2.5 & 0.4 \\
  $\dalphaHadMZ ~ [10^{-5}]$  & 22 (7)  & 22 (7)  & 22 (7)  & 22 (7) \\
  \noalign{\smallskip}\hline\noalign{\smallskip}
  $M_H(=120\gev) ~ [\gev] ~$         & $^{+\,  56}_{-\,  40}$ $\left(^{+\,  52}_{-\,  39}\right)$ $\left[^{+\,  39}_{-\,  31}\right]$
& $^{+\,  45}_{-\,  35}$ $\left(^{+\,  42}_{-\,  33}\right)$ $\left[^{+\,  30}_{-\,  25}\right]$
& $^{+\,  42}_{-\,  33}$ $\left(^{+\,  39}_{-\,  31}\right)$ $\left[^{+\,  28}_{-\,  23}\right]$
& $^{+\,  27}_{-\,  23}$ $\left(^{+\,  20}_{-\,  18}\right)$ $\left[^{+\,   8}_{-\,   7}\right]$
\\
$\asZ ~ [10^{-4}]$   & $  28$
 & $  28$
 & $  27$
 & $   6$
\\
\noalign{\smallskip}\hline
\end{tabular*} 
}
\caption{Measurement prospects at future accelerators for key observables used in the electroweak 
         fit, and their impact on the electroweak fit. The columns give, from the left to 
         the right: present errors, the expected uncertainties for the LHC with
         $10\invfb$ integrated luminosity, the ILC without and with the option to run 
         at the $Z$ resonance and along the $W$-pair production threshold (GigaZ)
         for one year of nominal running. The estimated improvement for \dalphaHadMZ (given in 
         parenthesis of the corresponding line) over the current uncertainty is unrelated to these 
         accelerators, and must come from new low-energy hadronic cross section measurements and 
         a more accurate theory (see text). The lower rows give the results obtained for $M_H$ and 
         $\asZ$. For $M_H$ are also given the results with improved \dalphaHadMZ precision 
         (parentheses -- this has no impact on \asZ), and when in 
         addition ignoring the theoretical uncertainties [brackets]. Note that all errors obtained 
         on $M_H$ are strongly central value dependent (see text). 
}
\label{tab:future}
\end{table}

\begin{figure}[!h]
  \centering
   \epsfig{file=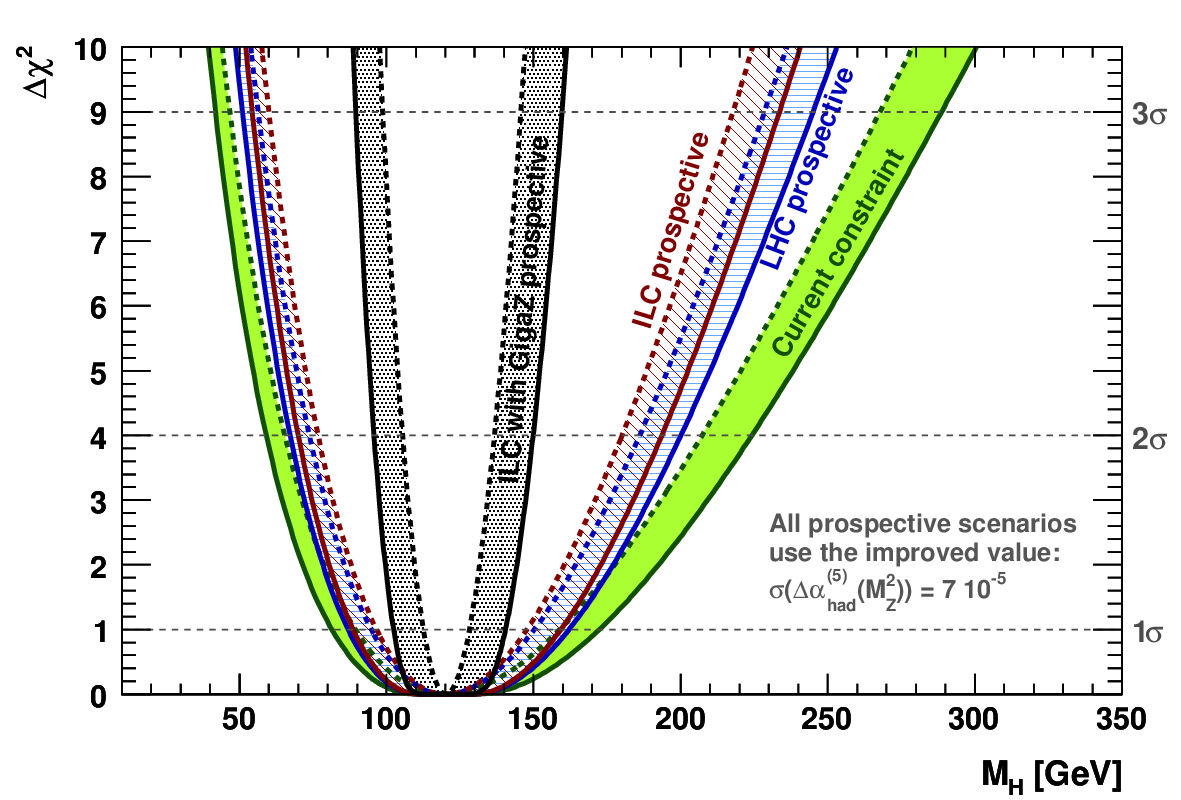, scale=\defaultFigureScale}
   \caption[.]{Constraints on $M_H$ obtained for the four scenarios given in Table~\ref{tab:future}, 
            assuming the improvement $\sigma(\dalphaHadMZ)=7\cdot 10^{-5}$ for all prospective curves. 
            Shown are, from wider to narrower \DeltaChi curves: present constraint, 
            LHC expectation, ILC expectations with and without GigaZ option. 
            The $1\sigma$ errors for $M_H$ given in Table~\ref{tab:future} correspond 
            to the $\DeltaChi=1$ intervals obtained from these graphs. The shaded bands 
            indicate the effect of theoretical uncertainties. 
          }
   \label{fig:future}
\end{figure}
At the time when the new measurements from the LHC experiments, and later the
ILC, become available, an improved determination of \dalphaHadMZ will be
needed to fully exploit the new precision data. This in turn requires a
significant improvement in the quality of the hadronic cross section data 
at energies around the $c\cbar$ resonances and below,
and a better knowledge of the $c$ and $b$ quark masses entering the
perturbative prediction of the cross sections where applicable, which serve as
input to the dispersion integral. Reference~\cite{Jegerlehner:2001ca} quotes
expected uncertainties of $\sigma(\dalphaHadMZ)\sim7\cdot10^{-5}$ and
$5\cdot10^{-5}$, compared to presently $22\cdot10^{-5}$, if the relative
precision on the cross sections attains $1\%$ below the $J/\psi$ and the
$\Upsilon$ resonances, respectively. The former estimate will be used for the
present study.  Since most of present data is dominated by systematic
uncertainties, measurements of state-of-the-art experiments with better
acceptance and control of systematics are needed. High-statistics ISR analyses
performed at the $B$ and $\Phi$ factories already provided promising results
on many exclusive hadronic channels.  New data will also come from the BESIII
experiment at the BEPCII $\ee$ collider that starts operation in Summer 2008.

The dominant theoretical uncertainties affecting the electroweak fit arise from the missing higher
order corrections in the predictions of $M_W$ and $\sinleff$ (\cf Section~\ref{sec:SMtheoerrors}),
which contribute similarly to the error on $M_H$. They 
amount to $10\gev$ ($13\gev$) at $M_H=120\gev$ ($150\gev$). Significant theoretical 
effort is needed to reduce these.

A summary of the current and anticipated future uncertainties on the
quantities $M_W$, $\mt$, $\sinleff$, $R_\ell^0$, and \dalphaHadMZ, for
the LHC, ILC, and the ILC with GigaZ option, is given in
Table~\ref{tab:future}.  By using these improved measurements the
global SM fit (not using the results from direct Higgs searches nor
measurements of \asZ) results in the constraints on the Higgs mass and
\asZ quoted in Table~\ref{tab:future}.  For all four scenarios the
true Higgs mass has been assumed to be $M_{H}=120\gev$ and the central
values for all observables are adjusted such that they are consistent
with this $M_H$ value. All fits
are performed using respectively the present uncertainty on
\dalphaHadMZ, and assuming the above-mentioned improvement. For the latter
case results for $M_H$ are given including (parentheses) and excluding [brackets] 
theory uncertainties. With the GigaZ option, the
uncertainty from \dalphaHadMZ would dominate the overall fit error on
$M_H$ if no improvement occurred. We emphasise that due (by part) to
the logarithmic dependence, the error obtained on $M_H$ is strongly
$M_H$ dependent: with the same precision on the observables, but central values 
that are consistent with a true value of $150\gev$, one would find
$M_{H}=150^{\,+66}_{\,-49}\gev$ in average, \ie, an error increase
over the $M_{H}=120\gev$ case of almost $30\%$.  With the GigaZ option
and the resulting improvement for $R_\l^0$ the uncertainty on \asZ from the 
fit is reduced by a factor of four.

The $M_H$ scans obtained for the four scenarios, assuming the improved \dalphaHadMZ precision
to be applicable for all future (LHC and beyond) scenarios, are shown in Fig.~\ref{fig:future}. 
The shaded bands indicate the effect of the current theoretical uncertainties.
As expected the theoretical errors included with the \Rfit scheme are visible by 
a broad plateau around the $\Delta\chi^2$ minimum.

A discovery of the Higgs boson at the LHC in the clean decay mode $H\to\gamma\gamma$
($H\to2\ell2\ell^\prime$) for a light (heavy) Higgs would soon allow a precision
measurement of $M_H$ beyond the percent level. Inserting the measurement into the 
global electroweak fit would lead to a prediction of the $W$ boson mass with $13\mev$ error, 
of which $5\mev$ is theoretical. Prediction and measurement could be directly confronted. 
More inclusively, the p-value of the data given the SM could be determined as a direct
test of the goodness of the SM fit.

\section{Extending the SM Higgs Sector -- The Two Higgs Doublet Model}
\label{sec:2hdmfit}

Two Higgs Doublet Models (2HDM)~\cite{Haber:1978jt} are simple extensions to the SM 
Higgs sector, only introducing an additional $SU(2)_L\times U(1)_Y$ Higgs doublet with 
hypercharge $Y=1$, leading to five physical Higgs bosons. 
Three Higgs bosons ($A^0$, $h^0$, $H^0$) are electrically neutral and the two remaining ones 
(${H^{\pm}}$) are electrically charged. The free parameters of the 2HDM are the Higgs boson 
masses $M_{A^0}, M_{h^0}, M_{H^0}$ and $\MHp$, the ratio of the vacuum expectation values 
of the two Higgs doublets $\tanb=v_2/v_1$, occurring in the mixing of charged and neutral Higgs
fields, and the angle $\alpha$, governing the mixing of the neutral \CP-even Higgs fields.
It should be noted that, in the most general 2HDM, $\tanb$ and hence the corresponding 
Higgs couplings and mass matrix elements depend on the choice of basis for the Higgs 
fields~\cite{Haber:2006ue,Davidson:2005cw}. 

Models with two Higgs doublets intrinsically fulfil the empirical equality 
$M_W^2\approx M_Z^2 \cos^2 \theta_W$. They also increase the maximum allowed 
mass of the lightest neutral Higgs boson for electroweak Baryogenesis scenarios to values not 
yet excluded by LEP (see, \eg, \cite{Cline:1996mga}), and introduce \CP violation in the Higgs 
sector. Flavour changing neutral currents can be suppressed with an appropriate choice of the 
Higgs-fermion couplings (see \eg, Ref.~\cite{Gunion:1989we,Gunion:1992hs}).
For example, in the {\em Type-I} 2HDM this is achieved by letting only one Higgs doublet
couple to the fermion sector. In the {\em Type-II} 2HDM~\cite{Abbott:1979dt}, which is 
chosen for this analysis, one Higgs doublet couples to the up-type quarks and leptons only, 
while the other one couples only to the down-type quarks and leptons. The Type-II 2HDM resembles 
the Higgs sector in the Minimal Supersymmetric Standard Model. It fixes the basis of the Higgs
fields and promotes \tanb to a physical parameter.

Our analysis is restricted to the \CP conserving 2HDM scalar potential and furthermore we only consider 
observables that are sensitive to corrections from the exchange of a charged Higgs boson. In the Type-II 
2HDM the charged Higgs-fermion interaction Lagrangian is given by~\cite{Gunion:1989we,Gunion:1992hs}
\beq
\label{eq:thdmlagr}
{\cal L}_{H^\pm\!ff} = \frac{g}{2\sqrt{2}m_W} \Big( H^+ \bar{U} \big(M_U K \left(1 - \gamma_5\right)\cotb 
                                                + K M_D \left(1 + \gamma_5\right)\tanb \big) D + {\rm h.c.} \Big)\,,
\eeq
where $U$ and $D$ are column matrices of three generation up-type and down-type quark fields, respectively,
$M_U$ and $M_D$ are the corresponding diagonal mass matrices, and $K$ is the Cabibbo-Kobayashi-Maskawa 
quark-mixing matrix. The charged Higgs interaction has the same structure as the charged current mediated 
by the $W$. Significant charged Higgs couplings to light quarks can occur for large values of $\tanb$.

By investigating observables that are sensitive to corrections from a charged Higgs exchange we derive 
constraints on the allowed charged-Higgs mass $\MHp$ and $\tanb$. Direct searches for the charged Higgs 
have been performed at LEP and the Tevatron. LEP has derived a lower limit of $\MHp>78.6\gev$ at 
95\%~CL~\cite{LEPHiggsWG}, for any value of \tanb.

\subsection{Input Observables}

The constraints on the charged Higgs are currently dominated by indirect measurements, as
opposed to direct searches at high-energy accelerators. A multitude of heavy flavour 
observables mainly from $B$-meson decays is available whose sensitivity to the 2HDM 
parameters varies however substantially, either due to limited experimental precision 
in case of rare decays, or because specific 2HDM contributions are strongly suppressed. 
The most relevant observables for the search of Type-II 2HDM signals
are the electroweak precision variable $R_b^0$, branching fractions of rare semileptonic 
$B$, $D$ and $K$ decays, and loop-induced radiative $B$ decays.\footnote
{
  Decays of $\tau$ and $\mu$ 
  leptons can also occur through charged-Higgs tree diagrams giving anomalous contributions 
  to the decay parameters ({\em Michel parameters}~\cite{Michel:1949qe}) measured in these 
  decays. Their present sensitivity is however not competitive with the other observables 
  (a 95\%~CL limit of $\MHp>1.9\gev\cdot\tanb$ is currently achieved from $\tau$ 
  decays~\cite{pdg_mureview}, see also~\cite{pdg_taureview} for a review of the $\mu$ decay 
  parameters).
} 
A summary of the experimental input used for this analysis is given in 
Table~\ref{tab:2HDMinput}. 

\subsubsection{Hadronic Branching Ratio of $\mathsf Z$ to $\mathsf b$ Quarks $\mathsf {R_b^0}$ }

\newcommand{\bm}[1]{\mbox{\boldmath{$#1$}}}

The sensitivity of $R_b^0$ to a charged Higgs boson arises from an exchange diagram
modifying the $Zb\bbar$ coupling. The corresponding corrections of the SM prediction 
have been calculated 
in Ref.~\cite{Haber:1999zh} and are given in Eqs. (6.3) and (6.4) thereof.\footnote
{
  These equations contain a misprint: the common factors $e/(s_W c_W)$
  should be removed. We are grateful to Pietro Slavich and Giuseppe Degrassi for 
  drawing our attention to this. 
}
The left- (right-) handed corrections to the effective couplings $\delta g^{L(R)}$ are 
proportional to $\cot^2\!\beta$ ($\tan^2\!\beta$) and to $R/(R-1)-R \log R/(R-1)^2$, 
where $R = \mt^2/\MHp^2$. The charged-Higgs exchange leads to a decrease of $R_b^0$. 
Neutral Higgs contributions can be neglected for small \tanb. For the SM prediction we use 
the result from the {\em complete} electroweak fit, $R_{b,{\rm SM}}^0= 0.21580 \pm 0.00006$, 
where the direct measurement of $R_b^0$ has been excluded (\cf Table~\ref{tab:results}).
It is confronted in the fit with the experimental value $R_{b,{\rm exp}}^0 = 0.21629 \pm 0.00066$, 
obtained at LEP~\cite{:2005ema}, giving 
$\Delta R_b^0 = R_{b,{\rm exp}}^0-R_{b,{\rm SM}}^0 = 0.00049 \pm 0.00066$.

\subsubsection{The Decay $\mathsf{B\to X_s\gamma}$}

The decay \btoxsg is an effective flavour changing neutral current process occurring
only at loop-level in the SM. The SM prediction for its branching fraction ($\BR$) at 
NNLO accuracy is $(3.15 \pm 0.23) \cdot 10^{-4}$~\cite{Misiak:2006zs}, where
the theoretical uncertainty is estimated by studying (in decreasing order of importance) 
nonperturbative, parametric, higher-order and $m_c$ interpolation ambiguity effects, and
where all errors have been added in quadrature. 
Averaging branching fraction measurements from the BABAR, Belle and CLEO Collaborations 
gives $\BR(B\to X_s\gamma)=(3.52 \pm 0.23 \pm 0.09)\cdot 10^{-4}$~\cite{hfag08}, where the
first error is experimental and the second stems from the modelling of the photon 
energy spectrum.
The improved NNLO calculation yields a branching fraction approximately 1.5 $\sigma$ lower 
than the NLO calculation~\cite{Gambino:2001ew}, resulting in a small tension with the 
experimental average and thus leading to less stringent constraints on the charged Higgs mass.
The 2HDM contribution to the $\BR(\btoxsg)$ arises from a charged Higgs replacing the 
$W^{\pm}$ in the loop from which the photon is radiated and is always positive in the type-II model. 
For the prediction of $\BR(\btoxsg)$ in the 2HDM we have used parametrised formulae~\cite{paolo} 
reproducing the result of~\cite{Misiak:2006zs} within 0.2\%. While the value of the branching 
fraction changes with \MHp and, to a lesser extent with \tanb, the associated 
theoretical uncertainty stays to good approximation constant at 7\%. Since it has been 
derived by quadratically combining several error estimates, we treat it as an additional 
Gaussian systematic error in the fit.

\subsubsection{Leptonic Decays of Charged Pseudoscalar Mesons}

In the SM the leptonic decay of charged pseudoscalar mesons proceeds via the 
annihilation of the heavy meson into a $W$ boson and its subsequent leptonic decay.
Angular momentum conservation leads to a helicity suppression factor that is squared
in the lepton mass. Competitive contributions from the charged Higgs sector can therefore 
occur. Neglecting photon radiation, the leptonic decay rate of a pseudoscalar meson $P$ 
has the form
\beq
\label{eq:Pdec}
   \Gamma\left(P \to \ell \nu\right) = \frac{\BR(P \to \ell \nu)}{\tau_P} =
           \frac{G_F^2}{8\pi}f_P^2 m_{\ell}^2 m_P
           \left(1-\frac{m_{\ell}^2}{m_P^2}\right)^{\!\!2}\left|V_{q_1 q_2}\right|^2 \, ,
\eeq
where $m_{P}$ ($m_{\ell}$) is the mass of the pseudoscalar meson (lepton), $|V_{q_1 q_2}|$ is 
the magnitude of the CKM matrix element of the constituent quarks in $P$, and $f_P$ is 
the weak decay constant. 

For $P=B$ (implying $B^{\pm}=B^{\pm}_u$) we use~\cite{hfag08}
$\tau_{B^{\pm}} = (1.639 \pm 0.009)\:{\rm ps}$ and $|V_{ub}| = (3.81 \pm 0.47) \cdot 10^{-3}$,
where the latter result has been averaged over inclusive and exclusive measurements.
For the $B$ decay constant we use the value $f_{B} = (216 \pm 22)\mev$, obtained 
by the HPQCD Collaboration from unquenched Lattice QCD calculations~\cite{Gray:2005ad}.
For meson and lepton masses we use the values of Ref.~\cite{pdg}.
With these inputs, we find the SM predictions
$\BR(B\to\tau\nu)=1.53^{\,+0.46}_{\,-0.38}\cdot10^{-4}$ and 
$\BR(B\to\mu\nu)=0.69^{\,+0.21}_{\,-0.17}\cdot10^{-6}$.

An alternative approach uses for the r.h.s. of Eq.~(\ref{eq:Pdec}) additional constraints from 
the global CKM fit enhancing the information on $|V_{ub}|$ beyond that of the direct 
measurement through the fit of the Wolfenstein parameters $\rhobar,\etabar$, and on $f_{B}$
through the measurement of the $\BzBzb$ mixing frequency. This assumes that the 
measurements entering the fit are free from significant new physics contributions. It is 
certainly the case for the charged Higgs, but cannot be excluded for the \CP-violation and 
neutral-$B$ mixing observables. Hence, albeit using the global CKM fit is an 
interesting test, it cannot replace the direct SM prediction of Eq.~(\ref{eq:Pdec})
based on tree-level quantities and lattice calculations only. Not using the direct 
measurements, the global CKM fit gives $|V_{ub}|=(3.44^{\,+0.22}_{\,-0.17}) \cdot 10^{-3}$, 
and for the complete prediction $\BR(B\to\tau\nu)=0.83^{\,+0.27}_{\,-0.10}$~\cite{ckmfitter:2008}. 
This latter result is about $1.9\sigma$ below the one from the ``tree-level'' determination, 
and a similar discrepancy is found for $B\to\mu\nu$ (\cf Table~\ref{tab:2HDMinput}).

The charged-Higgs amplitude contributes to the leptonic decays modifying Eq.~(\ref{eq:Pdec}) 
by a scaling factor $r_H$. In the Type-II 2HDM the $b$ quark couples only to one 
of the Higgs doublets at tree level so that the scaling factor for the decays 
\btotaunu and \btomunu reads~\cite{Hou:1992sy}
\beq
\label{eq:btaunu2hdm}
   r_H = \left( 1 - m_B^2 \, \frac{\tan^2\!\beta}{\MHp^2}\right )^{\!\!2} \, ,
\eeq
which can lead to both, an increase and a decrease in the branching fraction, depending 
on whether the $W^{\pm}$ and $H^{\pm}$ amplitudes interfere constructively or destructively.

The rare leptonic decay \btotaunu\ has been observed by the BABAR and Belle 
Collaborations~\cite{Aubert:2007xj,Aubert:2007bx,PaotiChang}, with an average branching 
fraction\footnote
{
   Updated results from BABAR and Belle have been presented at the recent workshops
   CKM 2008 and Tau 2008~\cite{MichaelMazur}, leading to the new average 
   $\BR(\btotaunu) = (1.73 \pm 0.35) \cdot 10^{-4}$. 
   They will be included in future updates of this analysis. 
} 
of $\BR(\btotaunu) = (1.51 \pm 0.33) \cdot 10^{-4}$. Only upper
limits are available for the muon channel so far, the tightest one, 
$\BR(\btomunu)<1.3 \cdot 10^{-6}$ at 90\% CL ($-12\pm20$ fitted events), being found
by BABAR~\cite{Aubert:2008ri}. 
For lack of an experimental likelihood we use the measured branching fraction of 
$(-0.57 \pm 0.71_{\rm stat} \pm 0.68_{\rm syst}) \times 10^{-6}$.

For $P=K$, contributions from a charged Higgs are suppressed by $(m_K/m_B)^2$ relative 
to leptonic $B$ decays. Moreover, due to the smaller phase space for hadronic final 
states, leptonic decays have large branching fractions, which -- on the other hand --
have been measured to an excellent 0.2\% relative accuracy for $\ell=\mu$. 
We follow the approach of Ref.~\cite{Antonelli:2008jg} and compare $|V_{us}|$
determined from helicity suppressed $K \rar \mu\nu$ decays and helicity allowed $K \rar \pi \mu\nu$
decays, considering the expression
\beq
  R_{\l 23} = \left| \frac{V_{us}(\ktomunu)}{V_{ud}(\pitomunu)} \frac{V_{ud}(0^+ \rar 0^+)}{V_{us}(K\rar\pi\mu\nu)} \right| \,
\eeq
which in the SM is equal to 1.
The ratio $\BR(\ktomunu)/\BR(\pitomunu) \sim (V_{us} f_K)/(V_{ud} f_{\pi})$ is used to reduce 
the theoretical uncertainties from the kaon decay constant $f_K$, 
and from electromagnetic corrections in the decay \ktomunu~\cite{Antonelli:2008jg}.
The dominant uncertainty in $V_{us}$ from $K\rar\pi\mu\nu$ decays stems from the $K \rar \pi$ 
vector form factor at zero momentum transfer, $f_+(0)$, while $V_{ud}$ determined from
super-allowed nuclear beta-decays ($0^+ \rar 0^+$) is known with very high precision~\cite{pdg_vusreview}.

In the 2HDM of Type-II the dependence of $R_{\l 23}$ due to charged Higgs 
exchange is given by~\cite{Antonelli:2008jg}
\beq
   R_{\l 23}^H = \left|1- \left(1 - \frac{m_d}{m_s}\right) \frac{m^2_{K^+}}{m^2_{H^+}}\tan^2\!\beta  \right|\,,
\eeq
where we use $m_s/m_d = 19.5 \pm 2.5$~\cite{pdg}. 
Experimentally, a value of $R_{\l 23}^{\rm exp}=1.004 \pm 0.007$ is found~\cite{Antonelli:2008jg}, 
where $(f_K/f_{\pi})/f_+(0)$ has been taken from lattice calculations. It dominates 
the uncertainty on $R_{\l 23}$.

\subsubsection{The Semileptonic Decay $\mathsf{B\to D\tau\nu}$}

Similar to the \btotaunu\ decay, the semileptonic decay \btodtaunu\ can be mediated a 
by charged Higgs. We follow the arguments of Ref.~\cite{Kamenik:2008tj} and use the
ratio $R_{D\tau/e} = \BR(\btodtaunu)/\BR(\btodenu)$ to reduce theoretical uncertainties 
from hadronic form factors occurring in the predictions of the individual branching 
fractions. In the Type-II 2HDM the ratio $R_{D\tau/e}$ can be expressed in the following 
compact form~\cite{Kamenik:2008tj}
\beq
\label{eq:bdtaunu}
   R_{D\tau/e}^H = 
          (0.28 \pm 0.02) \cdot \left[1 + (1.38 \pm 0.03)\cdot\Re(\Ctaunp) + 
                                       (0.88 \pm 0.02)\cdot\left|\Ctaunp\right|^2  \right]\,,
\eeq
where $\Ctaunp = - m_b m_{\tau} \tan^2\!\beta/m^2_{H^{\pm}}$. As for leptonic decays the 2HDM
contribution can either lead to an increase or decrease in the branching fraction.
Equation~\ref{eq:bdtaunu} is the result of an integration of the partial width 
$d\Gamma (B \to D\ell\nu)/dw$, assuming no Higgs contribution to \btodenu,
and where $w = v_B v_D$ with $v_B$ ($v_D$) being the four-velocity of the $B$ ($D$) meson. 

The ratio of branching 
fractions has been measured by BABAR to be 
$R_{D\tau/e}^{\rm exp} = 0.42 \pm 0.12_{\rm stat} \pm 0.05_{\rm syst}$~\cite{Aubert:2007dsa}.

\begin{table}[t]
\setlength{\tabcolsep}{0.0pc}
{\normalsize
\centering
\begin{tabular*}{\textwidth}{@{\extracolsep{\fill}}lclcc} 
   \hline\noalign{\smallskip}
   Parameter & Experimental value& Ref. & SM prediction & Ref. \\
   \noalign{\smallskip}\hline\noalign{\smallskip}
   $R_b^0$                           & $0.21629 \pm 0.00066$  &  \cite{:2005ema}
                                     & $0.21580 \pm 0.00006$                   & This work  \\
   $\BR(\btoxsg)$ $[10^{-4}]$ 
                                     &  $3.52 \pm 0.23 \pm 0.09$  & \cite{hfag08}   
                                     &  $3.15 \pm 0.23$              & \cite{Misiak:2006zs}   \\
                                     &     &                                  
                                     & $1.53^{\,+0.46}_{\,-0.38}$ {\ft$[f_B,|V_{ub}|]$}  &   This work              \\[-0.0cm]
   \rs{$\BR(\btotaunu)$ $[10^{-4}]$} &  \rs{$1.51 \pm 0.33$}  & \rs{\cite{PaotiChang}} 
                                     & $0.83^{\,+0.27}_{\,-0.10}$ {\ft[CKM fit]}      & {\cite{ckmfitter:2008}}\\
                                     &     &
                                     & $0.69^{\,+0.21}_{\,-0.17}$ {\ft$[f_B,|V_{ub}|]$} &  This work    \\[-0.0cm]
   \rs{$\BR(\btomunu)$ $[10^{-6}]$}  & \rs{$-0.57 \pm 0.68 \pm 0.71$}   & \rs{\cite{Aubert:2008ri}}  
                                     & $0.37^{\,+0.12}_{\,-0.04}$ {\ft[CKM fit]}      &  \cite{ckmfitter:2008}  \\
   $R_{D\tau/e}$                      & $0.42 \pm 0.12 \pm 0.05$ & \cite{Aubert:2007dsa}
                                     & $0.28\pm0.02$                                & \cite{Kamenik:2008tj} \\
   $R_{\ell23}$                        & $1.004 \pm 0.007$ & \cite{Antonelli:2008jg}
                                     & 1                                       &  -- \\
   \noalign{\smallskip}\hline
\end{tabular*} 
}
\caption{Experimental results and SM predictions for the input observables used in the analysis of 
         the charged-Higgs sector of the Type-II 2HDM.}
\label{tab:2HDMinput}
\end{table}
\begin{figure}[p]
\def\thdmfigscale{0.39}
   \centering
   \epsfig{file=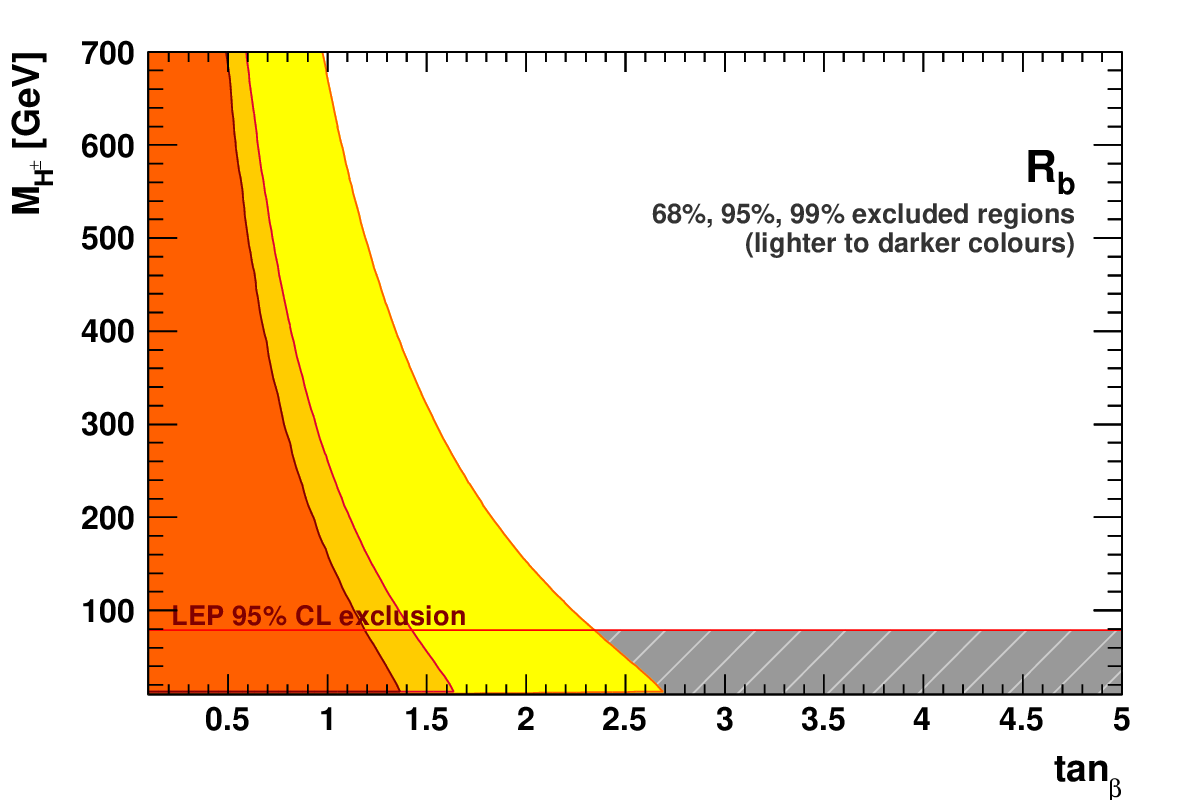, scale=\thdmfigscale}
   \epsfig{file=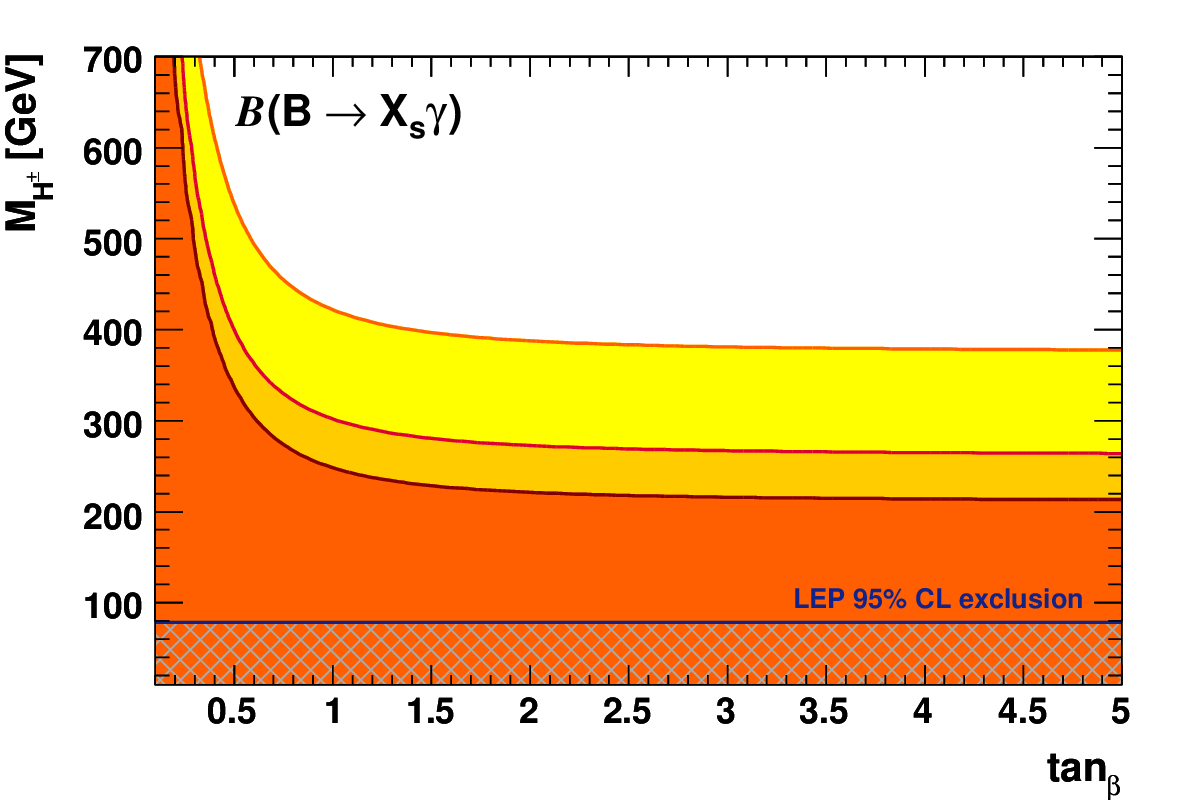, scale=\thdmfigscale}
   \vspace{-0.1cm}
   \epsfig{file=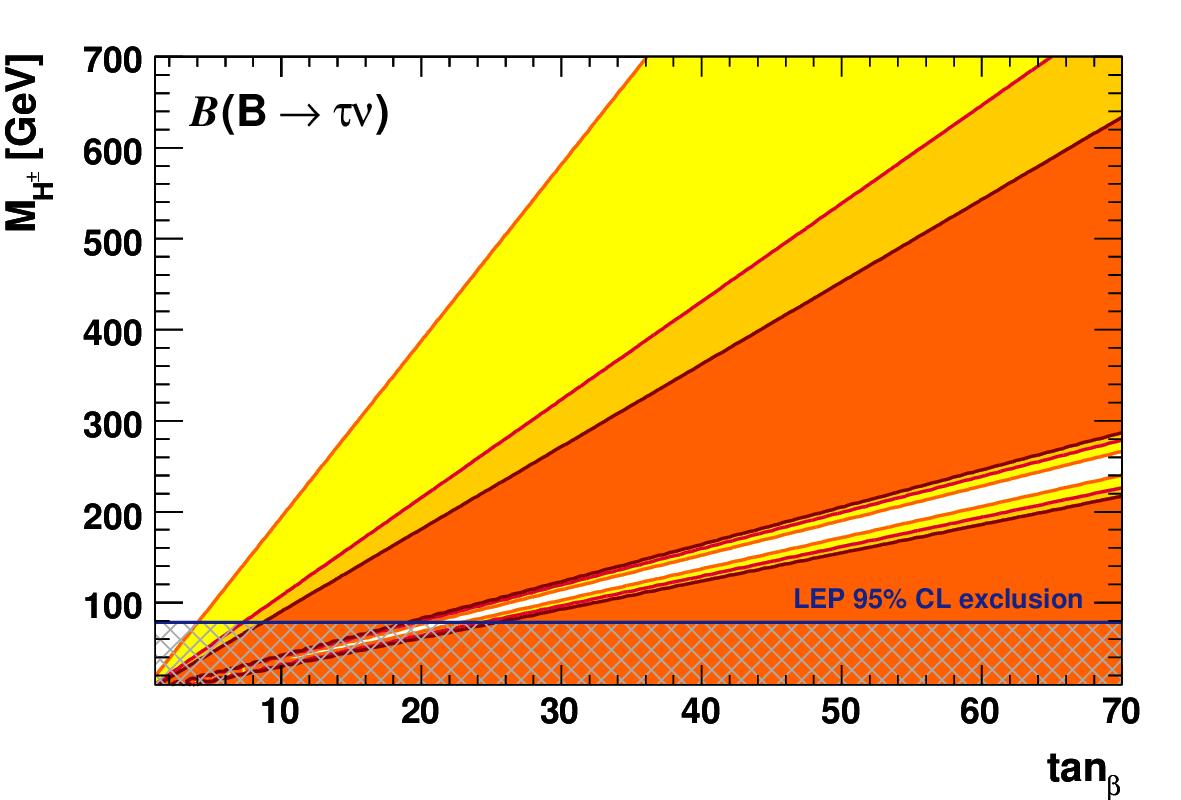, scale=\thdmfigscale}
   \epsfig{file=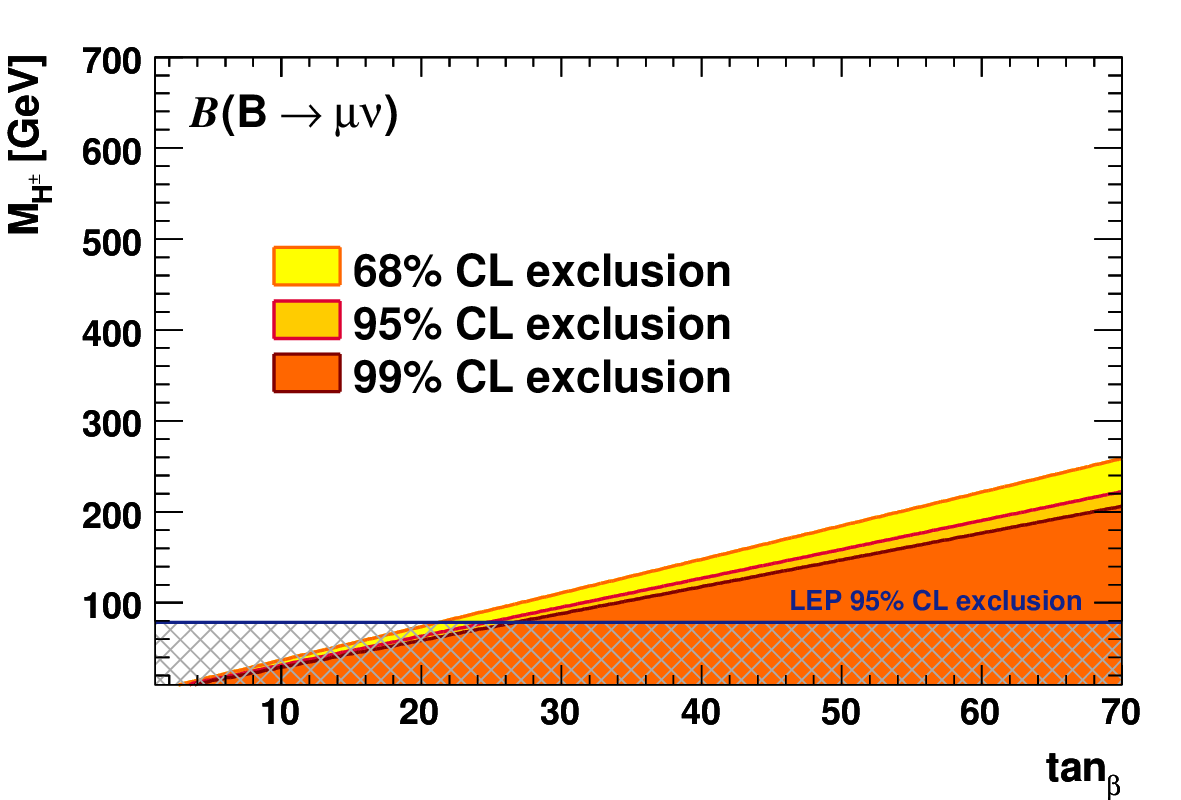, scale=\thdmfigscale}
   \vspace{-0.1cm}
   \epsfig{file=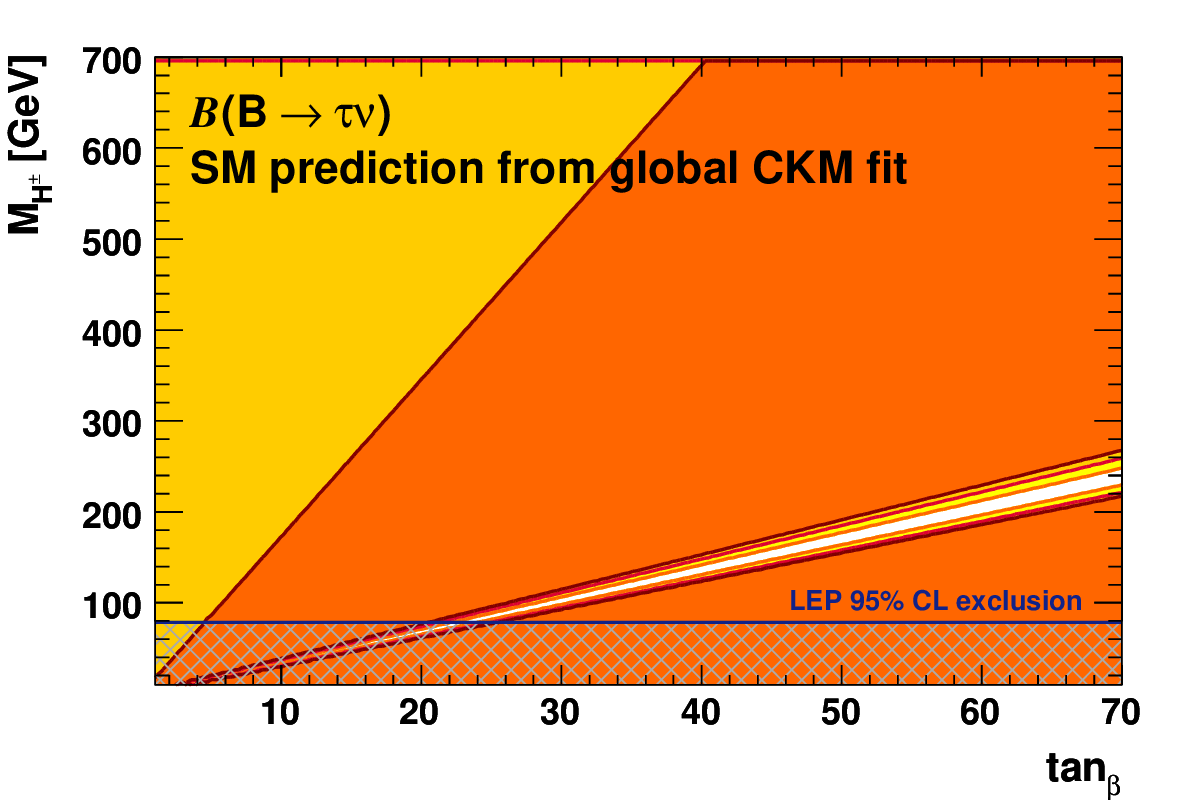, scale=\thdmfigscale}
   \epsfig{file=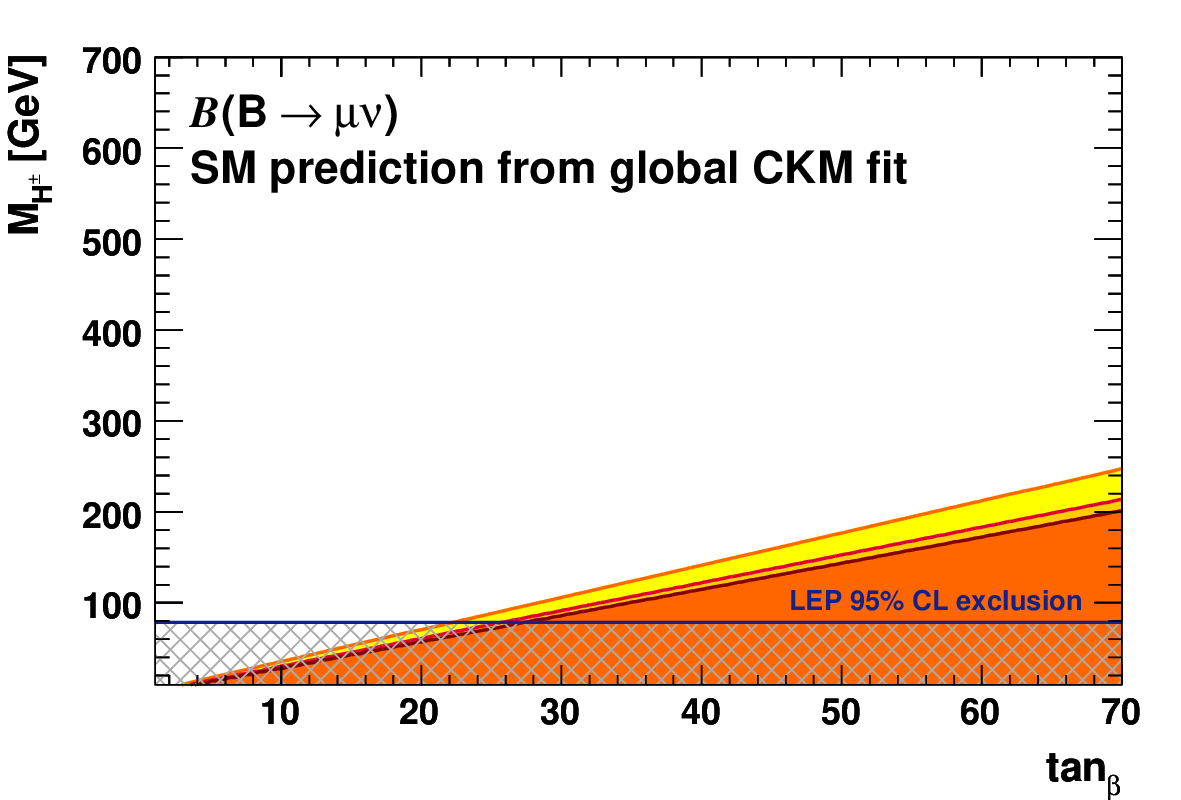, scale=\thdmfigscale}
   \vspace{-0.1cm}
   \epsfig{file=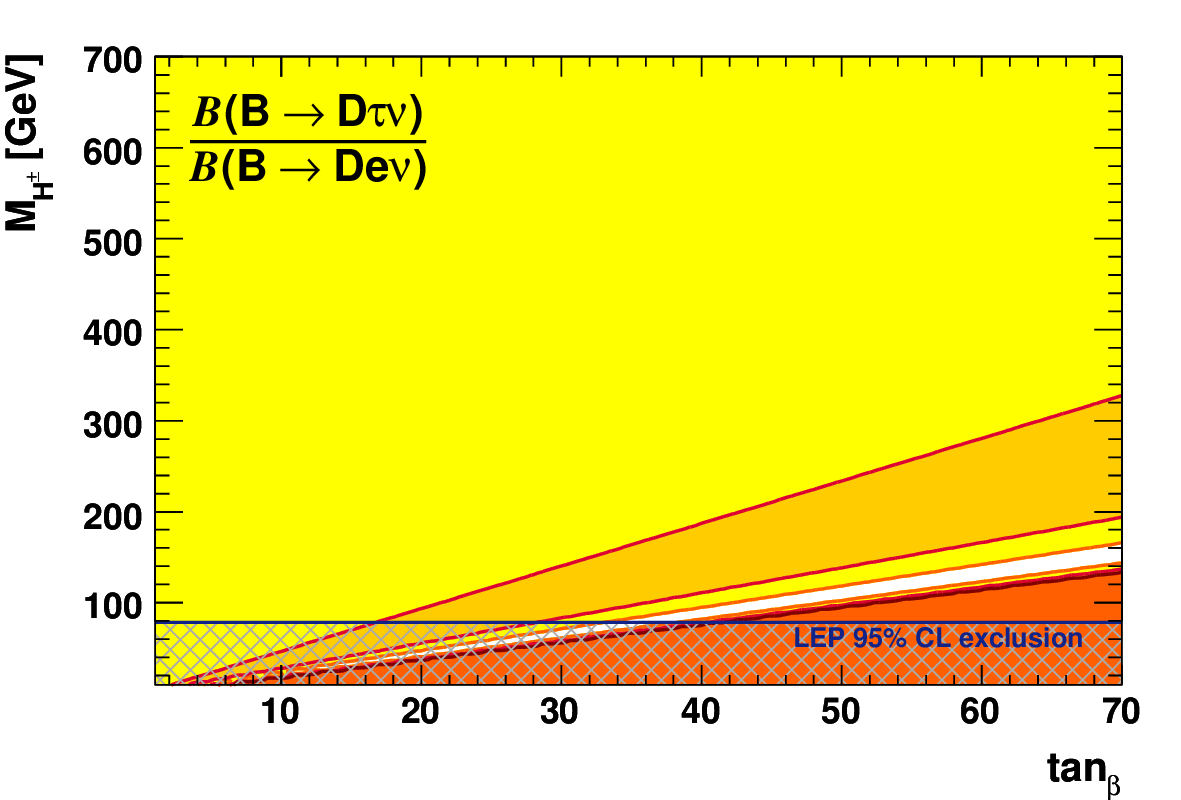, scale=\thdmfigscale}
   \epsfig{file=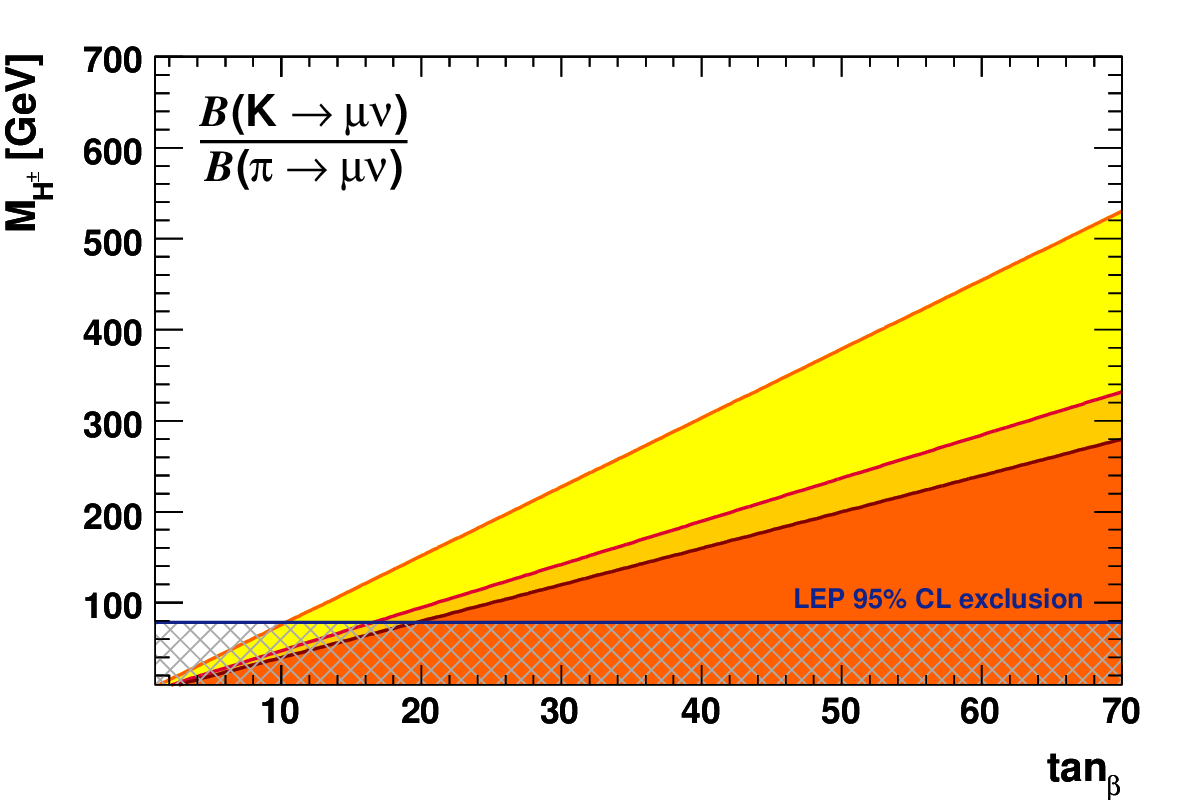, scale=\thdmfigscale}
\unitlength 1cm
   \caption{Two-sided 68\%, 95\% and 99\% CL {\em exclusion} regions obtained for the various observables 
            (see text) in the 2HDM parameter plane $\MHp$ versus $\tanb$.}
   \label{fig:2hdm-indiv}
\end{figure}

\subsection{Results and Discussion}
\label{subsec:2hdmfitResults}

The theoretical predictions of the Type-II 2HDM for the various observables
sensitive to corrections from the exchange of charged Higgs bosons have been 
implemented in a separate library integrated as a plug-in into the \Gfitter 
framework. Exclusion confidence levels have been derived in two ways: $(i)$ for 
each observable separately, and $(ii)$ in a combined fit.

\subsubsection{Separate Constraints from Individual Observables}

Constraints in the two-dimensional model parameter plane ($\tanb,\MHp$) 
have been derived using the individual experimental measurements and the corresponding 
theoretical predictions of the Type-II 2HDM. Figure~\ref{fig:2hdm-indiv} displays the resulting 
two-sided 68\% (yellow/light), 95\% (orange) and 99\% CL (red/dark) {\em excluded} regions 
separately for each of the observables given in Table~\ref{tab:2HDMinput}. The confidence
levels are derived assuming Gaussian behaviour of the test statistics, and using one degree
of freedom (\cf discussion in Footnote~\ref{ftn:dofdiscussion} on page~\pageref{ftn:dofdiscussion})
, \ie, ${\rm Prob}(\DeltaChi,1)$. 
Also indicated in the plots is the 95\% CL exclusion limit resulting from the direct searches 
for a charged Higgs at LEP~\cite{LEPHiggsWG} (hatched area).

The figures show that $R_b$ is mainly sensitive to \tanb excluding small values (below $\simeq$1). 
$\BR(\btoxsg)$ is only sensitive to \tanb for values below $\simeq$1. For larger \tanb it provides
an almost constant area of exclusion of a charged Higgs lighter than $\simeq$$260\gev$. (All 
exclusions at 95\% CL). The leptonic observables lead to triangle-shaped excluded areas in 
the region of large $\tanb$ and small $m_{H^\pm}$ values. $\BR(\btotaunu)$ gives the strongest 
constraint.\footnote
{
   The stronger constraint obtained from the global CKM fit for $\BR(\btotaunu)$ is a result of
   the increased theoretical precision and, more importantly, the $1.9\sigma$ deviation with respect to 
   the ``tree-level'' determination (\cf Table~\ref{tab:2HDMinput}).
}
For these observables the 2HDM contribution can be either positive or negative, because magnitudes 
of signed terms occur in the predictions of the branching fractions giving a two-fold 
ambiguity in the ($\tanb,\MHp$) plane. 

\subsubsection{Combined Fit}
\label{sec:2hdm_combinedFit}

\begin{figure}[p]
  \centering
   \centerline{\epsfig{file=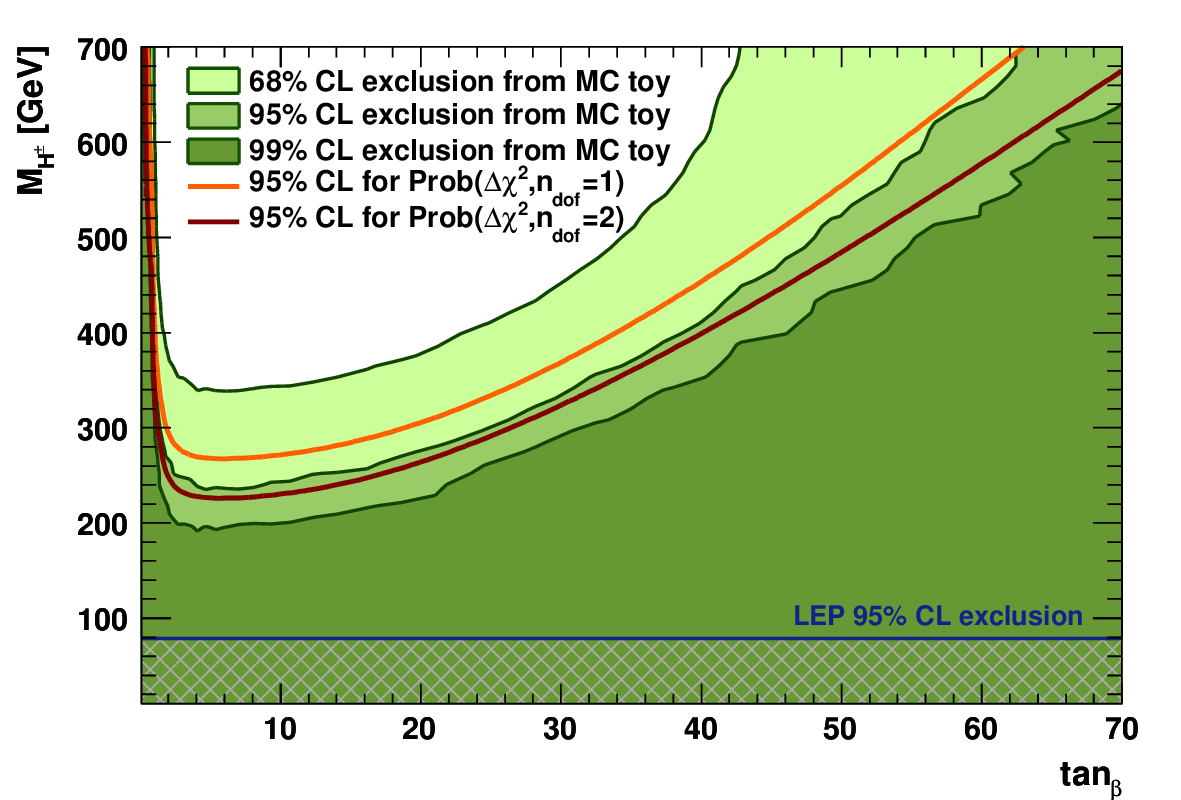, scale=\defaultFigureScale}}
   \vspace{0.3cm}
   \centerline{\epsfig{file=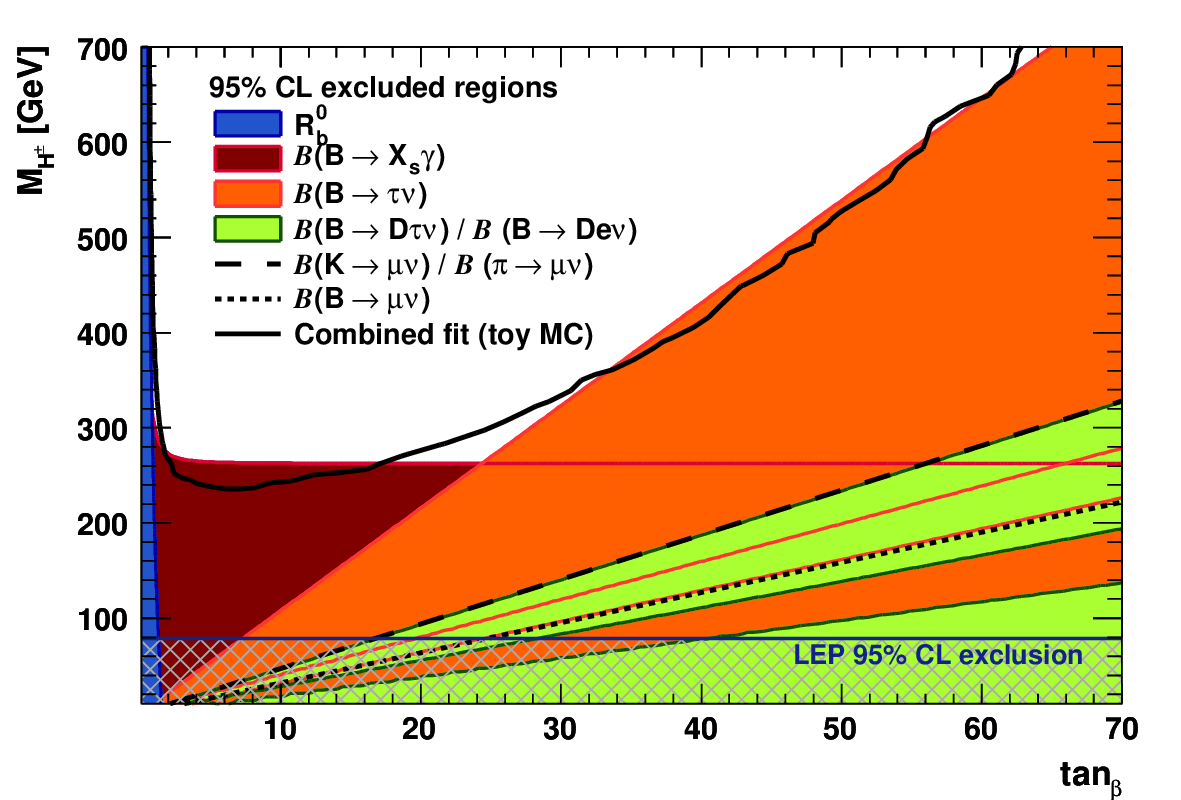, scale=\defaultFigureScale}}
   \vspace{0.2cm}
   \caption{Exclusion regions in the ($\tanb, \MHp$) plane. The top plot displays          
     the 68\%, 95\% and 99\% CL excluded regions obtained from the combined fit using 
     toy MC experiments. For comparison the 95\% CL contours using ${\rm Prob}(\Delta\chi^2,\ndof)$
     for $\ndof=1$ and $\ndof=2$ are also shown (see discussion in text).
     The bottom plot shows the 95\% CL excluded regions from the individual constraints given in 
     Table~\ref{tab:2HDMinput}, and the toy-MC-based result from the combined fit overlaid.}
   \label{fig:2hdm-combined}
\end{figure}
We have performed a global Type-II 2HDM fit combining all the
available observables (and using the tree-level SM predictions for the
leptonic $B$ decays). We find a global minimum $\ChiMin=3.9$ at $\MHp = 860
\gev$ and $\tanb = 7$.
Since the number of effective constraints
varies strongly across the $(\tanb,\MHp)$ plane, it is not
straightforward to determine the proper number of degrees of freedom
to be used in the calculation of the CL -- even if the test statistic
follows a $\chi^2$ distribution. According to the discussion in
Footnote~\ref{ftn:dofdiscussion} on page~\pageref{ftn:dofdiscussion}
we avoid this problem by performing 2\,000 toy-MC experiments in each
scan point to determine the associated p-value. 
The upper plot of Fig.~\ref{fig:2hdm-combined} shows the 68\%, 95\%
and 99\% CL excluded regions obtained from the toy-MC analysis of the
combined fit. For comparison the 95\% CL contours using ${\rm
Prob}(\Delta\chi^2,\ndof)$ for $\ndof=1$ and $\ndof=2$ are also
shown. As expected, the $\ndof=2$ approximation is more accurate in
regions where several observables contribute to the combined fit,
while $\ndof=1$ is better when a single constraint dominates over all
the others (very small and very large values of \tanb). 
For comparison the lower plot of Fig.~\ref{fig:2hdm-combined} shows 
again the 95\% CL excluded region obtained from the toy-MC analysis of the 
combined fit (hatched area) together with the corresponding regions obtained 
from the individual constraints. It can be seen that due to the increased 
number of effective degrees of freedom the combined fit does 
not necessarily lead to stronger constraints. 

The combination of the constraints excludes the high-\tanb, low-\MHp
region spared by the $B\to\tau\nu$ constraint. We can thus exclude a
charged-Higgs mass below $240\gev$ independently of $\tanb$ at 95\%
confidence level. This limit increases towards larger \tanb,
\eg, $\MHp<780\gev$ are excluded for $\tanb=70$ at 95\% CL.

\subsubsection{Perspectives}

Improvements on the low-energy $B$-meson observables are expected from the KEKB and Belle upgrade
program with an initial (final) target of $10\invab$ ($50\invab$) integrated 
luminosity~\cite{Akeroyd:2004mj,Hashimoto:2004sm,superkekb}. 
Parallel developments envision the construction of a new SuperB accelerator with similar target 
luminosities~\cite{Bona:2007qt}. With respect to the 2HDM analysis, these programs are particularly
interesting for the decays $B\to\tau\nu$, $B\to\mu\nu$ and \btodtaunu whose present branching fraction
measurements
are statistically dominated. Further improvement can also be expected for the measurement
of $\BR(B\to X_s\gamma)$ with however less prominent effect on the 2HDM parameter constraints 
due to the size of the theoretical uncertainties. The measurement of the ratio of partial $Z$ 
widths, $R_b^0$, could be improved at an ILC running at the $Z$ resonance (GigaZ, \cf
Section~\ref{sec:prospects}). The authors of Ref.~\cite{Hawkings:1999ac} estimate a factor of 
five increase over the current precision, mostly by virtue of the increased statistical yield, 
and the excellent impact parameter resolution suppressing background from charm quarks. 

The LHC experiments will attempt to directly detect signals from charged-Higgs production, either 
via $t\to b H^\pm$ decays, if $\MHp<\mt$, and/or via gluon-gluon and gluon-bottom fusion to 
$t(b)H^\pm$, and the subsequent decay $H^\pm\to\tau\nu$, or, if $\MHp>\mt$, via $H^\pm\to t b$.
The full \tanb parameter space is expected to be covered for $H^\pm$ lighter than top (a scenario
already strongly disfavoured by the current indirect constraints, especially the one from 
$\BR(B\to X_s\gamma)$), while the discovery of a heavy $H^\pm$ requires a large \tanb, 
which rapidly increases with rising \MHp~\cite{atlastdr,CMSTDR,Flechl:2007db}.

%
%
\section{Conclusions and Perspectives}
\label{sec:conclusions}

The wealth of available precision data at the electroweak scale requires consistent 
phenomenological interpretation via an overall ({\em global}) fit of the Standard Model
and beyond. Such fits, mainly determining the top-quark mass, the Higgs-boson mass, the 
strong coupling constant, and the overall consistency of the model, have been performed 
by several groups in the past. The fit has sensitivity to confirm electroweak unification 
and the Brout-Englert-Higgs mechanism~\cite{Higgs:1964pj,Englert:1964et} 
of spontaneous electroweak symmetry breaking for 
the dynamical generation of the fermion and boson masses, while posing problems for 
alternatives such as Technicolour in its simplest form~\cite{Ellis:1994pq}, requiring
more involved scenarios. 
Other theories, like Supersymmetry, are decoupling from the Standard Model if their masses 
are large. For such models the high energy precision data as well as constraints
obtained from rare decays, flavour mixing and \CP-violating asymmetries in the $B$ and $K$-meson 
sectors, the anomalous magnetic moment of the muon, and electric dipole moments of electron 
and neutron, exclude a significant part of the parameter space. However, the models can be 
adjusted to become consistent with the experimental data as long as these data agree with 
the Standard Model predictions. 

In this paper, we have revisited the global electroweak fit, and a simple extension of the Higgs
sector to two doublets, using the new generic fitting toolkit \Gfitter and its corresponding
electroweak and 2HDM libraries. We have included the constraints from direct Higgs searches by 
the LEP and Tevatron experiments in the former fit. Emphasis has been put on a consistent
treatment of theoretical uncertainties, using no assumptions other than their respective ranges, 
and a thorough frequentist statistical analysis and interpretation of the fit results.

\Gfitter is an entirely new fitting framework dedicated to model testing in high-energy physics. 
It features transparent interfaces to model parameters and measurements, theory libraries, and 
fitter implementations. Parameter caching significantly increases the execution speed of the 
fits. All results can be statistically interpreted with toy Monte Carlo methods, treating 
consistently correlations and rescaling due to parameter dependencies.

For the {\em complete fit}, including the results from direct Higgs searches, we find for the 
mass of the Higgs boson the $2\sigma$ and $3\sigma$ intervals $[114,\,145]\gev$ and 
$[[113,\,168]\,{\rm and}\,[180,\,225]]\gev$, respectively. The corresponding results without the direct 
Higgs searches in the {\em standard fit} are $[39,\,155]\gev$ and $[26,\,209]\gev$. 
Theoretical errors considered in the fit parametrise uncertainties in the perturbative predictions 
of $M_W$ and $\sinfeff$, and the renormalisation scheme ambiguity. They contribute with 
approximately $8\gev$ to the total fit error obtained for $M_H$ for the {\em standard fit}. 
In a fit excluding the measurement of the top quark mass (but including the direct Higgs
searches) we find $m_t=178.2^{\,+9.8}_{\,-4.2}\gev$, in fair agreement with the experimental 
world average. Finally, the strong coupling constant to \NNNLO order at the $Z$-mass scale 
is found to be $\asZ=0.1193^{\,+0.0028}_{\,-0.0027}$, with negligible theoretical error (0.0001) 
due to the good convergence of the perturbative series at that scale. 

We have probed the goodness of the Standard Model fit to describe the available data with toy 
Monte Carlo simulation. For the fit including the direct Higgs searches it results in a 
p-value of $0.22\pm0.01_{\,-0.02}$, where the first error accounts for the limited Monte Carlo 
statistics, and the second for the impact of theoretical uncertainties (without these, the 
p-value is reduced by 0.04). The p-value for the fit without direct Higgs searches
is similar (the reduced number of degrees of freedom approximately countervails the 
better $\chi^2$ value). The compatibility of the most sensitive measurements determining
$M_H$ has been estimated by evaluating the probability for a consistent set of measurements
to find a single measurement that increases the overall $\chi^2$ of the global fit by as much 
as is observed in data, when adding the least compatible measurement (here $A_{\rm FB}^{0,b}$). 
An analysis with toy MC experiments finds that this occurs in $(1.4\pm0.1)\%$ of the cases.

We have analysed the perspectives of the electroweak fit considering three future experimental 
scenarios, namely the LHC and an international linear collider (ILC) with and without high 
luminosity running at lower energies (GigaZ), all after years of data taking and assuming a good 
control over systematic effects. For a $120\gev$ Higgs boson, the improved $M_W$ and $m_t$ 
measurements expected from the LHC would reduce the error on the $M_H$ prediction by up to 20\% 
with respect to the present result. The ILC could further reduce the error by about 25\%
over the LHC, and -- if the hadronic contribution to $\alpha(M_Z^2)$ can be determined with 
better precision (requiring better hadronic cross section measurements at low and intermediate
energies) -- a 30\% improvement is possible. The largest impact on the fit accuracy can be 
expected from an ILC with GigaZ option. Together with an improved $\alpha(M_Z^2)$, the present
fit error on $M_H$ could be reduced by more than a factor of two. We point out however that, 
in order to fully exploit the experimental potential, in particular the anticipated improvements
in the accuracy of $M_W$, theoretical developments are mandatory. If the Higgs is discovered,
the improved electroweak fit will serve as a sensitive test for the Standard Model and its 
extensions.

By extending the Standard Model Higgs sector to two scalar doublets (2HDM of Type-II), we have 
studied the experimental constraints on the charged-Higgs mass $M_{H^{\pm}}$ and on \tanb, using 
as input branching fractions of the rare $B$ decays \btoxsg, \btotaunu, 
\btomunu, and \btodenu, the Kaon decay \ktomunu, and the electroweak precision observable $R_b^0$. 
Exclusion confidence levels have been derived by carrying out toy experiments for every point 
on a fine grid of the ($M_{H^{\pm}},\,\tanb$) parameter space. At 95\% confidence level we 
exclude charged Higgs masses $M_{H^{\pm}} < 240 \gev$ for any value of \tanb, and 
$M_{H^{\pm}} < 780 \gev$ for  $\tanb = 70$.

Inputs and numerical and graphical outputs of the \Gfitter Standard Model and 2HDM analyses 
are available on the \Gfitter web site: \url{http://cern.ch/gfitter}. They will be kept 
in line with the experimental and theoretical progress. Apart from these update commitments,
new theoretical libraries such as the minimal Supersymmetric extension of the Standard Model
will be included and analysed.

\subsubsection*{Acknowledgements}
\addcontentsline{toc}{section}{Acknowledgements}
\label{sec:Acknowledgments}

\begin{details}
We are indebted to the LEP-Higgs and Tevatron-NPH working groups for providing the numerical 
results of the direct Higgs-boson searches.
We thank Daisuke Nomura and Thomas Teubner for information on their analysis of the 
hadronic contribution to the running $\alphaMZ$.
We are grateful to Malgorzata Awramik for providing detailed information about the SM 
electroweak calculations, and to Bogdan Malaescu for help on the evaluation of 
theoretical errors affecting the determination of $\asZ$.
We are obliged to the DESY Summer Student Kieran Omahony for his work on the fit 
automation and the \Gfitter web page.
We thank the CKMfitter group for providing the best-effort predictions of the rare leptonic $B$ 
decays used in the 2HDM analysis, and Paolo Gambino, Ulrich Haisch and Mikolaj Misiak for 
valuable discussions and exchanges regarding the constraints on the 2HDM parameters.
Finally, we wish to thank J\'er\^ome Charles, Stefan Schmitt and St\'ephane T'Jampens 
for helpful discussions on statistical problems. 

\end{details}

\newpage
\begin{appendix}

\section{Standard Model Formulae}
\label{app:gsmFormulas}

This section gives the relevant formulae for the calculation of the electroweak observables used 
in the global electroweak fit. We discuss the scale evolution of the QED and QCD couplings
and quark masses, and give expressions for the electroweak form factors and radiator
functions.


\subsection{Running QED Coupling}
\label{app:alpha}

The electroweak fit requires the knowledge of the electromagnetic couping strength at the 
$Z$-mass scale to an accuracy of 1\% or better. The evolution of $\alpha(s)$ versus the 
mass scale-squared $s$ is conventionally parametrised by
\begin{align}
\label{eq:Dalpha}
   \alpha(s) \ = \ \frac{\alpha(0)} {1-\Delta\alpha(s)}\,,
\end{align}
following from an all-orders resummation of vacuum polarisation diagrams, sole contributors
to the running $\alpha$. Here $\alpha=\alpha(0)= 1/137.035\,999\,679(94)$ is the fine structure 
constant in the long-wavelength Thomson limit~\cite{Mohr:2008fa}, and the term 
$\Delta\alpha(s)$ controls the evolution. It is conveniently decomposed into leptonic and 
hadronic contributions
\begin{align}
\label{eq:alphaContributions}
   \Delta\alpha(s)\ =\ \dalphaLeps + \dalphaHads + \dalphaTops\,,
\end{align}
where the hadronic term has been further separated into contributions from the five light 
quarks (with respect to $M_Z$) and the top quark. The leptonic term in~(\ref{eq:alphaContributions}) 
is known up to three loops in the $q^2\gg m^2_\ell$ limit~\cite{Steinhauser:1998rq}. The dominant 
one-loop term at the $Z$-mass scale reads
\beq
   \Delta\alpha_{\rm lep}^{(1\mbox{-}{\rm loop})}(M_Z^2)=
   \alpha\sum_{\ell=e,\mu,\tau}\!\!
        \left(-\frac{5}{9}+\frac{1}{3}\ln\frac{M_Z^2}{m_\ell^2}-2\frac{m^2_\ell}{M_Z^2} 
        + \Order\!\left(\frac{m^4_\ell}{M_Z^4}\right)\right) \approx 314.19\cdot10^{-4}\,.
\eeq
Adding the sub-leading loops gives a total of $\dalphaLeps=314.97\cdot10^{-4}$, 
with negligible uncertainty.\footnote
{
   While the two-loop leptonic contribution of $0.78\cdot10^{-4}$ is significant (roughly one 
   third of the uncertainty in the hadronic contribution), the third order term, 
   $0.01\cdot10^{-4}$, is very small,
}

The hadronic contribution for quarks with masses smaller than $M_Z$ cannot be 
obtained from perturbative QCD alone because of the low energy scale involved. 
Its computation relies on analyticity and unitarity to express the photon vacuum polarisation 
function as a dispersion integral involving the total cross section for \ee\ annihilation to 
hadrons at all time-like energies above the two-pion threshold. In energy regions where perturbative 
QCD fails to locally predict the inclusive hadronic cross section, experimental data is used. 
The accuracy of the calculations has therefore followed the progress in the quality of the 
corresponding data. Recent calculations improved the precision by extending the use of 
perturbative QCD to energy regions of relatively low scales, benefiting from global quark-hadron
duality. For the fits in this paper we use the most recent value, $\dalphaHadMZ=(276.8\pm2.2)\cdot10^{-4}$, 
from Ref.~\cite{Hagiwara:2006jt}. The error is dominated by systematic uncertainties in the 
experimental data used to calculate the dispersion integral. A small part of the error, $0.14\cdot10^{-4}$, 
is introduced by the uncertainty in \ass (the authors of~\cite{Hagiwara:2006jt} used the value 
$\asZ=0.1176\pm0.0020$~\cite{Teubner:private2007}). We include this dependence in the fits via
the parameter rescaling mechanism implemented in \Gfitter (\cf Section~\ref{sec:gfitter}).

The small top-quark contribution at $M_Z^2$ up to second order in $\as$ 
reads~\cite{Kuhn:1998ze,Chetyrkin:1995ii,Chetyrkin:1996cf,Chetyrkin:1997mb} to
\begin{align}
   \dalphaTopMZ \ =\  &
   -\frac{4}{45}\frac{\alpha}{\pi}\frac{M_Z^2}{\mt^2}
   \left\{ 1 + 5.062\, \aas^{(5)}(\mu^2)
          + \left(28.220 + 9.702\,\ln\frac{\mu^2}{\mt^2}\right)
            \left(\aas^{(5)}(\mu^2)\right)^{\!\!2}
   \label{eqtop}
   \right.\\&\mbox{}\hspace{1.0cm}\left.
   +\;\frac{M_Z^2}{\mt^2}\left[
   0.1071 + 0.8315 \,\aas^{(5)}(\mu^2)
          + \left(6.924 + 1.594\,\ln\frac{\mu^2}{\mt^2}\right)
            \left(\aas^{(5)}(\mu^2)\right)^{\!\!2}
   \right]
   \right\}\,,
   \nonumber \\
            \ \approx\ & -0.7\cdot10^{-4}\,,\nonumber
\end{align}
where the short-hand notation $\aas=\as/\pi$ is used, and where $\as^{(5)}$ is the strong coupling constant 
for five active quark flavors, and $\mu$ is an arbitrary renormalisation scale, chosen to be $\mu=M_Z$ 
in the fit.

The uncertainty on $\alpha(M_Z^2)$ is dominated by the hadronic contribution \dalphaHadMZ,
which is a floating parameter of the fit constrained to its phenomenological value. The 
errors due to uncertainties in $M_Z$, $\mt$ and $\as$ are properly propagated throughout 
the fit. Other uncertainties are neglected.


\subsection{QCD Renormalisation}
\label{app:QCD}

Like in QED, the subtraction of logarithmic divergences in QCD is equivalent to renormalising the 
coupling strength ($\as\equiv g_s^2/4\pi$), the quark masses ($m_q$), etc., and the fields in the 
bare (superscript $B$) Lagrangian such as $\as^B=s^{\varepsilon}Z_{\as}\as$, $m_q^B=s^{\varepsilon}Z_mm_q$, 
etc. Here $s$ is the renormalisation scale-squared, $\varepsilon$ the dimensional regularisation 
parameter, and $Z$ denotes a series of renormalisation constants obtained from the generating 
functional of the bare Green's function. Renormalisation at scale $\mu$ introduces a differential 
renormalisation group equation (RGE) for each renormalised quantity, governing its running. 
All formulae given below are for the modified minimal subtraction renormalisation scheme 
(\MSb)~\cite{tHooft:1972t2,Bardeen:1978yd}.

\subsubsection{The Running Strong Coupling}
\label{app:alphas}

The RGE for $\as(\mu^2)$ reads
\beq
\label{eq:RGE}
   \frac{d\as}{d\ln \mu^2} \ =\ \beta(\as) 
                           \ =\  - \beta_0 \as^2 - \beta_1 \as^3 - \beta_2 \as^4
                                 - \beta_3 \as^5 - \ldots\,,
\eeq
The perturbative expansion of the $\beta$-function is known up to four 
loops~\cite{vanRitbergen:1997va,Czakon:2004bu}
(and references therein), with the coefficients
\begin{align}
\renewcommand{\arraystretch}{ 1.3}
   \beta_0 \ =\ & \frac{1}{4\pi}\left[ 11 - \frac{2}{3} n_f \right]\,,   \\
   \beta_1 \ =\ & \frac{1}{(4\pi)^2} \left[102 - \frac{38}{3} n_f \right]\,,  \\
   \beta_2 \ =\ & \frac{1}{(4\pi)^3} \left[ \frac{2857}{2} - \frac{5033}{18} n_f 
      + \frac{325}{54}n_f^2 \right]\,,  \\
   \beta_3 \ =\ & \frac{1}{(4\pi)^4} \left[  \left( \frac{149753}{6} 
               + 3564 \,\zeta_3 \right)
               - \left( \frac{1078361}{162} + \frac{6508}{27} \zeta_3 \right) n_f
   \right. \nonumber \\ & \hspace{1.15cm}
               \left. + \left( \frac{50065}{162} + \frac{6472}{81} \zeta_3 \right) n_f^2
               + \frac{1093}{729}  n_f^3 \right]\,,
\label{eq:beta}
\end{align}
where $n_f$ is the number of active quark flavours with masses smaller than $\mu$, and
where $\zeta_3\simeq 1.2020569$. Solving Eq.~(\ref{eq:RGE}) for $\as$ introduces a constant 
of integration, $\Lambda^{(n_f)}$, with dimension of energy. The solution in the ultraviolet limit 
reads~\cite{Chetyrkin:1997sg,prosperi-2007-58}
\beq
\label{eq:alphas}
\begin{aligned}
   \as (\mu^2)\ =\ &\frac{1}{\beta_0\,L}\,
   \left\{1-\frac{\beta_1}{\beta_0^2}\frac{\ln L}{L}+
   \frac{1}{\beta_0^2 L^2}\left[\frac{\beta_1^2}{\beta_0^2}
   \left(\ln^2\!L-\ln L-1\right)+\frac{\beta_2}{\beta_0}\right]\right.\\
   &\hspace{1.08cm}
   \left.+\:\frac{1}{\beta_0^3L^3}\left[\frac{\beta_1^3}{\beta_0^3}\left(
   -\ln^3\!L+\frac{5}{2}\ln^2\!L+2\ln L -\frac{1}{2}\right)-3\frac{\beta_1\beta_2}
   {\beta_0^2}\ln L+\frac{\beta_3}{2\beta_0}\right]\right\}\,,
\end{aligned}
\eeq
where $L = 2\ln (\mu/\Lambda^{(n_f)})\gg1$. 

As \as evolves it passes across quark-flavour thresholds. Matching conditions at these 
thresholds connect $\as^{(n_f)}$ of the full theory with $n_f$ flavors to the effective 
strong coupling constant  $\as^{(n_f-1)}$, where the heaviest quark decouples.
The coupling constant of the full theory is developed in a power series of the coupling 
constant of the effective theory with coefficients that depend on 
$x=2\ln(\mu/\mq)$~\cite{Bernreuther:1981sg,Wetzel:1981qg,Chetyrkin:1997sg,Rodrigo:1997zd}: 
\beq
   \aas^{(n_f)}\ =\ \aas^{(n_f-1)}\left[1+
   C_1(x)\left(\aas^{(n_f-1)}\right) +
   C_2(x)\left(\aas^{(n_f-1)}\right)^{\!\!2} +
   C_3(x)\left(\aas^{(n_f-1)}\right)^{\!\!3}\right]\,,
\label{eq:matching}
\eeq
with $\aas=\as/\pi$ (recalled), and
\beq
\begin{aligned}
   C_{1}(x)&\ =\ \frac{x}{6}\,,\qquad C_{2}(x)\ =\ c_{2,0}+\frac{19}{24}x+\frac{x^{2}}{36}\,,\\
   C_{3}(x)&\ =\ c_{3,0}+\left(\frac{241}{54} + \frac{13}{4}c_{2,0} - \left(\frac{325}{1728} 
            + \frac{c_{2,0}}{6}\right) n_f \right)x + \frac{511}{576}x^{2} + \frac{x^{3}}{216}\,.
\end{aligned}
\eeq
The integration coefficients $c_{i,0}$ computed in the \MSb scheme at the scale of the 
quark masses are
\beq
   c_{2,0}\ =\ -\frac{11}{72}\,,\qquad c_{3,0}\ =\ \frac{82043}{27648}\zeta_3 - \frac{575263}{124416}
       + \frac{2633}{31104} n_f\,.
\eeq

The solution of the RGE~(\ref{eq:RGE}) at arbitrary scale requires $\as$ to be known at some 
reference scale, for which the $Z$ pole is commonly chosen. Three evolution procedures are 
implemented in \Gfitter, which lead to insignificant differences in the result. The first uses 
numerical integration of the RGE with a fourth-order Runge-Kutta method. The second 
(the one chosen for this paper) determines $\Lambda^{(5)}$ at $M_Z$ by numerically evaluating
the root of Eq.~(\ref{eq:alphas}), and the values for $\Lambda^{(n_f\ne5)}$ are obtained 
via the matching conditions. Both methods use \asZ as floating parameter in the fit. 
In the third approach, $\Lambda^{(5)}$ is directly determined by the fit without explicit use 
of \asZ.

\subsubsection{Running Quark Masses}
\label{app:runningmasses}

The \MSb RGE for massive quarks is governed by the $\gamma$-function defined by
\beq
   \frac{1}{\mq}\frac{d\mq}{\ln\mu^2}\ =\ \gamma(\as)
                           \ =\  - \gamma_0 \as - \gamma_1 \as^2 - \gamma_2 \as^3
                                 - \gamma_3 \as^4 - \ldots\,.
\eeq
Its perturbative expansion has been computed to four loops~\cite{Vermaseren:1997fq} (and 
references therein), which for the $c$ and $b$-quark flavours reads~\cite{Vermaseren:1997fq}
\beq
\label{evolution}
\begin{aligned}
   \mc (\mu^2) &\ =\ \hat{m}_c \aas^{12/25}
      \left[1 + 1.0141\, \aas
      + 1.3892\, \aas^2  
      + 1.0905\, \aas^3   \right]\,,\\
   \mb (\mu^2) &\ =\  \hat{m}_b \aas^{12/23}
      \left[1 + 1.1755\, \aas
      + 1.5007\, \aas^2  
      + 0.1725\, \aas^3 \right]\,.
\end{aligned}
\eeq
The scale dependence of $\mq(\mu^2)$ is given by the scale dependence of $\aas=\aas(\mu^2)$. 
The renormalisation group independent mass parameters $\hat{m}_q$ are determined from the measured 
quark masses at fixed scales (\cf Table~\ref{tab:results}).


\subsection{Electroweak Form Factors}
\label{app:ewformfactors}

The electroweak form factors for lepton or quark flavours $f$, $\rZ{f}$
and $\kZ{f}$, absorbing the radiative corrections, are used in the
Gfitter software for the calculation of the partial and total widths
of the $Z$ boson and of the total width of the $W$ boson. 
The relevant implementations have been integrated
from the ZFITTER package~\cite{Arbuzov:2005ma,Bardin:1999yd} (cf. Footnote~\ref{fn}) 
and are co-authored by both groups~\cite{zfittergfitter}. It includes up to
two-loop electroweak
corrections~\cite{Akhundov:1985fc,Arbuzov:2005ma,Bardin:1986fi,Barbieri:1992dq,Fleischer:1993ub,Bardin:1999yd,Degrassi:1994tf,Degrassi:1995mc,Degrassi:1996mg,Degrassi:1999jd,Bardin:1997xq,Bardin:1999ak}
and all known QCD
corrections~\cite{Arbuzov:2005ma,Bardin:1999yd,Kniehl:1989yc}. In
these calculations the intermediate on-shell mass
scheme~\cite{Bardin:1999yd} is used, which lies between OMS-I and
OMS-II. These latter two schemes are used to estimate the uncertainty 
arising from the renormalisation scheme ambiguity (see~\cite{goebel} for more information).
The form factors in the intermediate scheme are given by
\begin{align}
\label{eq:rho}
\rZ{f} &\ =\ \displaystyle{\frac{1+\drhov^{f,[G]}_{\rm rem}}
            {1+\drhovb^{(G)}\,\left( 1-\Delta{\bar{r}}^{[G]}_{{\rm rem}}\right)} 
               +\drhov^{f,G^2}_{\rm rem}}\,,\\
\label{eq:kappa}
\kZ{f} &\ =\ \left( 1+\dkapv^{f,[G]}_{{\rm rem}}\right)
               \left[
               1-\frac{c_{W}^{2}}{s_{W}^{2}}\drhovb^{(G)}\left( 1-\Delta{\bar{r}}^{[G]}_{{\rm rem}}\right)
               \right]
               +\dkapv^{f,G^2}_{{\rm rem}}\,,
\end{align}
where the superscript ($G$) stands for the inclusion of all known terms, whereas $[G]= G + \as G$ 
includes the electroweak one-loop corrections together with all known orders in the strong coupling 
constant. These QCD corrections are taken from~\cite{Kniehl:1989yc}. The parameter $\drhovb^{(G)}$ 
contains all known corrections to the Veltman parameter, defined by the ratio of effective couplings of 
neutral to charged currents~\cite{Ross:1975fq,Veltman:1977kh}.
The subscript ``rem'' stands for ``remainder''. The correction $\Delta{\bar{r}}^{[G]}_{{\rm rem}}$ 
is given by
\begin{align}
   \Delta{\bar{r}}^{[G]}_{{\rm rem}}\ =\ \Delta{\bar{r}}^{G}_{{\rm rem}} + \Delta r^{G\as}_{{\rm rem}},
\end{align}
with the one-loop remainder
\begin{align}
   \Delta{\bar{r}}^{G}_{{\rm rem}}\ =\ & \frac{\sqrt{2} \GF M_Z^2 \sw \cw}{4\pi^2} \Biggl\{ -\frac{2}{3}
         +\frac{1}{\sw} \biggl( \frac{1}{6} \NCf{f}-\frac{1}{6}-7\cw\biggr)
         \ln{\cw} \nonumber \\ 
         & \hspace{2.9cm}
           +\,\frac{1}{\sw } 
           \biggl[ \Delta\rho^{\sss{F}}_{\sss{W}}+\frac{11}{2}-\frac{5}{8}\cw
                   \left( 1+\cw \right) +\frac{9\cw}{4\sw}\ln{\cw} \biggr] \Biggr\}\,,
   \label{drrembar}
\end{align}
where $\NCf{f=q(\ell)}=3(1)$ is the colour factor, $\sw=\sin^{2}\!\tzw$ 
and $c_{W}^{2}=\cos^{2}\!\tzw$, and where $\Delta\rho^{\sss{F}}_{\sss{W}}$ is given by 
\beq
   \Delta\rho^{F}_{_{W}} \ =\ \frac{1}{M_W^2} \left[\Sigma^{F}_{WW}(0) -\Sigma^{F}_{WW}(M_W^2)\right]\,.
\eeq
The terms $\Sigma^{F}_{WW}(0)$ and $\Sigma^{F}_{WW}(M_W^2)$ are the $W$ boson self energies 
discussed below. 
 
For the purpose of illustration we give the formulae for the one-loop corrections of the 
electroweak form factors at the $Z$ pole for vanishing external fermion 
masses~\cite{Akhundov:1985fc}:
\begin{align}
\label{eq:rho_Z_f}
   \drhov^{f,[G]}_{{\rm rem}}  &\ =\  \frac{\alpha}{4\pi\stws} \left[ \zmz -\delrho{F}_{\sss{\zb}}
                   -\frac{11}{2}+\frac{5}{8} \ctws(1+\ctws)
                   -\frac{9}{4}\frac{\ctws}{\stws} \ln \ctws + 2 u_f\right]\,, \\
\label{eq:kappa_Z_f}
   \dkapv^{f,[G]}_{{\rm rem}} &\ =\  \frac{\alpha}{4\pi\stws}\left[
              - \frac{\ctws}{\stws} \delrho{F}
              + \Pzg^{F}(\mzs)
              + \frac{\stwf}{\ctws}Q_f^2{V}_{1Z}({M}_{Z}^2) - u_f\right]\,,
\end{align}
where 
\begin{align}
   u_f \ =\ & \frac{1}{4\ctws} \left[1-6|Q_f|\stws + 12 Q_f^2\stwf\right]{V}_{1Z}({M}_{Z}^2)
        + \left[\frac{1}{2}-\ctws -~|Q_f|\stws\right]{V}_{1W}({M}_{Z}^2)+\ctws{V}_{2W}({M}_{Z}^2)\,, \nonumber\\
    \delrho{F}_{\sss{\zb}}\ =\ & \frac{1}{\mws} 
                                \left[ \Sigma^{F}_{\sss{\wb\wb}}(0)-\Sigma^{F}_{\sss{\zb\zb}}(\mzs) \right]\,.\nonumber
\end{align} 
The term $\Sigma^{F}_{ZZ}(M_W^2)$ is the $Z$ boson self energy.
The vertex functions in the chiral limit are given by~\cite{Bardin:1997xq}
\begin{align}
\label{eq:V_1V} 
   {V}_{1V}(s) &\ =\  -\frac{7}{2}-2 R_{V}-(3+2R_{V})\ln(-\tilde{R}_{V} )
                      + 2(1+R_{V})^2 \left[ \mbox{Li}_2(1+\tilde{R}_{V})-\mbox{Li}_2(1)\right] \; , \\                      
\label{eq:V_2W} 
   V_{{2W}}(s) & \ =\ -\frac{1}{6}-2R_{W}-\left(\frac{7}{6}+R_{W}\right)
                     \frac{L_{{WW}}(s)}{s}+2R_{W} \left( R_{W} +2 \right){\cal F}_3(s,M_{W}^2) \; ,
\end{align}
where ${\rm Li}_2$ is the dilogarithm function, and where 
$\tilde{R}_{V}=R_{V} -i\gamma_{_V}$, $\gamma_{V} =M_{V} \Gamma_{V}/s$, and 
$R_{V} =M_{V}^2/s$. The $Z$--$\gamma$ mixing function in Eq.~(\ref{eq:kappa_Z_f}) is 
given by
\begin{align}
\label{eq:ssss8}
   \Pzg^{F}(\mzs) &\ =\  2 \sum_f \NCf{f} |Q_f| v_f {\bf I}_3(-s;m_f^2,m_f^2) \; ,
\end{align}
where $v_f = 1 - 4Q_f\sw$, and where the index $f$ runs over all fundamental fermions.
The integrals ${\cal F}_3$ in (\ref{eq:V_2W}) and ${\bf I}_3(Q^2;M_1^2,M_2^2)$ in (\ref{eq:ssss8}) 
are given in Appendices~C and D of Ref.~\cite{Bardin:1989di}.

For the two-loop corrections to the electroweak form factors, $\drhov^{f,G^2}_{\rm rem}$ 
and $\dkapv^{f,G^2}_{{\rm rem}}$ in Eqs.~(\ref{eq:rho}) and (\ref{eq:kappa}), the interested 
reader is referred to the original 
literature~\cite{Bardin:1999yd,Bardin:1997xq,Bardin:1999ak,Degrassi:1994tf,Degrassi:1995mc,Degrassi:1996mg,Degrassi:1999jd}.
Because of missing two-loop corrections to the form factors $\rho_{Z}^{b}$ and $\kappa_{Z}^{b}$ 
occurring in $Z \to b\bbar$, an approximate expression is used, which includes the full 
one-loop correction and the known leading two-loop terms $\propto \mt^4$. Non-universal top 
contributions~\cite{Bardin:1999yd,Barbieri:1992dq,Fleischer:1993ub} must be taken into account 
in this channel due to a CKM factor close to one and the large mass difference of 
bottom and top quarks
\begin{align}
\label{eq:taub}
   \tau_{b} & \ =\  -2 x_t\left[1-\frac{\pi}{3}\alpha_{s}(m^2_t) 
                        + x_t \,\tau^{(2)}\!\!\left(\frac{m^2_t}{M_H^2}\right) \right]\, , 
\end{align}
where $x_t=G_F m_t^2/(8\pi^2\sqrt{2})$ and 
the function $\tau^{(2)}(m^2_t/M_Z^2)$ is given in~\cite{Barbieri:1992dq}.
Since the first term in Eq.~(\ref{eq:taub}) represents one-loop corrections, it must be subtracted 
from the universal form factors to avoid double-counting. Let $\rho_b^\prime$ and $\kappa_b^\prime$ 
be these corrected form factors (\cf Refs.~\cite{Bardin:1999yd,Barbieri:1992dq,Fleischer:1993ub} 
for the correction procedure), the form factors beyond one-loop are obtained by
$\rho_{b}=\rho_b^\prime \left(1 + \tau_{b} \right)^{2}$ and
$\kappa_{b}=\kappa_b^\prime (1 + \tau_{b})^{-1}$.

\subsubsection{Self-Energies of  $\mathsf W$ and $\mathsf Z$ Boson}

The $W$ and $Z$ boson self-energies ${\sum}_{\sss{WW}}^{F}$ and ${\sum}_{\sss{ZZ}}^{F}$ and 
on-shell derivative ${\sum}_{\sss{ZZ}}^{\prime F}$ are the sums of bosonic and fermionic parts. 
The bosonic parts read~\cite{Bardin:1997xq,Bardin:1981sv}
\begin{align}
   \frac{\sum_{\sss{WW}}^{{\rm Bos},F}(0)}{M_W^2} \ =\ &  
      \frac{5\ctws(1+\ctws)}{8}  - \frac{17}{4} + \frac{5}{8\ctws} -\frac{r_W}{8}
      +\left( \frac{9}{4} +\frac{3}{4\ctws} - \frac{3}{\stws}\right)\ln \ctws 
      + \frac{3r_W}{4(1-r_W)} \ln r_W\, , \\
\frac{\sum_{\sss{WW}}^{{\rm Bos},F}(M_W^2)}{M_W^2} \ =\ & - \frac{157}{9} + \frac{23}{12\ctws} + \frac{1}{12\ctwf}
           - \frac{r_W}{2} +\frac{r_W^2}{12}
     +\frac{1}{\ctws}\left( - \frac{7}{2} + \frac{7}{12\ctws}
           + \frac{1}{24\ctwf}\right) \ln \ctws
&\nonumber    \\
& +\, r_W \left( -\frac{3}{4} + \frac{r_W}{4} - \frac{r_W^2}{24} \right) \ln r_W
    +\left( \frac{1}{2}   - \frac{r_W}{6} + \frac{r_W^2}{24} \right)
            \frac{L_{WH}(M^2_{W})}{M^2_{W}}
&\nonumber  \\
& +\, \left(- 2\ctws - \frac{17}{6} + \frac{2}{3\ctws} + \frac{1}{24\ctwf} \right)
   \frac{ L_{WZ}(M^2_{W})}{M^2_{W}} \,, & \\
\frac{\sum_{\sss{ZZ}}^{{\rm Bos},F}(M_Z^2)}{M_W^2} \ =\ &
-8\ctwf -\frac{34\ctws}{3} + \frac{35}{18}\left(1 + \frac{1}{\ctws} \right)
 -\frac{r_W}{2} + \frac{r_Z^2}{12\ctws} + r_W \left( -\frac{3}{4} + \frac{r_Z}{4}
 - \frac{r_Z^2}{24} \right) \ln r_Z
&\nonumber  \\
&  +\, \frac{5\ln \ctws}{6\ctws} +\left( \frac{1}{2}
-\frac{r_Z}{6}+ \frac{r_Z^2}{24} \right) \frac{L_{ZH}(M^2_{Z})}{M_{W}^2}
&\nonumber  \\
& +\,\left(-2\ctwsix -\frac{17}{6}\ctwf + \frac{2}{3}\ctws   + \frac{1}{24} \right)
    \frac{{L}_{WW}(M^2_{Z})}{M^2_{W}} \; ,& \\
{\sum}_{\sss{ZZ}}^{\prime{\rm Bos},F}(M_Z^2) \ =\ &
   -4\ctwf+ \frac{17\ctws}{3}-\frac{23}{9}+\frac{5}{18\ctws}-\frac{r_W}{2}
   +\frac{r_Wr_Z}{6}
   +r_W\left(-\frac{3}{4}+\frac{3r_Z}{8}-\frac{r_Z^2}{12} \right) \ln r_Z
& \nonumber  \\
&-\,\frac{1}{12\ctws}\ln \ctws + \frac{\ln r_Z}{2\ctws}
   +\left(-\ctwsix+\frac{7\ctwf}{6}-\frac{17\ctws}{12}-\frac{1}{8}
   \right)\frac{L_{{WW}}(M_Z^2)}{M_W^2}
& \nonumber \\
&+\,\left(\frac{1}{2}-\frac{5r_Z}{24}+\frac{r_Z^2}{12}+\frac{1}{2(r_Z-4)}
   \right) \frac{ L_{ZH} (M_Z^2) } {{M_W}^2} \; ,
\end{align}
where the shorthand notation $r_W=M_H^2/\mws$ and $r_Z=M_H^2/\mzs$ has been used. The function 
$L_{V_1V_2}(s) \equiv L(-s;M_{V_1}^2,M_{V_2}^2)$ is defined in Eq.~(2.14) of Ref.~\cite{Bardin:1980fe}. 

The fermionic parts read~\cite{Bardin:1997xq,Bardin:1981sv}
\begin{align}
\label{eq:fermparts1}
   \frac{\sum_{\sss{WW}}^{{\rm Fer},F}(M_W^2)}{M_W^2} \ =\ & \!\!\sum_{f=f_u,f_d} \!\!\NCf{f}
      \left[ -\frac{2s}{M_{W}^2} {\bf I}_3(\dots)
      + \frac{m_{f_u}^2}{M_{W}^2} {\bf I}_1(\dots)
      \right. \left. +\frac{m_{f_d}^2}{M_{W}^2} {\bf I}_1(-s;m_{f_d}^2,m_{f_u}^2) \right], &  \\
   \frac{\sum_{\sss{ZZ}}^{{\rm Fer},F}(M_Z^2)}{M_W^2} \ =\ & \frac{1}{2\ctws} \sum_{f} \NCf{f}
       \left[ -\frac{s}{M_{Z}^2} \left(1+v_f^{2} \right)
       {\bf I}_3(-s;m_f^2,m_f^2) + \frac{m_f^2}{M_{Z}^2}
       {\bf I}_0(-s;m_f^2,m_f^2) \right] ,& \\
   {\sum}_{\sss{ZZ}}^{\prime{\rm Fer},F}(M_Z^2) \ =\ & -\sum_f \NCf{f}
      \left\{ \frac{r_f}{2} \left[ 1- r_f M_W^2 {\cal F}(-M_Z^2,m_f^2,m_f^2) \right]
      +\frac{1}{6\ctws} \left(1+v_f^2\right) \right. & \\
   &  \left. \hspace{1.0cm}
      \times\left[ \frac{1}{2}\ln(r_f\ctws) + r_f\ctws + (-\frac{1}{4\ctws} + \frac{r_f}{2}
      -r_f^2\ctws) M_W^2 {\cal F}(-M_Z^2,m_f^2,m_f^2) \right] \right\} ,\nonumber  &   
\end{align}
with $r_f=m_f^2 / M_W^2$ and $v_f = 1 - 4Q_f\sw$ (recalled from above), and where $(\dots)$ in 
Eq.~(\ref{eq:fermparts1}) stands for $(-s;m_{f_u}^2,m_{f_d}^2)$. The sums are taken over 
all fundamental up-type and down-type fermions of all $SU(2) \otimes U(1)$ doublets with masses 
$m_{f_u}$ and $m_{f_d}$, respectively. The integrals ${\bf I}_n(Q^2;M_1^2,M_2^2)$ and 
${\cal F}$ are given in Appendix~D of Ref.~\cite{Bardin:1989di}.

\subsection{Radiator Functions}
\label{app:radfunctions}

The radiator functions $R^{q}_{\sss{V}}(s)$ and $R^{q}_{\sss{A}}(s)$ absorb the final state QED 
and QCD corrections to the vector and axial-vector currents in hadronic $Z$ decays. They
also contain mixed ${\rm QED}\otimes{\rm QCD}$ corrections and finite quark-mass corrections 
expressed in terms of running masses. 
The following formulae as implemented in the Gfitter subpackage GSM are taken
from~\cite{Bardin:1999ak} and the ZFITTER package~\cite{Arbuzov:2005ma,Bardin:1999yd} 
(cf. Footnote~\ref{fn}). They have been updated to take into account results from the 
recent \NNNLO calculation of the massless QCD Adler function~\cite{Baikov:2008jh} 
(represented by the coefficient $C_{04}$).
\begin{flalign}
R^{q}_{\sss{V}}(s)= 1 & + \frac{3}{4} Q^2_q \frac{\alpha(s)}{\pi}
              +\aas(s)
              -\frac{1}{4}Q^2_q\frac{\alpha(s)}{\pi}\aas(s)
&\nonumber \\
        &   + \left[C_{02}+C^t_2\left(\frac{s}{\mts}\right)\right]
                \aas^2(s)         
              + C_{03}\aas^3(s)
              + C_{04}\aas^4(s)              
&\nonumber\\
        &   +\, \delta_{C05}\aas^5(s) 
            + \frac{\mcS(s)+\mbS(s)}{s} C_{23}
                                \aas^3(s)
&\nonumber\\
        &   +\, \frac{\mqS(s)}{s} \left[ 
                        C^V_{21}      \aas(s)
                      + C^V_{22}\aas^2(s)
                      + C^V_{23}\aas^3(s)
                                  \right]
&\nonumber\\
    &+\, \frac{\mcQ(s)}{s^2}\left[ C_{42}-\ln\frac{\mcS(s)}{s}\right]
                                \aas^2(s) 
      +\frac{\mbQ(s)}{s^2}\left[ C_{42}-\ln\frac{\mbS(s)}{s}\right]
                                \aas^2(s) 
&\nonumber\\
    &   +\, \frac{\mqQ(s)}{s^2} \left\{
                       C^V_{41}       \aas(s)
                 + \left[C^V_{42}+C^{V,L}_{42}\ln\frac{\mqS(s)}{s}\right]
                                \aas^2(s)
                                    \right\}
&\nonumber\\
    &   +\, 12\frac{\mqpQ(s)}{s^2}
               \aas^2(s)
              -\frac{\mqX(s)}{s^3}\left\{8+\frac{16}{27}
               \left[155+6\ln\frac{\mqS(s)}{s}\right]
                                      \aas(s)\right\}\,,&
\label{eq:rvfact}
\end{flalign}

\begin{flalign}
R^{q}_{\sss{A}}(s) = 1& + \frac{3}{4} Q^2_q \frac{\alpha(s)}{\pi}
             + \aas(s)
             - \frac{1}{4}Q^2_q\frac{\alpha(s)}{\pi}\aas(s)
&\nonumber\\
         & + \left[C_{02}+C^t_2\left(\frac{s}{\mts}\right)
            - \left(2I_{3}^q\right)
              {\cal I}^{(2)}\left(\frac{s}{\mts}\right)\right]
              \aas^2(s)         
&\nonumber\\
        & + \left[C_{03} - 
              \left(2I_{3}^q\right) 
              {\cal I}^{(3)}\left(\frac{s}{\mts}\right)\right] 
              \aas^3(s) 
              + \left[ C_{04} - \left(2I_{3}^q\right)\delta_{{\cal I}^{(4)}} \right] 
              \aas^4(s)
&\nonumber\\
       &    +\, \delta_{C05}\aas^5(s) 
            + \frac{\mcS(s)+\mbS(s)}{s} C_{23}
                                \aas^3(s)
&\nonumber\\
       & +\, \frac{\mqS(s)}{s} \left[C^A_{20} + C^A_{21}\aas(s)  +\, C^A_{22}\aas^2(s)
                      + 6\left(3 + \ln\frac{\mts}{s}\right)
                                \aas^2(s) 
                      + C^A_{23}\aas^3(s)
                                \right]
&\nonumber\\
       & -\, 10\frac{\mqS(s)}{\mts}
              \left[\frac{8}{81}+\frac{1}{54}\ln\frac{\mts}{s}\right]
                                \aas^2(s)
&\nonumber\\
       & +\, \frac{\mcQ(s)}{s^2}\left[ C_{42}-\ln\frac{\mcS(s)}{s} \right]
                                \aas^2(s)
         + \frac{\mbQ(s)}{s^2}\left[ C_{42}-\ln\frac{\mbS(s)}{s} \right]
                                \aas^2(s)
&\nonumber\\
        &   +\, \frac{\mqQ(s)}{s^2} \left\{C^A_{40} 
              + C^A_{41}       \aas(s)
              + \left[C^A_{42}+C^{A,L}_{42}\ln\frac{\mqS(s)}{s}\right]
                                \aas^2(s)
                                    \right\}
&\nonumber\\
       &   -\,12\frac{\mqpQ(s)}{s^2}
                                \aas^2(s)\,, &  
\label{eq:rafact}
\end{flalign}
where the finite quark-mass corrections are retained for charm and bottom quarks only, \ie, all 
lighter quarks are taken to be massless. This restricts the validity of the above formula to energies 
well above the strange-pair and below the top-pair production thresholds, which is sufficient for our use.
The mass $\mqP$ denotes the other quark mass, \ie, it is $\mb$ if $q=c$ and $\mc$ if $q=b$.
The running of the quark masses is computed in the \MSb scheme according to Eq.~(\ref{evolution}). 
The two parameters $\delta_{{\cal I}^{(4)}}$ and $\delta_{C_{05}}$ represent the next unknown 
coefficients in the perturbative expansion. They are treated as theoretical errors within
the \Rfit scheme, and vary within the bounds obtained when assuming a geometric growth
of the perturbative coefficients with the perturbative order, \ie, for a coefficient $H$ one has
$\delta_{H_n} = (H_{n-1}/H_{n-2})\cdot H_{n-1}$. 

The expressions for the fixed-order perturbative 
coefficients $C_{ij}^{(V/A)}$ in Eqs.~(\ref{eq:rvfact}) and (\ref{eq:rafact}) are given below.

{\em Massless non-singlet corrections~\cite{Baikov:2008jh,Chetyrkin:1979bj,Dine:1979qh,Celmaster:1979xr,Gorishnii:1991hw}:}
\begin{flalign}
C_{02} \ =\ &\frac{365}{24}-11\,\ztri
                         +\left(-\frac{11}{12}+\frac{2}{3}\ztri\right)\nf\,,
&  \\
C_{03} \ =\ &\frac{87029}{288}-\frac{121}{8}\ztwo
                         -\frac{1103}{4}\ztri +\frac{275}{6}\zfiv \nonumber
& \\
        &+\,\left(-\frac{7847}{216}+\frac{11}{6}\ztwo
          +\frac{262}{9}\ztri-\frac{25}{9}\zfiv\right)\nf
         +\left(\frac{151}{162}-\frac{1}{18}\ztwo
          -\frac{19}{27}\ztri\right)\nf^2\,,
& \\
C_{04} \ =\ & -156.61 + 18.77\, \nf -0.7974\, \nf^2 + 0.0215\, \nf^3\,,&
\label{mass0}
\end{flalign}
which for $\nf=5$ take the values
$C_{02} = 1.40923$,
$C_{03} = -12.7671$ and
$C_{04} = -80.0075$, exhibiting satisfactory convergence given that $\asZ/\pi\simeq0.04$.

\noindent 
{\em Quadratic massive corrections~\cite{Chetyrkin:1994js}:}
\begin{flalign}
C_{23} \ =\ &-80+60\,\ztri+\left[\frac{32}{9}-\frac{8}{3}\ztri\right]\nf\,,
&  \\
C^V_{21}\ =\ &12\,,
&  \\
C^V_{22} \ =\ &\frac{253}{2} - \frac{13}{3}\nf\,,
&  \\
C^V_{23} \ =\ & 
2522-\frac{855}{2}\ztwo+\frac{310}{3}\ztri-\frac{5225}{6}\zfiv
&\nonumber \\ 
         &+\,\left[-\frac{4942}{27}+34\,\ztwo
           -\frac{394}{27}\ztri+\frac{1045}{27}\zfiv\right]\nf
            +\left[\frac{125}{54}-\frac{2}{3}\ztwo\right]\nf^2\,,
&  \\
C^A_{20} \ =\ &-6\,,
&  \\
C^A_{21} \ =\ &-22\,,
&  \\
C^A_{22} \ =\ &-\frac{8221}{24}+57\,\ztwo+117\,\ztri
           +\left[\frac{151}{12}-2\,\ztwo-4\,\ztri\right]\nf\,,
&  \\
C^A_{23} \ =\ &-\frac{4544045}{864}+1340\,\ztwo+\frac{118915}{36}\ztri
                                -127\,\zfiv
&\nonumber \\ 
         &+\,\left[\frac{71621}{162}-\frac{209}{2}\ztwo-216\,\ztri
                                +5\,\zfor+55\,\zfiv\right]\nf
&\nonumber \\ 
         &+\,\left[-\frac{13171}{1944}+\frac{16}{9}\ztwo
                                +\frac{26}{9}\ztri\right]\nf^2\,.&
\label{mass2}
\end{flalign}
\noindent 
{\em Quartic massive corrections~\cite{Chetyrkin:1994js}:}
\begin{flalign}
C_{42} \ =\ &\frac{13}{3}-4\,\ztri\,,
&  \\
C^V_{40} \ =\ &-6\,,
&  \\ 
C^V_{41} \ =\ &-22\,,
& \\
C^V_{42} \ =\ &-\frac{3029}{12}+162\,\ztwo+112\,\ztri
           +\left[\frac{143}{18}-4\,\ztwo-\frac{8}{3}\ztri\right]\nf\,,
&  \\
C^{V,L}_{42} \ =\ &-\frac{11}{2}+\frac{1}{3}\nf\,,
& \\
C^A_{40} \ =\ &6\,,
&  \\
C^A_{41} \ =\ &10\,,
&  \\
C^A_{42} \ =\ &\frac{3389}{12}-162\,\ztwo-220\,\ztri
           +\left[-\frac{41}{6}+4\,\ztwo+\frac{16}{3}\ztri\right]\nf\,,
& \\
C^{A,L}_{42} \ =\ &\frac{77}{2}-\frac{7}{3}\nf\,. & 
\label{mass4}
\end{flalign}
\noindent
{\em Power suppressed top-mass correction~\cite{Chetyrkin:1994js}:}
\begin{flalign}
C^t_2(x) \ =\ &x\left(\frac{44}{675} - \frac{2}{135}\ln x \right)\,.&
\label{powsup}
\end{flalign}
\noindent 
{\em Singlet axial-vector corrections~\cite{Chetyrkin:1994js}:}
\begin{flalign}
{\cal I}^{(2)}(x) \ =\ &-\frac{37}{12} + \ln x + \frac{7}{81}x
                       + \mbox{$0.0132$}x^2\,,
&  \\   
{\cal I}^{(3)}(x) \ =\ &-\frac{5075}{216} 
                      + \frac {23}{6}\ztwo + \ztri + \frac{67}{18}\ln x
                      + \frac{23}{12} \ln^2\!x\,.&
\end{flalign}
\noindent 
{\em Singlet vector correction~\cite{Chetyrkin:1994js}:}
\begin{flalign}
R^h_{\sss{V}}(s) \ =\ & 
\left(\sum_f v_f\right)^{\!\!\!2}\left(-0.41317\right)
               \aas^3(s)\,.&
\label{singlet}
\end{flalign} 

\end{appendix}

%
%
\newpage

\addcontentsline{toc}{section}{References}
\bibliography{References}{}

\begin{thebibliography}{100}
\bibitem{GfitterSW}
The Gfitter code is available at the CERN CVS repository. Contact the authors
  for a copy.
\end{thebibliography}

\end{document}